\documentclass[
superscriptaddress,
showpacs,
preprintnumbers,
nofootinbib,
amsmath,amssymb,
aps,
prd,
floatfix,
twocolumn,
showkeys,
reprint,
10pt,
]{revtex4-1}
\usepackage{color}
\usepackage{amssymb}   
\usepackage{graphicx}
\usepackage{dcolumn}
\usepackage{bm}
\usepackage{hyperref}
\usepackage{xspace}
 \usepackage[T1]{fontenc}
\usepackage{ae,aecompl}
\usepackage{extarrows}
\graphicspath{{./figs/}}

\newcommand{\REM}[1]{}

\newcommand{\Fermi}{\emph{Fermi}\xspace}
\newcommand{\eqg}{E_{\rm QG}}
\newcommand{\eqgl}{E_{\rm QG,1}}
\newcommand{\eqgq}{E_{\rm QG,2}}
\newcommand{\epl}{E_{\rm Pl}}

\newcommand{\Err}{\mathcal{E}}
\newcommand{\err}{\epsilon}
\newcommand{\tint}{\tau_{\rm int}}
\newcommand{\tildetint}{\tilde{\tau}_{\rm int}}
\newcommand{\tliv}{\tau_{\rm LIV}}
\newcommand{\ttot}{\tau_n}
\newcommand{\spm}{\mathit{s_\pm}}
\newcommand{\tne}{\hat{\tau}_n}

\newcommand{\ptint}{P_{\rm \tau_{int}}}

\begin{document}

\title{Constraints on Lorentz Invariance Violation from \Fermi-Large
Area Telescope Observations of Gamma-Ray Bursts}

\author{V.~Vasileiou}
\email{vlasisva@gmail.com}
\affiliation{Laboratoire Univers et Particules de Montpellier, Universit\'e Montpellier 2, CNRS/IN2P3,  CC 72, Place Eug\`ene Bataillon, F-34095 Montpellier Cedex 5, France}
\author{A.~Jacholkowska}
\email{agnieszka.jacholkowska@cern.ch}
\affiliation{LPNHE, Universit\'e Pierre et Marie Curie Paris 6, Universit\'e Denis Diderot Paris 7, CNRS/IN2P3, 4 Place Jussieu, F-75252, Paris Cedex 5, France}
\author{F.~Piron}
\affiliation{Laboratoire Univers et Particules de Montpellier, Universit\'e Montpellier 2, CNRS/IN2P3,  CC 72, Place Eug\`ene Bataillon, F-34095 Montpellier Cedex 5, France}
\author{J.~Bolmont}
\affiliation{LPNHE, Universit\'e Pierre et Marie Curie Paris 6, Universit\'e Denis Diderot Paris 7, CNRS/IN2P3, 4 Place Jussieu, F-75252, Paris Cedex 5, France}
\author{C.~Couturier}
\affiliation{LPNHE, Universit\'e Pierre et Marie Curie Paris 6, Universit\'e Denis Diderot Paris 7, CNRS/IN2P3, 4 Place Jussieu, F-75252, Paris Cedex 5, France}
\author{J.~Granot}
\affiliation{Department of Natural Sciences, The Open University of Israel, 1 University Road, POB 808, Ra'anana 43537, Israel}
\author{F.~W.~Stecker}
\affiliation{Astrophysics Science Division, NASA/Goddard  Space   Flight  Center, Greenbelt,     MD     20771, U.S.A.}
\affiliation{Department of Physics  and Astronomy, University of California, Los  Angeles,  CA 90095-1547, U.S.A.}
\author{J.~Cohen-Tanugi}
\affiliation{Laboratoire Univers et Particules de Montpellier, Universit\'e Montpellier 2, CNRS/IN2P3,  CC 72, Place Eug\`ene Bataillon, F-34095 Montpellier Cedex 5, France}
\author{F.~Longo}
\affiliation{Istituto Nazionale di Fisica Nucleare, Sezione di Trieste, I-34127 Trieste, Italy}
\affiliation{Dipartimento di Fisica, Universit\`a di Trieste, I-34127 Trieste, Italy}


\begin{abstract}
We analyze the MeV/GeV emission from four bright Gamma-Ray Bursts (GRBs) observed by the \textit{Fermi}-Large Area Telescope to produce robust, stringent constraints on a dependence of the speed of light \textit{in vacuo} on the photon energy (vacuum dispersion), a form of Lorentz invariance violation (LIV) allowed by some Quantum Gravity (QG) theories. First, we use three different and complementary techniques to constrain the total degree of dispersion observed in the data. Additionally, using a maximally conservative set of assumptions on possible source-intrinsic spectral-evolution effects, we constrain any vacuum dispersion solely attributed to LIV. We then derive limits on the ``QG energy scale'' (the energy scale that LIV-inducing QG effects become important, $E_{\rm QG}$) and the coefficients of the Standard Model Extension. For the subluminal case (where high energy photons propagate more slowly than lower energy photons) and without taking into account any source-intrinsic dispersion, our most stringent limits (at 95\% CL) are obtained from GRB~090510 and are $\eqgl>7.6$ times the Planck energy ($\epl$) and $\eqgq>1.3\times10^{11}$~GeV for linear and quadratic leading order LIV-induced vacuum dispersion, respectively. These limits improve the latest constraints by \textit{Fermi} and H.E.S.S. by a factor of $\sim2$. Our results disfavor any class of models requiring $\eqgl\lesssim\epl$.
\end{abstract}

\maketitle

\section{Introduction}

While general relativity and Quantum Field Theory have each enjoyed impressive success so far, their formulations are currently inconsistent, hence motivating searches for unification schemes that can collectively be subsumed under the name of Quantum Gravity (QG) theories. These theories generally predict the existence of a natural scale at which the physics of space-time, as predicted by relativity theory, is expected to break down, hence requiring modifications or the creation of a new paradigm to avoid singularity problems. This scale, referred to as the ``Quantum Gravity energy scale'' $E_{\rm QG}$, is in general expected to be of the order of the Planck scale~\citep{Planck1899}, $\epl\equiv \sqrt{(\hbar c^5)/G} \simeq 1.22 \times 10^{19}$~GeV, or in some cases lower (e.g., for some QG scenarios such as loop quantum gravity).

Since relativity precludes an invariant length, the introduction of such a constant scale violates Lorentz Invariance (LI). Thus, tests of LI are strongly motivated by the search for a theory of QG. Additional motivations for testing LI are the need to cut off high-energy (ultraviolet) divergences in quantum field theory and the need for a consistent theory of black holes~\cite{2011LRR....14....8S,2008LRR....11....5R}.


The idea that LI may be only approximate has been explored within the context of a wide variety of suggested Planck-scale physics scenarios. These
include the concepts of deformed relativity, loop quantum gravity, non-commutative geometry, spin foam models, and some string theory (M theory) models (for reviews see, e.g., Refs.~\citep{Smolin2001, Mattingly:05, 2006AnPhy.321..150J}). These theoretical explorations and their possible consequences, such as observable modifications in the energy-momentum dispersion relations for free particles and photons, have been discussed under the general heading of ``Planck scale phenomenology''.

There is also the motivation of testing LI in order to extent or limit its domain of applicability to the highest observable energies. Since the group of pure LI transformations is unbounded at high energies, one should look for its breakdown at high energy scales, possibly through effects of Planck scale physics but perhaps through the effects of other unknown phenomena. To accomplish such a program, tests of the kinematics of LI violation (LIV) within the context of physical interaction dynamics such as quantum electrodynamics or standard model physics (e.g.,
\cite{1999PhRvD..59k6008C,2004PhRvL..93b1101J,2009NJPh...11h5003S,2011APh....35...95S,2011PhRvD..83e6012A}) have been proposed. Fruitful frameworks for this kind of analysis, useful for testing the effects of LIV at energies well below the Planck scale, are the Taylor series expansion originally proposed in the seminal paper by Amelino-Camelia et al.~\cite{1998Natur.393..768} and the more comprehensive Standard  Model Extension (SME) parametrization of
Kosteleck\'{y} and collaborators~\cite{1997PhRvD..55.6760C,2009PhRvD..80a5020K}. These phenomenological parameterizations can be viewed as low-energy effective field theories, holding at energies $E \ll \epl$ and providing an effective framework to search for LIV at energies far below the Planck scale.

One manifestation of LIV is the existence of an energy-dependent ``maximum attainable velocity'' of a particle and its effect on the thresholds for various particle interactions, particle decays, and neutrino oscillations~\cite{1999PhRvD..59k6008C}. Assuming that the mass of the photon is zero, its maximum attainable velocity can be determined  by measuring its velocity at the highest possible observable energy. This energy is, of necessity, in the gamma-ray range. Since we know that LI is accurate at accelerator energies, and even at cosmic-ray energies~\cite{2001APh....16...97S}, any deviation of the velocity of a photon from its low energy value, $c$, must be very small at these energies. Thus, a sensitive test of LI requires high-energy photons (i.e., gamma rays) and entails searching for dependencies of the speed of light \textit{in vacuo} on the photon energy. The method used in this study to search for such an energy dependence consists in comparing the time of flight between photons of different energies emitted together by a distant astrophysical source. As will be shown in the next section, the magnitude of a LIV-induced difference on the time of flight is predicted to be an increasing function of the photon energy and the distance of source. Thus, because of the high-energy extend of their emission (up to tens of GeV), their large distances (redshifts up to a value of $\sim$8), and their rapid (down to ms scale) variabilities, Gamma-Ray Bursts (GRBs) are very effective probes for searching for such LIV-induced spectral dispersions~\citep{1998Natur.393..768}.


There have been several searches for LIV applying a variety of analysis techniques on GRB observations. Some of the pre-\textit{Fermi} studies are those by Lamon et al.~\cite{LamonIntegral} using INTEGRAL GRBs; by Bolmont et al.~\cite{2008ApJ...676..532B} using HETE-2 GRBs; by Ellis et al.~\cite{2006APh....25..402E} using HETE, BATSE, and Swift GRBs; and by Rodr\'iguez-Mart\'inez et al. using Swift and Konus-Wind observations of GRB~051221A~\cite{2006JCAP...05..017R}. The most stringent constraints, however, have been placed using $\Fermi$ observations, mainly thanks to the unprecedented sensitivity for detecting the prompt MeV/GeV GRB emission by the $\Fermi$ Large Area Telescope (LAT)~\cite{Piron2012,2009ApJ...697.1071A}. These constraints include those by the \Fermi LAT and Gamma-Ray Burst Monitor (GBM) collaborations using GRBs 080916C~\cite{2009Sci...323.1688A} and 090510~\cite{2009Natur.462..331A}, and by Shao et al.~\cite{2010APh....33..312S} and Nemiroff et al.~\cite{2012PhRvL.108w1103N} using multiple \textit{Fermi} GRBs. In addition to these GRB-based studies, there have been some results using TeV observations of bright
Active Galactic Nuclei flares, including the MAGIC analysis of the flares of Mrk~501~\cite{Magic_PKS_2008PhLB..668..253M, Martinez:08} and the H.E.S.S. analysis on the exceptional flare of PKS~2155-304~\cite{2008PhRvL.101q0402A,hesslike}. It should be noted that the studies above did not assume any dependence of LIV on the polarization of the photons, manifesting as birefringence. In the case that such a dependence exists, constraints on LIV effects can be produced~\cite{2012PhRvL.109x1104T,2011PhRvD..83l1301L,2011APh....35...95S} that are 13 orders of magnitude stronger than the dispersion-only constraints placed with time-of-flight considerations (as in this work). It should be added that there is a class of theories that allow for photon dispersion without birefringence that can be directly constrained by our results (e.g.,~\cite{2008PhLB..665..412E}).

The aim of this study is to produce a robust and competitive constraint on the dependence of the velocity of light \textit{in vacuo} on its energy. Our analysis is performed on a selection of \Fermi-LAT~\cite{2009ApJ...697.1071A} GRBs with measured redshifts and bright GeV emission. We first apply three different analysis techniques to constrain the total degree of spectral dispersion observed in the data. Then, using a set of maximally conservative assumptions on the possible source-intrinsic spectral evolution (which can masquerade as LIV dispersion), we produce constraints on the degree of LIV-induced spectral dispersion. The latter constraints are weaker than those on the total degree of dispersion, yet considerably more robust with respect to the presence of a source-intrinsic effects. Finally, we convert our constraints to limits on LIV-model-specific quantities, such as $\eqg$ and the coefficients of the SME.

The first method used to constrain the degree of dispersion in the data, named ``PairView'' (PV), is created as part of this study, and performs a statistical analysis on all the pairs of photons in the data to find a common spectral lag. The second method, named ``Sharpness-Maximization Method'' (SMM), is a modification of existing techniques (e.g., DisCan~\cite{0004-637X-673-2-972}) and is based on the fact that any spectral dispersion will smear the structure of the light curve, reducing its sharpness. SMM's best estimate is equal to the negative of a trial degree of dispersion that, when applied to the actual light curve, restores its assumed-as-initially-maximal sharpness. Finally, the third method employs an unbinned maximum likelihood (ML) analysis to compare the data as observed at energies low enough for the LIV delays to be negligible to the data at higher energies. The three methods were tested using an extensive set of simulations and cross-checks, described in several appendices.


Our constraints apply only to classes of LIV models that possess the following properties. First, the magnitude of the LIV-induced time delay depends either linearly or quadratically on the photon energy. Second, this dependence is deterministic, i.e., the degree of LIV-induced increase or decrease in the photon propagation speed does not have a stochastic (or ``fuzzy'') nature as postulated by some of the LI models~(see Ref.~\cite{2009PhRvD..80h4017A} and references therein). Finally, the sign of the effect does not depend on the photon polarization -- the velocities of all photons of the same energy are \textit{either} increased \textit{or} decreased due to LIV by the same exact amount.

In Sec.~\ref{ch_formalism} we describe the LIV formalism and notation used in the paper, in Sec.~\ref{sec:Data} we describe the data sets used for the analysis, in Sec.~\ref{sec:Methods} we describe the three analysis methods and the procedure we used to take into account      possible intrinsic spectral-evolution effects, in Sec.~\ref{sec:Results} we present the results, in Sec.~\ref{sec:systematics} we report and discuss their associated systematic uncertainties and caveats, and finally in Sec.~\ref{sec:Conclusion} we compare our results to previous measurements and discuss their physical implications. We present our Monte Carlo simulations used for verifying PV and SMM in Appendix~\ref{appendix:PVSMM}, the calibrations and verification tests of the ML method in Appendix~\ref{appendix:likelihood}, a direct one-to-one comparison of the results of the three methods after their application on a common set of simulated data in Appendix~\ref{appendix:comparison}, and cross-checks of the results in Appendix~\ref{appendix:Checks}.

\section{Formalism}
\label{ch_formalism}

In QG scenarios, the LIV-induced modifications to the photon dispersion relation can be described using a series expansion in the form
\begin{equation}
\label{dispersion}
\ensuremath{E^{2}\simeq p^{2}c^{2}\times\left[1 -\overset{\infty}{\underset{n=1}{\sum}}\mathit{s_{\pm}}\left(\frac{E}{\eqg}\right)^{n}\right]},
\end{equation}
where $c$ is the constant speed of light (at the limit of zero photon energy), $\mathit{s_\pm}$ is the ``sign of LIV'', a theory-dependent factor equal to $+$1 ($-$1) for a decrease (increase) in photon speed with an increasing photon energy (also referred to as the ``subluminal'' and ``superluminal'' cases). For $E \ll \eqg$, the lowest order term in the series not suppressed by theory is expected to dominate the sum. In case the $n=1$ term is suppressed, something that can happen if a symmetry law is involved, the next term $n = 2$ will dominate. We note that in effective field theory $n=d-4$, where $d$ is the mass dimension of the dominant operator.
Therefore, the $n = 1$ term arises from a dimension 5 operator~\cite{2003PhRvL..90u1601M}. It has been shown that odd mass-dimension terms violate $\cal{CPT}$~\cite{1997PhRvD..55.6760C,1998PhRvD..58k6002C}. Thus, {\it if} $\cal{CPT}$ is preserved, {\it then} the $n = 2$ term is expected to dominate.  In this study, we only consider the $n=1$ and $n=2$ cases, since the \Fermi results are not sensitive to higher order terms.

Using Eq.~\ref{dispersion} and keeping only the lowest-order dominant term, it can be found that the photon propagation speed $u_{\rm ph}$, given by its group velocity, is
\begin{equation}
\label{eq:uph}
u_{\rm ph}(E)=\frac{\partial E}{\partial p}\simeq c \times\left[1-\mathit{s_{\pm}}\frac{n+1}{2}\left(\frac{E}{\eqg}\right)^{n}\right],
\end{equation}
where $c\equiv \underset{E\rightarrow 0}{\rm lim}u_{\rm ph}(E)$. Because of the energy dependence of $u_{\rm ph}(E)$, two photons of different energies $E_h>E_l$ emitted by a distant source at the same time and from the same location will arrive on Earth with a time delay
$\Delta t$ which depends on their energies. We define the ``LIV parameter'' $\ttot$ as the ratio of this delay over $E_h^n-E_l^n$~\cite{2008JCAP...01..031J}:
\begin{equation}
\label{deltatn}
\ttot 	\equiv \frac{\Delta t}{E_{h}^n - E_{l}^n} \simeq \mathit{s_\pm} \frac{(1+n)}{2H_0}\frac{1}{\eqg^n}\times \kappa_n,
\end{equation}
where
\begin{equation}
\label{eq:kappa}
 \kappa_n\equiv\int\limits_0^z\frac{(1+z')^n}{\sqrt{\Omega_{\rm\Lambda} + \Omega_{\rm M} (1 + z')^3}}dz'
\end{equation}
is a comoving distance that also depends on the order of LIV ($n$), $z$ is redshift, $H_0$ is the Hubble constant, and $\Omega_{\rm\Lambda}$ and $\Omega_{\rm M}$ are the cosmological constant and the total matter density (parameters of the $\mathrm{\Lambda}$CDM model).

In the SME framework~\cite{2009PhRvD..80a5020K}, the slight modifications induced by LIV effects are also described by a series expansion with respect to powers of the photon energy. In this framework, LIV can also be dependent on the direction of the source. Including only the single term assumed to dominate the sum, $\ttot$ is given by:
\begin{equation}
\label{eq:SME}
\ttot \simeq \frac{1}{H_0}\left(\sum_{jm}{_0Y_{jm}}(\hat{\boldsymbol{n}})c^{(n+4)}_{(I)jm}\right)\times \kappa_n,
\end{equation}
where $\hat{\boldsymbol{n}}$ is the direction of the source, $_0Y_{jm}(\hat{\boldsymbol{n}})$ are spin-weighted spherical harmonics, and $c^{(n+4)}_{(I)jm}$ are coefficients of the framework that describe the strength of LIV. In the SME case of a direction-dependent LIV, we constrain the sum (enclosed in parentheses) as a whole. For the alternative possibility of direction independence, all the terms in the sum become zero except $_0Y_{00}=Y_{00}=\sqrt{1/(4\pi)}$. In that case, we constrain a single $c^{(n+4)}_{(I)00}$ coefficient.

The coordinates of $\hat{\boldsymbol{n}}$ are in a Sun-centered celestial equatorial frame described in Section V of Ref.~\cite{2009PhRvD..80a5020K}, directly related to the equatorial coordinates of the source such that its first coordinate is equal to $90^\circ-\mathrm{Declination}$ and the second
being equal to the Right Ascension. Finally, the coefficients $c^{(n+4)}_{(I)jm}$ can be either positive or negative depending on whether LIV-induced dispersion corresponds to a decrease or increase in photon speed with an increasing energy respectively (i.e., the sign of the SME coefficients plays the role of the $\mathit{s_\pm}$ factor of the series-expansion framework).

In the important case of a $d = 5$ modification of the free photon Lagrangian in
effective field theory, \mbox{Myers} and \mbox{Pospelov} have shown that the only $d=5$ ($n=1$) operator that preserves both gauge invariance and rotational symmetry implies vacuum birefringence~\cite{2003PhRvL..90u1601M}. In such a case, and as was mentioned in the Introduction, significantly stronger constraints can be placed using the existence of birefringence than with just dispersion (as in this work). For this reason, in this paper and when working within the given assumptions of the SME framework, we proceed assuming that the $d=5$ terms are either zero or dominated by the higher-order terms, and proceed to constrain the $d=6$ terms, which are not expected to come with birefringence.

Our aim is to constrain the LIV-related parameters involved in the above two parametrizations: the quantum gravity energy $\eqg$ (for $n=\{1,2\}$ and $\spm=\pm$1) and the coefficients of the SME framework (for $d=6$ for both the direction dependent and independent cases) . To accomplish this, we first constrain the total degree of spectral dispersion in the data, $\ttot$, and then using the measured distance of the GRB and the cosmological constants, we calculate lower limits on $\eqg$ through Eq.~\ref{deltatn} and confidence intervals for the SME coefficients using Eq.~\ref{eq:SME}. We also produce an additional set of constraints after accounting for GRB-intrinsic spectral evolution effects (which can masquerade as LIV). In that case, we first treat $\ttot$ as being the sum of the GRB-intrinsic dispersions $\tint$ and the LIV-induced dispersion $\tliv$, then we constrain $\tliv$ assuming a model for $\tint$, and finally constrain $\eqg$ and the SME coefficients using the constraints on $\tliv$.

We employ the cosmological parameters determined using WMAP 7-year data $\Omega_{M}=0.272$ and $\Omega_{\Lambda}=0.728$~\cite{2011ApJS..192...18K}, and a value of \mbox{$H_0=73.8\pm2.4$~km\,s$^{-1}$\,Mpc${^{-1}}$} as measured by the Hubble Space Telescope~\cite{2011ApJ...730..119R}.

\section{\label{sec:Data}The Data}

We analyze the data from the four \Fermi-LAT GRBs having bright GeV prompt emission and measured redshifts, namely GRBs 080916C, 090510, 090902B, and 090926A.
We analyze events passing the P7\_TRANSIENT\_V6 selection, optimized to provide increased statistics for signal-limited analyses~\cite{FermiPass7}. Its main difference from the earlier P6\_V3\_TRANSIENT selection used to produce previous \textit{Fermi} constraints on LIV~\cite{2009Sci...323.1688A, 2009Natur.462..331A} consists in improvements in the classification algorithms, which brought an increase in the instrument's acceptance mostly below $\sim$300~MeV\footnote{For a detailed list of differences see http://fermi.gsfc.nasa.gov/ ssc/data/analysis/documentation/Pass7\_usage.html}.

We reject events with reconstructed energies less than 30~MeV because of their limited energy and angular reconstruction accuracy. We do not apply a maximum-energy cut. In the case of GRB~080916C, however, we removed an 106~GeV event detected during the prompt emission, since detailed analyses by the LAT collaboration\footnote{not published} showed that it was actually a cosmic-ray event misclassified as a photon.

We keep events reconstructed within a circular region of interest (ROI) centered on the GRB direction and of a radius large enough to accept 95\% of the GRB events according to the LAT instrument response functions, i.e., a radius equal to the 95\% containment radius of the LAT point spread function (PSF). Because the LAT PSF is a function of the true photon energy and off-axis angle (the angle between the photon true incoming direction and the LAT boresight), the PSF containment radius is calculated on a per-photon basis. In this calculation, we approximate the (unknown) true off-axis angles and energies with their reconstructed values, something that induces a slight error at low energies. Below $\sim$100~MeV, the LAT angular reconstruction accuracy deteriorates and the 95\% containment radius becomes very large. To limit the inclusion of background events due to a very large ROI radius and also reject some of the least accurately reconstructed events, we limit the ROI radius to be less than 12$^{\circ}$. The GRB direction used for the ROI's center is obtained by follow-up ground-based observations (see citations in Tab.~\ref{tab_app_distances}) and can be practically assumed to coincide with the true direction of the GRB.

\begin{figure*}[ht]
\includegraphics[width=0.25\textwidth]{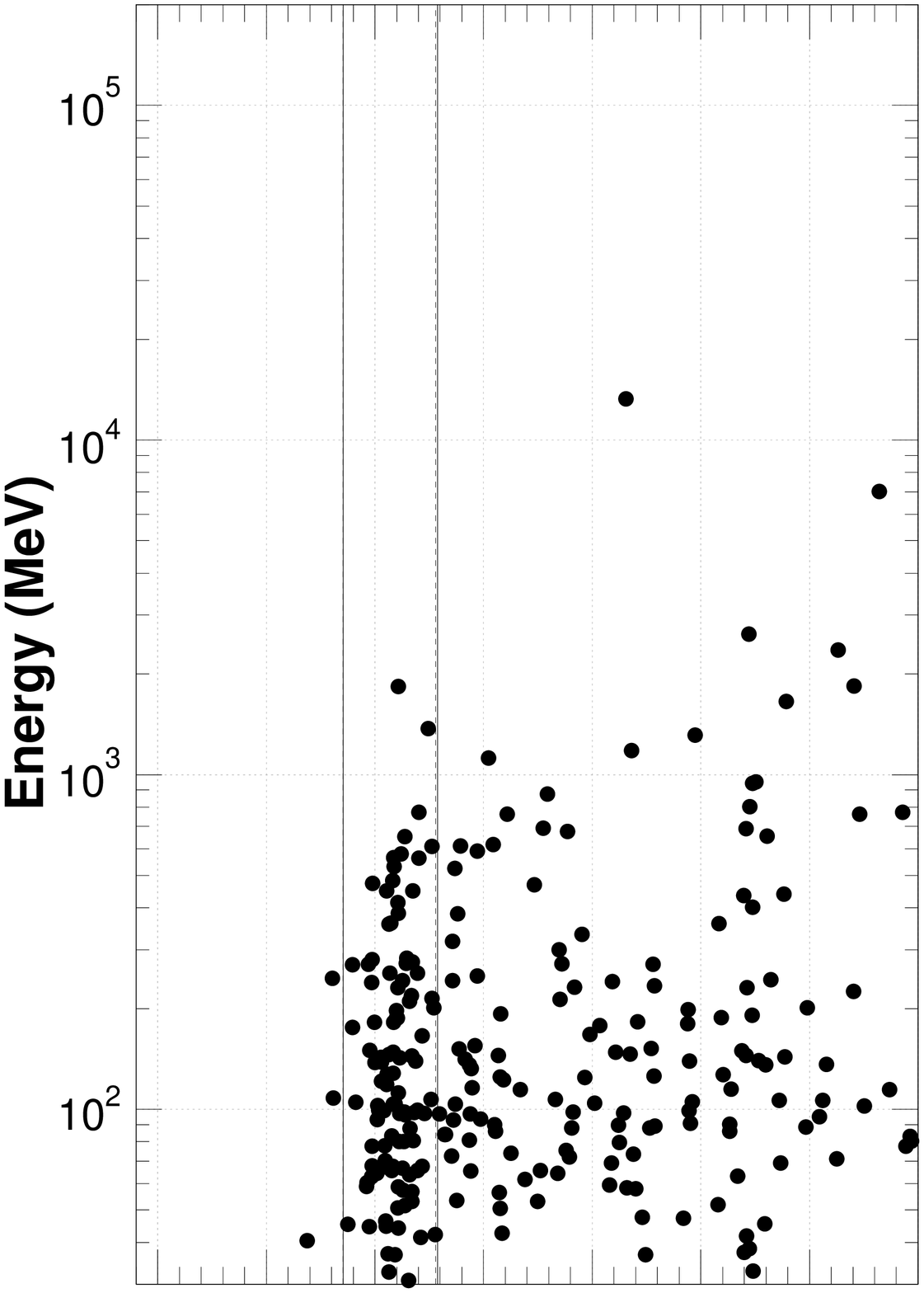}\includegraphics[width=0.25\textwidth]{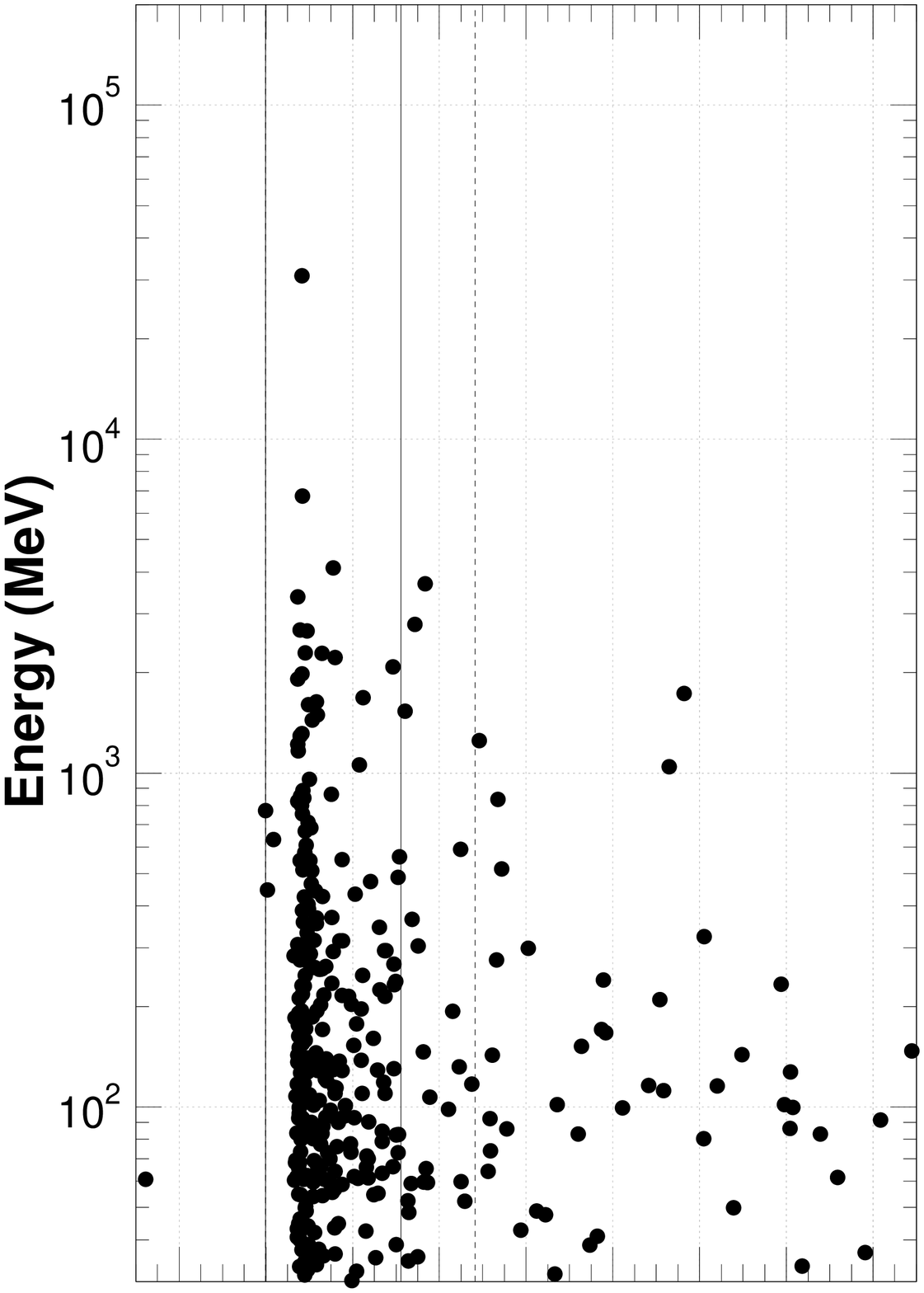}\includegraphics[width=0.25\textwidth]{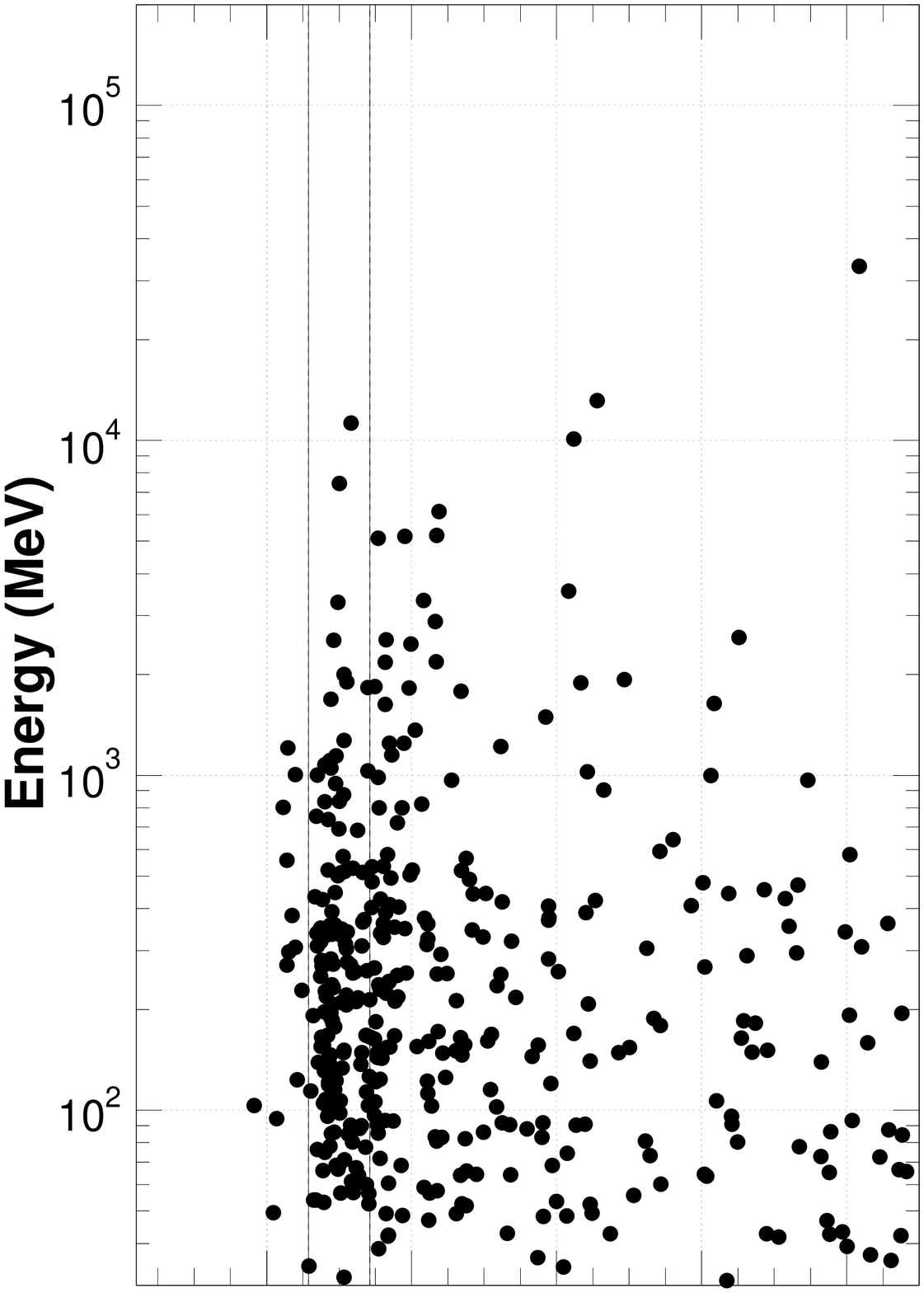}\includegraphics[width=0.25\textwidth]{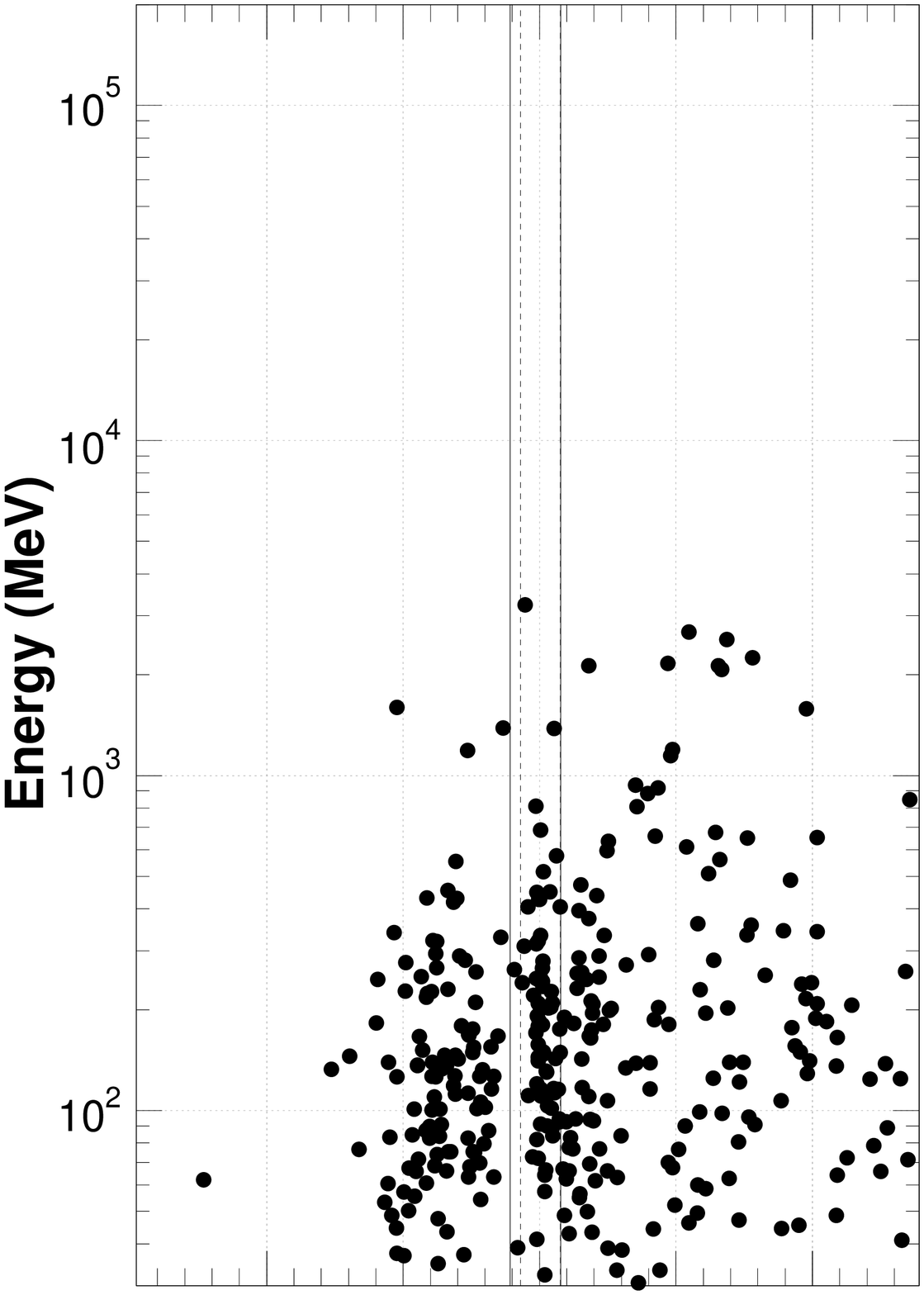}
\includegraphics[width=0.25\textwidth]{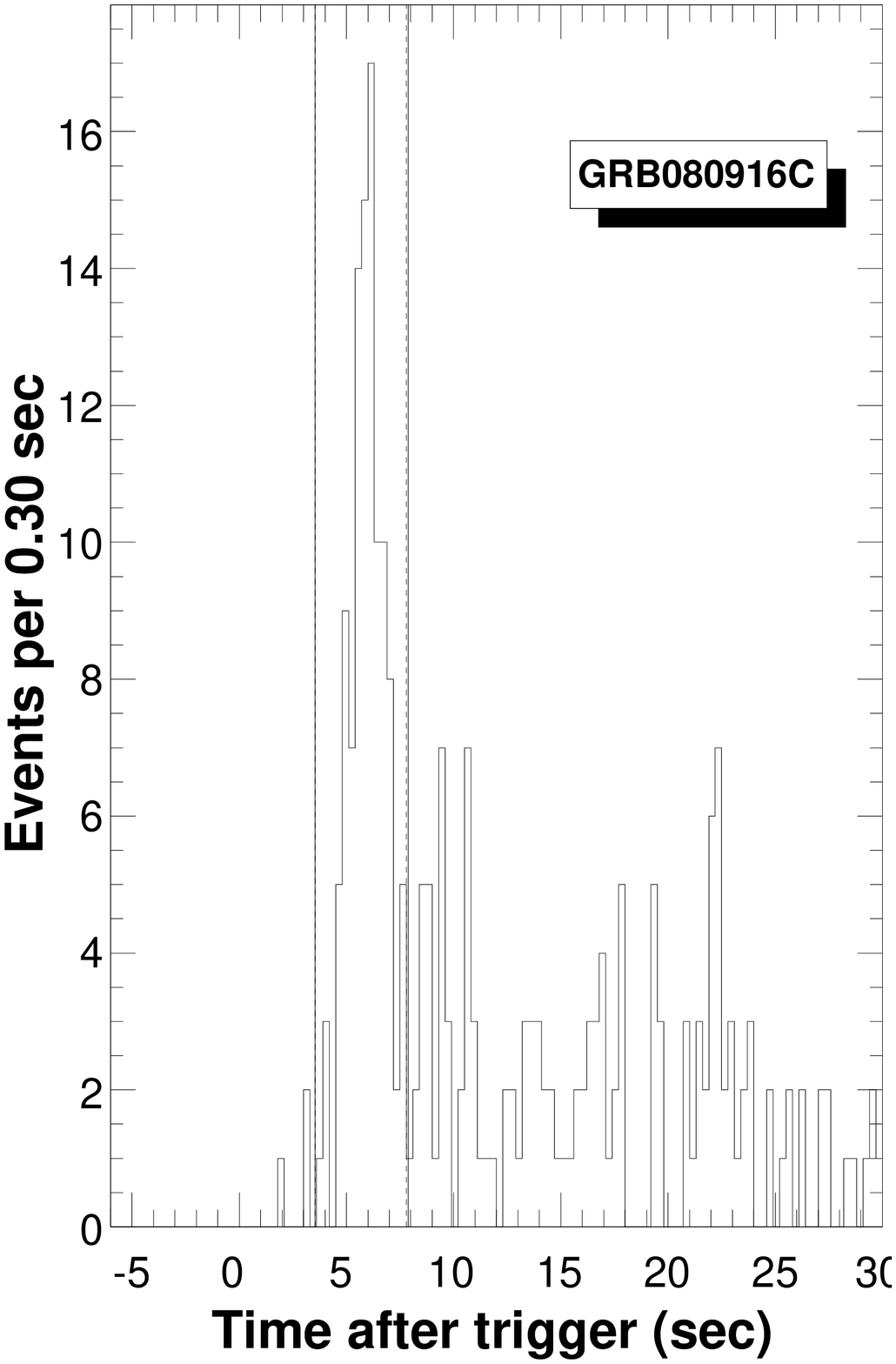}\includegraphics[width=0.25\textwidth]{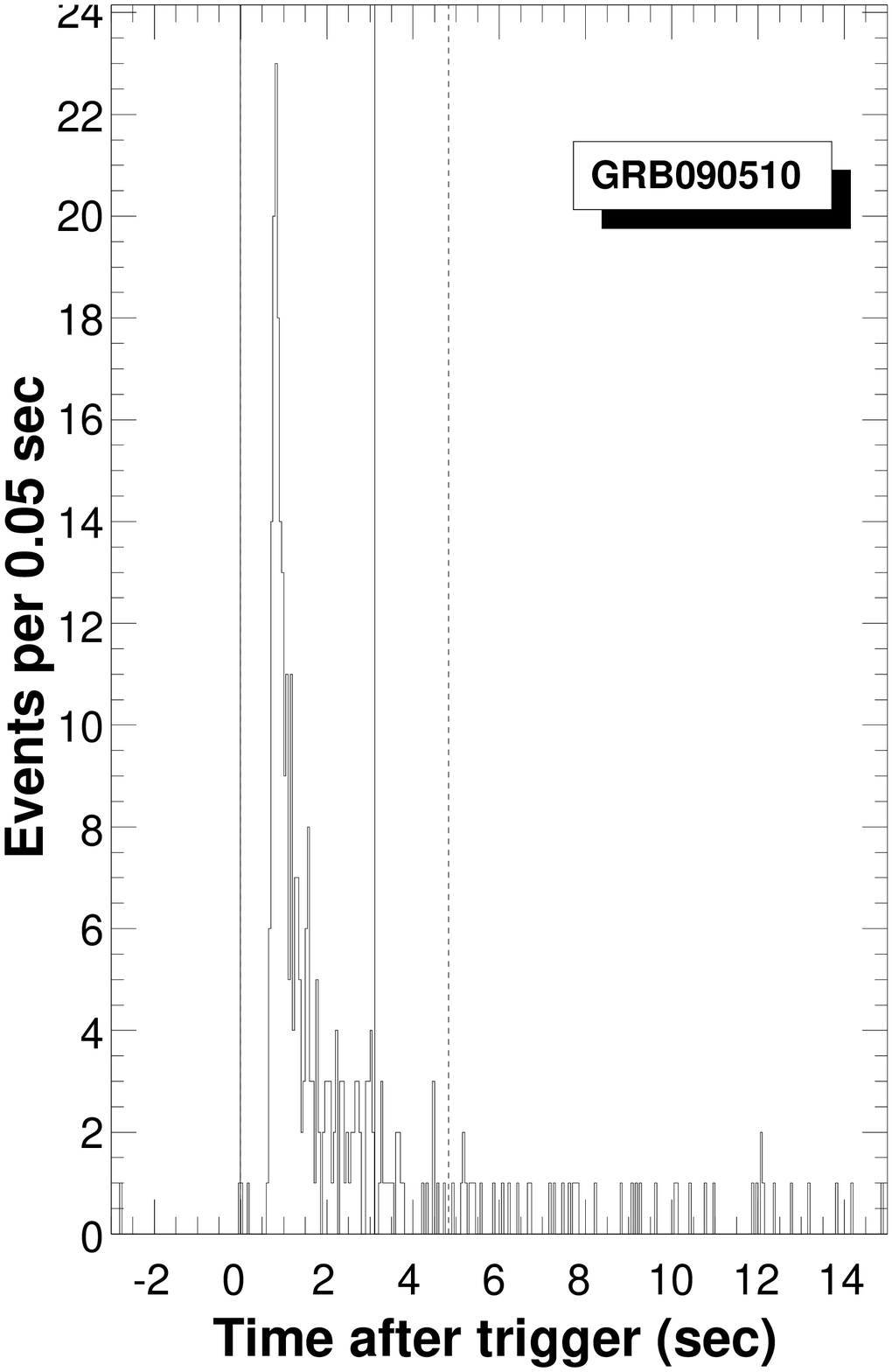}\includegraphics[width=0.25\textwidth]{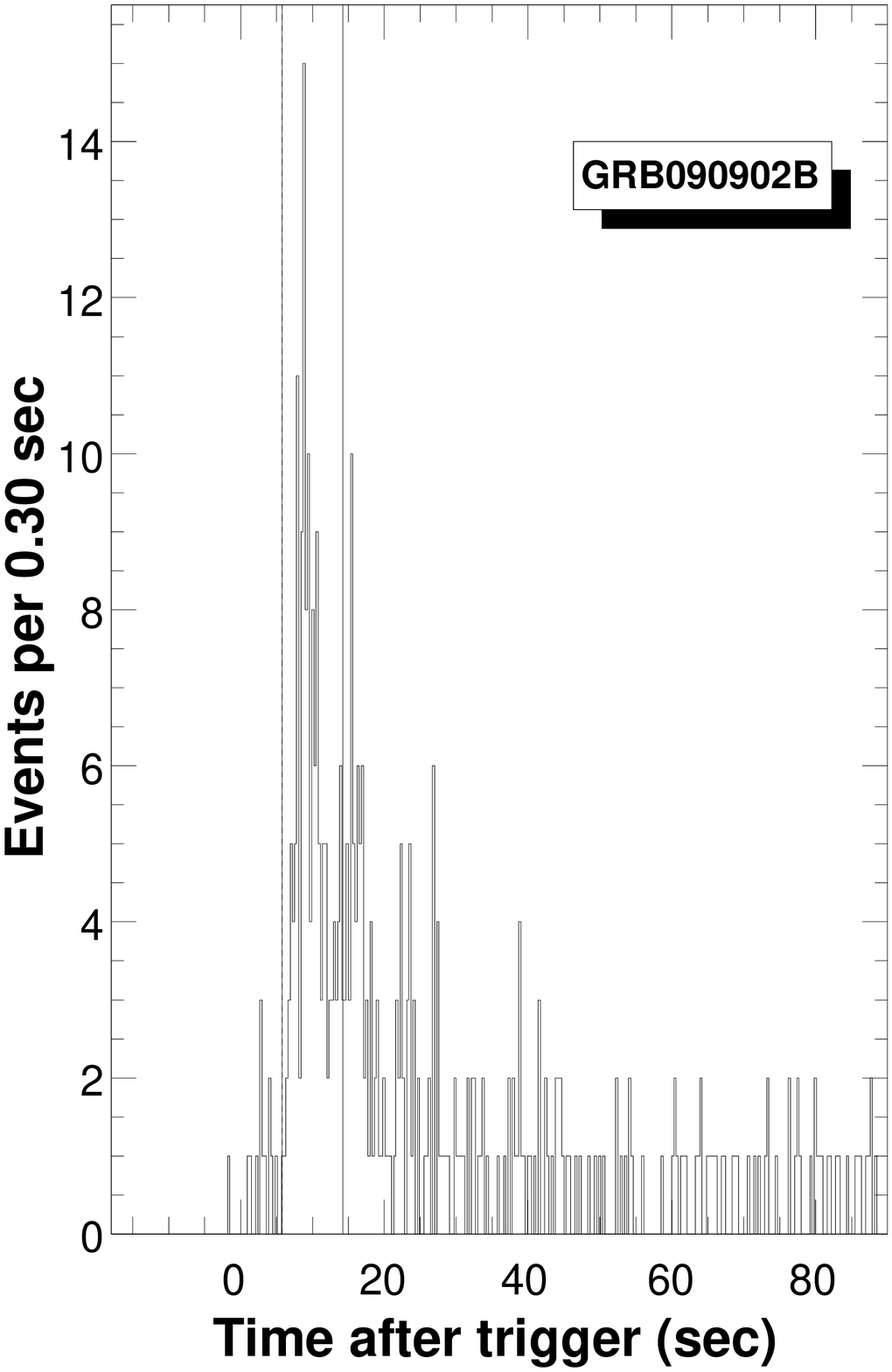}\includegraphics[width=0.25\textwidth]{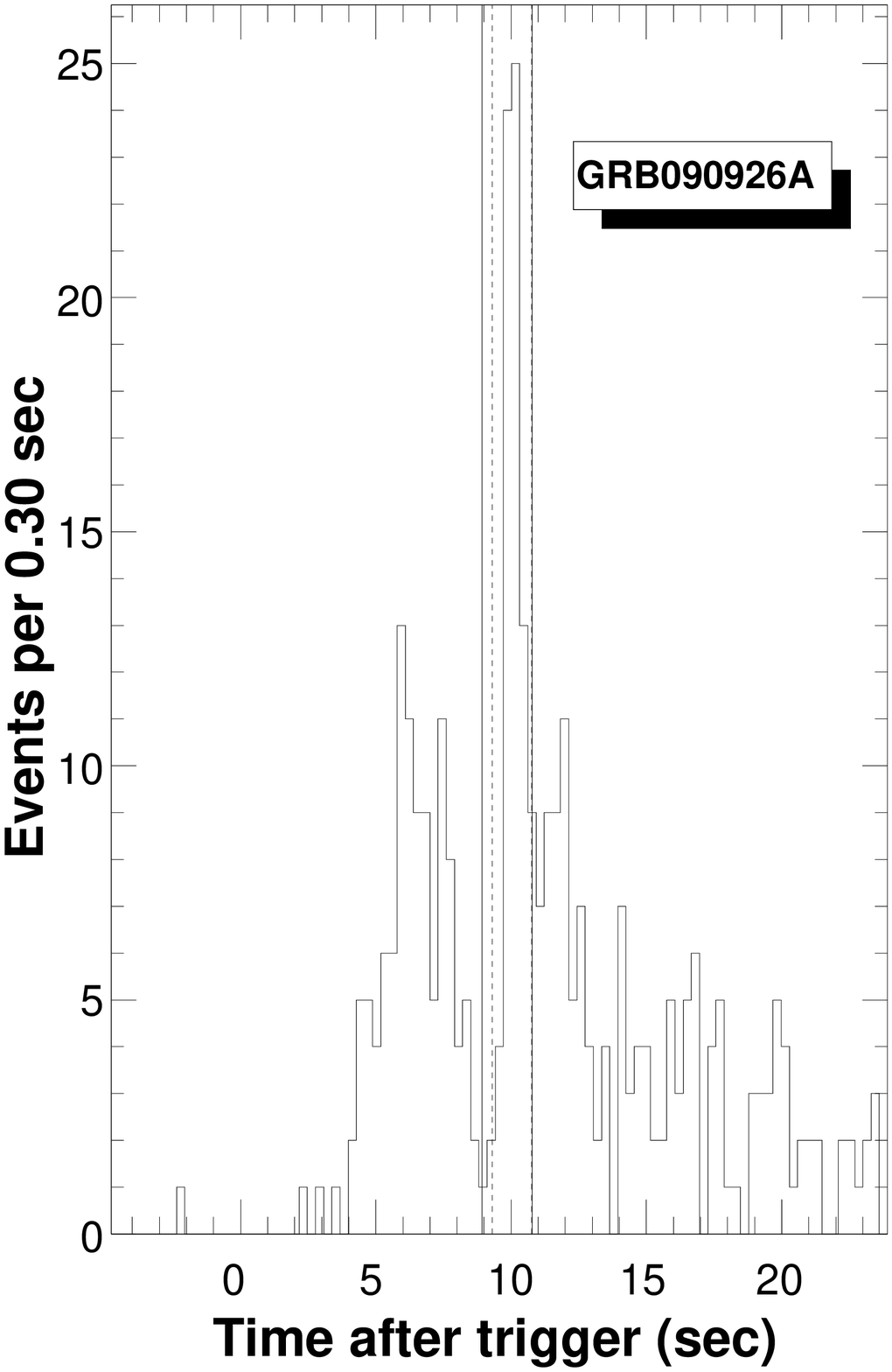}

\caption{\label{fig:grb_lcs}Time and energy profiles of the detected events from the four GRBs in our sample. Each column shows an event energy versus event time scatter plot (top) and a light curve (bottom). The vertical lines denote the time intervals analyzed (solid line for $n=1$ and dashed line for $n=2$), the choice of which is described in Sec.~\ref{sec:Methods}. If a dashed line is not visible, it approximately coincides with the solid one.}
\end{figure*}

The above data set are further split and cut depending on the requirements of each of the three analysis methods (as described below). Figure~\ref{fig:grb_lcs} shows the light curves and the event time versus energy scatter plots of the GRBs in our sample, and Tab.~\ref{tab_app_distances} shows the GRB redshifts and $\kappa_1$ and $\kappa_2$ distances (defined in Eq.~\ref{eq:kappa}).

\begin{table}
\caption{\label{tab_app_distances}Distances of Analyzed GRBs}
\begin{ruledtabular}
\begin{tabular}{cccc}
GRB & Redshift & $\kappa_1$ & $\kappa_2$   \\ \hline
  080916C & 4.35 $\pm$ 0.15~\citep{2009AA...498...89G} & 4.44 & 13.50   \\
  090510 & 0.903 $\pm$ 0.003~\citep{2009GCN..9352....1O}& 1.03&  1.50  \\
  090902B & 1.822~\citep{Z_090902B}\footnote{This GRB had a spectroscopically-measured redshift, which implies an error at the 10$^{-3}$ level.} & 2.07& 3.96    \\
  090926A & 2.1071 $\pm $0.0001~\citep{2010AA...523A..36D}& 2.37 & 4.85    \\
\end{tabular}
\end{ruledtabular}
\end{table}

\subsection*{Time Interval Selection}
The analyzed time intervals are chosen to correspond to the period with the highest temporal variability, focusing on the brightest pulse of each GRB. This choice is dictated by the fact that GRB emission typically exhibits spectral variability, which can potentially manifest as a LIV-dispersion effect (see discussion in Sec.~\ref{sec:systematics} for details on GRB spectral variability). By focusing on a narrow snapshot of the burst's emission, we aim to obtain constraints that are affected as little as possible by such GRB-intrinsic effects. Starting from this requirement, we select the time intervals to analyze, hereafter referred to as the ``default'' time intervals, using a procedure we devised $\textit{a priori}$ and applied identically on all four GRBs.

We start by characterizing the brightest pulse in each GRB by fitting its time profile with the flexible model used by Norris et al.~\citep{1996ApJ...459..393N} to successfully fit more than 400 pulses of bright BATSE bursts:
\begin{equation}
\label{eq:pulse}
I(t)=\begin{cases}
A\exp[-(|t-t_\mathrm{max}|/\sigma_{r})^{v}] & t<t_\mathrm{max}\\
A\exp[-(|t-t_\mathrm{max}|/\sigma_{d})^{v}] & t\geq t_\mathrm{max},
\end{cases}`
\end{equation}
where $t_\mathrm{max}$ is the time of the pulse's maximum intensity A, $v$ is a parameter that controls the shape of the pulse, and $\sigma_r$ and $\sigma_d$ are the rise and decay time constants. For $v=\{1,2\}$ the equation describes a two-sided exponential or Gaussian function respectively. We use the best fit parameters (as obtained from a maximum likelihood analysis) to define a ``pulse interval'' extending from the time instant that the pulse height rises to 5\% of its amplitude to the time instant that it fells to 15\% of its amplitude. We choose such an asymmetric cut because of the long falling-side tails of GRB pulses.

We then expand this initial ``pulse interval'' until no photons that were generated outside of it (at the source) could have been detected inside of it (at the Earth) due to LIV dispersion, and also until no photons that were generated inside of it (at the source) could have been detected outside of it (at the Earth) due to LIV dispersion. We use conservative values of $\eqgl= 0.5 \times \epl$ and $\eqgq=1.5\times10^{10}$~GeV for the maximum degree of LIV dispersion considered in extending the time interval, values which correspond to roughly one half of the stringent and robust limits obtained by $\Fermi$~\cite{2009Natur.462..331A} and H.E.S.S.~\cite{2008PhRvL.101q0402A,hesslike}. The interval resulting from this expansion is the one chosen for the analysis (hereafter referred to as the ``default'' interval).
The main reason for extending the interval is to avoid constraining the possible emission time of the highest-energy photons in the initial ``pulse interval'' to a degree that would imply an artificially small level of dispersion.

\begin{figure}[ht]
\includegraphics[width=0.8\columnwidth]{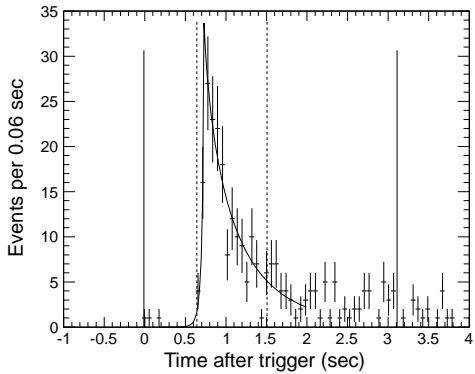}
\caption{\label{fig:090510_interval}Demonstration of the calculation of the default time interval for GRB~090510 and linear LIV. The data points
  show the event rate for energies greater than 30~MeV focusing on its main pulse, the thick curve shows the fit on the pulse using Eq.~\ref{eq:pulse}, the
  dashed vertical lines denote the ``pulse interval'', and the solid vertical lines denote the extended time interval (the default interval) chosen for the analysis.
}
\end{figure}

The choice of time interval for GRB~090510 and $n=1$ is demonstrated in Fig.~\ref{fig:090510_interval}. The (default) time intervals for all GRBs are shown in Fig.~\ref{fig:grb_lcs} with the vertical solid ($n=1$) and dashed lines ($n=2$), and are also reported in Tab.~\ref{tab_app_details}.

\section{\label{sec:Methods}Data Analysis}

\subsection{\label{subsec:PV_and_SMM}PairView and Sharpness-Maximization Methods}

Because the way we calculate confidence intervals is identical between PV and SMM, we first describe how the best estimate of the LIV parameter is
calculated by each of these two methods, and then proceed to describe their common confidence-interval calculation procedure.

\subsubsection*{Best-Estimate Calculation: PairView}

The PV method calculates the spectral lags $l_{i,j}$ between all pairs of photons in a data set and uses the distribution of their values to estimate the LIV parameter. Specifically, for a data set consisting of $N$ photons with detection times $t_{1...N}$ and energies $E_{1...N}$, the method starts by calculating the $N\times (N-1)/2$ photon-pair spectral lags $l_{i,j}$ for each $i>j$:
\begin{equation}
l_{i,j}\equiv \frac{t_i-t_j}{E_i^{n}-E_j^{n}}
\end{equation}
(where $n$ is the order of LIV), and creates a distribution of their values.

Let us examine how the distribution of $l_{i,j}$ values depends on the properties of the data and LIV dispersion.
For a light curve comprising \textit{at emission} a single $\delta$-function pulse and for a dispersion $\ttot$, the $l_{i,j}$ distribution will consist of a single $\delta$-function peak at a value of exactly $\ttot$. For a light curve comprising (at emission) a finite-width pulse, the now non-zero time differences between the emission times of the events behave as noise inducing a non-zero width to the distribution of $l_{i,j}$. Similarly to the previous ideal case however, the $l_{i,j}$ distribution will be peaked at approximately $\ttot$. For a realistic light curve consisting of one or more peaks superimposed on a smoothly varying emission, the distribution of $l_{i,j}$ will be composed of a signal peak centered at $\sim \ttot$ (consisting of $l_{i,j}$ values created primarily by events ${i,j}$ emitted temporally close and with not too similar energies) and a smoother underlying wide background (consisting of the rest of the $l_{i,j}$ values).

Following the above picture, the estimator $\tne$ of $\ttot$ is taken as the location of the most prominent peak in the $l_{i,j}$ distribution. This peak becomes taller and narrower, thus more easily detectable, as the variability time scale decreases and as the width of the energy range increases.

Searching for the peak using a histogram of the $l_{i,j}$ values would require us to first bin the data, a procedure that would include choosing a bin width fine enough to allow for identifying the peak with good sensitivity but also wide enough to allow for good statistical accuracy in the bin contents. We decided not to use a histogram to avoid the subjective choice of bin width. Instead, we use a kernel density estimate (KDE), as it provides a way to perform peak finding on
unbinned data, and as it is readily implemented in easy to use tools with the
ROOT TKDE method\footnote{http://root.cern.ch/root/html/TKDE.html}. We use a Gaussian kernel for the KDE and a bandwidth chosen such as to minimize the Mean Integrated Squared Error calculated between the KDE and a very finely binned histogram of the photon-pair lags.

\subsubsection*{Best-Estimate Calculation: Sharpness-Maximization Method}

SMM is based on the fact that the application of any form of spectral dispersion to the data (e.g., by LIV) will smear the light curve decreasing its sharpness. Based on this, SMM tries to identify the degree of dispersion that when removed from the data (i.e., when the negative value of it is applied to the data) maximizes its sharpness. This approach is similar to the ``Dispersion Cancellation'' (DisCan) technique~\cite{0004-637X-673-2-972}, the ``Minimal Dispersion'' method~\cite{2008PhRvD..78c3013E}, and the ``Energy Cost Function'' method ~\cite{Magic_PKS_2008PhLB..668..253M,2008PhRvD..78c3013E}. The most important difference between these approaches is the way the sharpness of the light curve is measured.

We start the application of SMM by analyzing a data set consisting of photons with detection times $t_i$ and energies $E_i$ to produce a collection of ``inversely smeared'' data sets, each corresponding to a trial LIV parameter $\ttot$, by subtracting $E_i^n\times \ttot$ from the detection times $t_i$. For each of the resulting data sets, the modified photon detection times are first sorted to create a new set $t'_i$, and then the sharpness of its light curve is measured using $t'_i$ and $E_i$. After this procedure has been applied on a range of trial LIV parameters, we find the inversely smeared data set with the sharpest light curve, and select the trial $\ttot$ value used to produce it as the best estimate of $\tne$.

In their analysis of the data from a flare of the blazar Mrk~501, the MAGIC collaboration~\cite{Magic_PKS_2008PhLB..668..253M} quantified the sharpness of the light curve using an ``Energy Cost Function'', which was essentially the sum of the photon energies detected in some predefined time interval chosen to correspond to the most active part of the flare.
Scargle et al.~\cite{0004-637X-673-2-972} explored a range of different cost functions to measure the sharpness of the light curve, including Shannon, Renyi, and Fisher information, variance, total variation, and self-entropy, finding that the Shannon information is the most sensitive. In this study, we use a function $\mathcal{S}$ that is similar to the Shannon information and is defined as:
\begin{equation}
\label{eq:sharpness}
\mathcal{S}(\ttot)=\underset{i=1}{\overset{N-\rho}{\sum}} \log \left( \frac{\rho}{t'_{i+\rho}-t'_{i}} \right),
\end{equation}
where $\rho$ is a configurable parameter of the method.

\begin{figure}[ht!]
 \includegraphics[width=1.0\columnwidth]{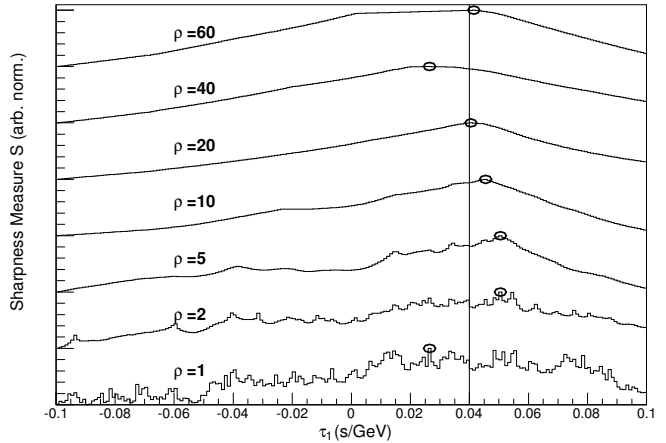}
 \caption{\label{fig:DC_different_smooth}Curves of SMM's sharpness measure $\mathcal{S}(\ttot)$ versus the trial value of the LIV parameter $\tau_1$, each produced using a different $\rho$ value. These curves were generated using the GRB~090510-inspired data set described in Appendix A after the application of a dispersion equal to +0.04~s/GeV (value denoted with the vertical line). The circles denote the maxima of the curves, the positions of which are used to produce $\hat{\tau}_1$. As can be seen, too small or too large values of $\rho$ correspond to a reduced accuracy for measuring the position of the peak.} \end{figure}

Different values of $\rho$ will tune the algorithm to evaluate the sharpness of the light curve focusing on intervals consisting of different numbers of events (i.e., of $\rho$ events) or equivalently focusing on different time scales. As a result, the choice of $\rho$ affects the performance of the algorithm in two ways. For a small value of $\rho$ (up to $\sim$3), some of the durations in the denominator of $\mathcal{S}(\ttot)$ can become relatively very small, making some of the
$1/(t'_{i+\rho}-t'_{i})$ terms very large. In this case, $\mathcal{S}(\ttot)$ can fluctuate significantly as a function of the trial lag,
decreasing the accuracy with which the best LIV parameter can be measured. For too large values of $\rho$ the algorithm essentially tries to minimize the total duration of the analyzed data, focusing on time scales larger than the variability time scale, ending up with a diminished sensitivity (in practice the peak of $\mathcal{S}$ becomes flatter). These effects are demonstrated in Fig.~\ref{fig:DC_different_smooth}.

To choose the value of $\rho$ we first generate a large number of simulated data sets inspired by the GRB under study, we then apply the method using a series of different $\rho$ values, and finally we choose the $\rho$ value that produces the most constraining median upper limit on $\tau_n$ (for $\spm=+$1).
These simulated data sets are constructed similarly to the procedure described in Appendix \ref{appendix:PVSMM} using a light-curve template produced by a KDE of the actual light curve, with the same statistics as the data, a spectrum similar to that in the data, and without any spectral dispersion applied.

Finally, it should be noted that the method's description above was for the case of zero source-intrinsic spectral evolution effects since the light curve of the GRB mission \textit{at the source} was treated as being maximally sharp. This picture is equivalent to assuming that there is an initial (imaginary) maximally-sharp signal that is first distorted by GRB-intrinsic effects and then by LIV. In that case, the constraints provided by SMM will be on the aggregate effect.

\subsubsection*{Confidence-Interval Calculation}
\label{sec:pvsmm_conf}
The PV and SMM methods produce a confidence interval on the best LIV parameter by means of a randomization analysis.

We start by producing one hundred thousand randomized data sets by shuffling the association between energies and times of the detected events. Because the total number of events and the distributions of energies and times are identical between the actually detected and the randomized data sets, their statistical power (i.e., their ability to constrain the dispersion) is similar. However, because of the randomization, any dispersion potentially present in the actual data is lost. After the set of randomized data sets is constructed, the best LIV parameter is measured on each one of them and the measurements are used to create a (normalized to unity) distribution $f_r$.

We then define the measurement error on $\ttot$ (for the general case of any $\ttot$) as $\Err=\tne-\ttot$ and the probability distribution function (PDF) of $\Err$ as $P_\Err(\err)$, where $\err$ is a random realization of $\Err$. We assume that $P_\Err$ has a negligible dependence on $\ttot$ (at least for the range of values of $\ttot$ expected to be present in the data) and approximate:
\begin{equation}
P_\Err(\err) \simeq P_\Err(\err | \ttot=0).
\end{equation}
The PDF $P_\Err$ for the case of a zero $\ttot$, $P_\Err(\err | \ttot=0)$, can be identified as
the normalized distribution $f_r$ produced using our randomization simulations. Thus,
\begin{equation}
P_\Err(\err) \simeq f_r(\err).
\end{equation}

Since $\Err$ is a quantity with a known PDF and since it depends on the unknown parameter $\ttot$, it can be used as a pivotal quantity to construct a two-sided Confidence Interval (CI) of Confidence Level (CL) for $\ttot$ as:
\begin{eqnarray}
\label{eq:pvsmm_conf}
CL & = & Pr(q_{(1-CL)/2}<\Err<q_{(1+CL)/2}) \label{eq:CLs} \\
  & = & Pr(q_{(1-CL)/2}<\hat{\tau}_{n}-\tau_{n}<q_{(1+CL)/2}) \nonumber\\
 & = & Pr(\hat{\tau}_{n}-q_{(1+CL)/2}<\tau_{n}<\hat{\tau}_{n}-q_{(1-CL)/2}) \nonumber\\
  & = & Pr(LL<\tau_{n}<UL) \nonumber,
\end{eqnarray}
where $LL=\hat{\tau}_{n}-q_{(1+CL)/2}$ and $UL=\hat{\tau}_{n}-q_{(1-CL)/2}$ are the lower and upper limits defining the CI, and $q_{(1-CL)/2}$ and $q_{(1+CL)/2}$ are the $(1-CL)/2$ and $(1+CL)/2$ quantiles of $f_r$.

To produce a lower limit on $\eqg$ for the subluminal or the superluminal case, we use eq.~\ref{deltatn} substituting $\ttot$ with its lower or upper limit, respectively, and solve for $\eqg$.

\newcommand{\disp}{\displaystyle}
\newcommand{\eqn}[1]{\begin{equation}\\#1\\\end{equation}}
\newcommand{\eqc}[1]{Eq.~(\ref{#1})}
\newcommand{\arr}{\rightarrow}
\newcommand{\Ninf}{N\arr +\infty}

\newcommand{\nfit}{N_\mathrm{fit}}
\newcommand{\like}{\mathcal{L}}
\newcommand{\emin}{E_\mathrm{cut}}
\newcommand{\emax}{E_\mathrm{max}}
\newcommand{\tmin}{t_1}
\newcommand{\tmax}{t_2}
\newcommand{\mysum}{\sum_{i=1}^{N}}
\newcommand{\myint}{\int_{\emin}^{\emax}\int_{\tmin}^{\tmax}}
\newcommand{\rate}{R}
\newcommand{\aeff}{A_\mathrm{eff}}
\newcommand{\flux}{F}
\newcommand{\npred}{N_\mathrm{pred}}
\newcommand{\pdf}{P}
\newcommand{\tn}{\tau_n}
\newcommand{\bn}{b_n}
\newcommand{\sn}{\tau_n E^n}
\newcommand{\sni}{\tau_n E_i^n}
\newcommand{\atn}{|\tn|}
\newcommand{\stn}{\sigma[\tne]}
\newcommand{\amp}{\Omega}
\newcommand{\epiv}{E_\mathrm{piv}}
\newcommand{\ecut}{E_\mathrm{f}}
\newcommand{\ind}{\Gamma}
\newcommand{\SP}{\mathcal{S}}
\newcommand{\LC}{\mathcal{C}}
\newcommand{\lup}{\Lambda_h}
\newcommand{\llo}{\Lambda_l}
\newcommand{\mean}[1]{\langle #1 \rangle}
\newcommand{\mom}[1]{E[#1]}
\newcommand{\momm}[1]{V[#1]}
\newcommand{\ord}[1]{{\cal{O}}(#1)}
\newcommand{\elow}{E_\mathrm{min}}
\newcommand{\emlo}{\mean{E^n}_l}
\newcommand{\emlol}{\mean{E}_l}
\newcommand{\emloq}{\mean{E^2}_l}
\newcommand{\eml}{\mean{E^{n}}_h}
\newcommand{\emq}{\mean{E^{2n}}_h}
\newcommand{\gmq}{\mean{g^2}}

\subsection{Likelihood Method}
\label{subsec_Likelihood}
The ML fit procedure used in this work has been developed and applied by Martinez and Errando~\cite{Martinez:08} to MAGIC data for the 2005 flare
of Mkn~501 and by Abramowski et al.~\cite{hesslike} to H.E.S.S. data for the gigantic flare of PKS 2155$-$304 in 2006.
This section describes its key aspects, its underlying assumptions, and the details of its application to GRB data.

The ML method consists in comparing the arrival time of each detected photon with a template light curve which is shifted in time by an
amount depending linearly or quadratically on the event's energy.
For a fixed number of independent events $\nfit$ with energies and times $\{E_i,t_i\}_{i=1,\nfit}$ observed in the energy and time intervals
$[\emin,\emax]$ and $[\tmin,\tmax]$, the unbinned likelihood function is:
\eqn{\label{eq:L}\like=\prod_{i=1}^{\nfit}\pdf(E_i,t_i|\tn),}
where $\pdf$ is the PDF of observing one event at energy $E$ and time $t$, given $\tn$. For an astrophysical
source observed by a gamma-ray telescope, it is
 $\disp{\pdf(E_i,t_i|\tn)=\rate(E_i,t_i|\tn)/\npred}$, where $\rate$ is the expected differential count
rate at energy $E$ and time $t$ and $\disp\npred=\myint\rate(E,t|\tn)\;dE\;dt$ is the total number of events
predicted by the model. For a point-like source observed by the \Fermi-LAT:
\eqn{\rate(E,t|\tn)=\int_0^\infty \flux(E_t,t|\tn)\;\aeff(E_t)\;\mathcal{D}(E_t,E)\;dE_t,}
where $\flux(E_t,t|\tn)$ is the model for the photon flux which is incident on the LAT at the photon (true) energy $E_t$ and
time $t$, whereas $\aeff(E_t)$\footnote{The effective area also depends on the direction of the source in instrument coordinates, a typically continuously varying quantity. We can drop the time dependence by approximating $\aeff(E_t,t)$ with its averaged over the observation value $\aeff(E_t)$.} and $\mathcal{D}(E_t,E)$ are the LAT effective area and energy redistribution functions, respectively.
As the energy resolution with the LAT is better than 15\% above 100~MeV~\cite{FermiPass7} we can neglect any energy mis-reconstruction effects.
Assuming no spectral variability and that the flux spectrum follows a power law with possible attenuation at the highest energies, then:
\eqn{\label{eq:flux}\flux(E,t|\tn)=\phi_0\;E^{-\ind}\;e^{-E/\ecut}\;f(t-\sn),}
where $\ind$ is the time-independent spectral index, $\ecut$ is the cutoff energy, and the function $f(t)$ is the time profile of the emission that would be received by the LAT in case of a null LIV-induced lag $\sn$. We explain further below how the function $f(t)$ is derived from the data in practice.
Finally, defining the observed spectral profile as $\disp\Lambda(E)=\phi_0\;E^{-\ind}\;e^{-E/\ecut}\;\aeff(E)$, we obtain:
\eqn{\pdf(E,t|\tn)=\Lambda(E)\;f(t-\sn)/\npred}.
Thus, the ML estimator $\tne$ of the LIV parameter $\tn$ satisfies:
\eqn{\label{eq:DL}\left[\mysum\frac{\partial\log
      f(t_i-\sni)}{\partial\tn}-\frac{\nfit}{\npred}\frac{\partial\npred}{\partial\tn}\right]_{\tn=\tne}=0.}
For the brightest LAT-detected GRBs, $\nfit\gtrsim50$ typically (see Tab.~\ref{tab_app_details}) thus a good estimate of the sensitivity offered by
the estimator $\tne$ can be obtained by considering the ideal case of the large sample limit. In this regime, $\tne$ is unbiased and efficient like
any ML estimator. Namely, its variance reaches the Cram\'er-Rao bound, e.g.,
given by Eq.~(9.34) page 217 of \cite{frodesen79}:
\eqn{\momm{\tne}  =  \left[\nfit\myint\frac{1}{\pdf}\left(\frac{\partial \pdf}{\partial \tn}\right)^2dE\;dt\right]^{-1}.}
As the time profile can be measured up to very large times in case of large photon statistics, one can show that the standard deviation of $\tne$ is
simply given by:
\eqn{\label{eq:stn}\stn=\frac{1}{\sqrt{\nfit\gmq\emq}},}
where $\disp\gmq=\int_{-\infty}^{+\infty}f'(t)^2/f(t)\;dt$ ($=1/\sigma^2$ for a Gaussian time profile of standard deviation $\sigma$),
$\disp\mean{E^{m\times n}}_h=\int_{\emin}^{\emax}E^{m\times n}\;\Lambda(E)\;dE\;/\;\lup$ and $\disp\lup=\int_{\emin}^{\emax} \Lambda(E)\;dE$.
The above expression for $\stn$ is a good approximation (within a factor 2 to 3) of the actual standard deviation of $\tne$, and it gives a useful
estimate of the expected sensitivity. However, our final results are based on a proper derivation of confidence intervals as described further below
in this section.

The spectral profile $\Lambda(E)$ is constant with time since $\Gamma$ is assumed to be constant during the considered time interval (see further
discussions on possible spectral evolution effects in Sec.~\ref{sec:systematics}). The spectral profile is also independent of the LIV
parameter, and is only used as a weighting function in the PDF normalization $\npred$.
For these reasons, we approximate the spectral profile by a power-law function (with a fixed attenuation when needed):
\eqn{\Lambda(E)\propto E^{-\gamma}\;e^{-E/\ecut}.}
The spectral index $\gamma$ is obtained from the fit of the above function to the time-integrated spectrum $\SP(E)$ observed by the LAT:
\eqn{\SP(E)=\int_{\tmin}^{\tmax}\flux(E,t|\tn)\;\aeff(E)\;dt.}

In practice, we define a $\emin\in[100,150]$\,MeV and we fit the spectrum $\SP(E)$ above $\emin$ (see Fig.~\ref{080916C_spec_example} for an example)
in order to obtain a fairly good estimate of the spectral index, namely $\gamma\simeq\Gamma$ within errors (see discussion in
Sec.~\ref{sec:systematics} regarding this approximation).
For the case of GRB~090926A, we use a power-law function that has an exponential break, in accordance with the findings of Ackermann et
al.~\cite{GRB090926A:Fermi}.

Knowledge of the time profile $f(t)$ is crucial for the ML analysis.
Typically, $\emin$ divides the LAT data set in two samples of roughly equal statistics.
The ML analysis is performed using events with energies above $\emin$, whereas the fit of the light curve $\LC(t)$ observed by
the LAT below $\emin$ is used to derive the time profile:
\eqn{\label{eq:LC}\LC(t)=\int_{\elow}^{\emin}\Lambda(E)\;f(t-\sn)\;dE\simeq\llo\;f[t-\tn\emlo],}
where $\elow=30$~MeV, $\disp\llo=\int_{\elow}^{\emin} \Lambda(E)\;dE$ and $\disp\emlo=\int_{\elow}^{\emin}E^n\;\Lambda(E)\;dE\;/\;\llo$.
The Taylor expansion used in \eqc{eq:LC} is justified as LIV-induced lags are effectively negligible for low-energy events, and it yields the time profile:
\eqn{\label{eq:f}f(t)=\LC[t+\tn\emlo]\;/\;\llo\simeq\LC(t)\;/\;\llo.}
In practice, we fit the light curve $\LC(t)$ with a function comprising up to three Gaussian functions (see for example
Fig.~\ref{fig:likelihood_templates}).
The fit is performed on events detected in a time interval somewhat wider than the default time intervals (defined in
the beginning of Sec.~\ref{sec:Methods}) to allow for better statistics and because the calculations need an estimate of the GRB flux at times that
are also external to the default time intervals.

We then proceed with calculating the likelihood function $\like$ for a series of trial values of the LIV parameter $\tau_n$, and
plotting the curve of ${-2\Delta \rm{ln}(\like)=-2\ln\left[\like(\tn)/\like(\tne)\right]}$ as a function of $\tau_n$. We first produce a best estimate of $\ttot$, $\tne$, equal to the location of the minimum of the \mbox{$-2\Delta \rm{ln}(\like)$} curve. We also produce a CI on $\tau_n$ for an approximately two-sided CL (90\%, 99\%) using the two values of $\tau_n$ around the global minimum
at $\tne$ for which the curve reaches a values of 2.71 and 6.63, respectively\footnote{These two values correspond to the (90\%, 99\%) CL quantile of a $\chi^2_1$ distribution.}. Hereafter we refer to these CIs as being obtained ``directly from the data''. In addition, we produce a set of ``calibrated'' CIs on $\tau_n$ using Monte Carlo simulations and as described in Appendix~\ref{appendix:likelihood}. The calibrated CIs take into account intrinsic uncertainties arising from the ML technique (e.g., due to biases from the finite size of the event sample or from an imperfect characterization of the GRB's light curve), and are, most importantly, constructed to have proper coverage. Our final constraints on the LIV parameter and the LIV energy scale are produced using the calibrated CIs.

As a final note, we would like to stress that the time shift $\tn\emlo$ in \eqc{eq:f} has been set to zero following Refs.~\cite{Martinez:08,
  hesslike}. This implies that the time correction of any event entering the likelihood function is overestimated by a factor $1/\eta_n$, with
$\disp\eta_n=1-\emlo/E^n\in[0.5,1.0]$ for $E\in[0.1,30]$\,GeV, $n=1$ and, e.g., $\emlol=50$\,MeV.
In principle, ignoring this time shift would thus produce an additional uncertainty $\tne-\tau_n$ which is negative on average. This would also
slightly distort the likelihood function since $\eta_n$ varies with photon energy, possibly causing a reduction in sensitivity.
In the large sample limit, one can show that the bias of the estimator takes the form $b_n\simeq -\tn\emlo\eml/\emq$, namely the fractional bias
$b_n/\tn$ is negative and decreases with increasing hardness of the spectrum. In practice, it ranges from $\sim$0.5\% to $\sim$8\% for spectral
hardnesses similar to the ones of bursts we analyzed.
In addition, due to the limited photon statistics available in our analysis and to the relatively small values of $\tn$ likely to be present in
the data, the ratio $b_n/\stn$ is also negligible (a few percent at most).
One should, however, keep this effect in mind for future analyses of much brighter sources and/or in case of significant detections of LIV effects.

\subsection{Estimating the Systematic Uncertainty due to Intrinsic Spectral Evolution}
\label{subsec:intrinsic_corrections}

 So far we have concentrated on characterizing the statistical
 uncertainties of our measurements.  However, systematic uncertainties
 can also be very important and should be taken into account, if
 possible. Here, we describe how we model the dominant
 systematic uncertainty in our data, namely the intrinsic spectral
 evolution observed in GRB prompt-emission light curves. A detailed
 discussion of this phenomenon is given in Sec.~\ref{sec:systematics}.

For simplicity, in our introduction of LIV formalism (Sec.~\ref{ch_formalism}), we implicitly ignored the presence of GRB-intrinsic effects (which can in general masquerade as a LIV-induced dispersion), and instead just used the quantity $\ttot$ to describe the degree of dispersion in the data. However, $\ttot$ describes the \textit{total} degree of dispersion and is, in general, the sum of the LIV-induced degree of dispersion, described by a parameter $\tliv$, and the GRB-intrinsic degree of dispersion, described by a parameter $\tint$, i.e.,
\begin{equation}
 \ttot=\tint+\tliv \nonumber.
\end{equation}

Our methods do not differentiate between the different sources of dispersion. Instead, they directly measure and constraint their sum $\ttot$. We can either ignore any intrinsic effects (i.e., assume $\tint\simeq0$) and proceed directly to constrain LIV using the obtained CI on $\ttot$ or we can first assume a model for $\tint$, proceed to constrain $\tliv$, and finally constrain LIV using the CI on $\tliv$. The second approach is more appropriate for constraining LIV, since its results are more robust with respect to the presence of GRB-intrinsic effects\footnote{Since the majority of previously published LIV constraints have not taken into account GRB-intrinsic effects, limits of the first approach are still useful for comparing experimental results across different studies.}.

In principle, one could try to model $\tint$ using some knowledge
 of the physical processes generating the detected GRB emission or possibly
using phenomenological models constructed from large sets of MeV/GeV observations
of GRBs. Unfortunately, because of the scarcity of GRB observations at LAT energies, neither
 approach has reached a mature enough stage to produce trustworthy and robust predictions of GRB spectral lags (at such energies). Thus, any attempts to model $\tint$ would, at this point, likely end up producing unreliable constraints on LIV. However, a more robust and conservative approach can be adopted, as follows.

Since we do not have a model for $\tint$ that reliably predicts GRB-intrinsic lags, we instead choose to model it in a way that produces the most reasonably conservative constraints on $\tliv$.

One of the main considerations behind modeling $\tint$ is the reasonable assumption that our measurements of $\ttot$ are dominated by GRB intrinsic effects or in other words that our constraints on $\ttot$ also apply to $\tint$. We start with the fact that we already have obtained a coarse measurement of the possible magnitude of $\tint$, provided by our constraints on $\ttot$. Specifically, we know that the value of $\tint$ (for a particular observation) is not likely larger than the width of the allowed range of $\ttot$, as described by its CIs\footnote{The alternative case of a large $\tint$ being approximately canceled by an oppositely large $\tliv$ seems extremely unlikely since it would require the improbable coincidence of LIV actually existing, that the sign of the dispersion due to LIV being opposite of the sign of the dispersion due to intrinsic effects, and that the magnitudes of the two effects be comparable for each of the four GRBs (a ``conspiracy of Nature'').}. Thus, we start with the working assumption that the width of the possible range of $\tint$ is equal to the possible range of $\ttot$ (as inferred by our CIs on it).

Second, we assume that the $\tint$ has a zero value on average. This is a reasonable assumption given that we analyze in this study only cases where there is no clear detection of a
spectral lag signal (i.e., $\ttot$ is consistent with zero within the
uncertainty of its measured value). Moreover, this also avoids the need
for introducing by hand a preferred sign for
$\mean{\tint}$.

In principle, there are infinite choices for a particular shape of $\tint$ given our constraint for its width and (zero) mean value. We choose the one that produces the least stringent (the most conservative) overall constraints on $\tliv$, by modeling $\tint$ so that it reproduces the allowed range of possibilities of $\ttot$. For example, if our measurements imply that the data are compatible with (i.e., they cannot exclude) a positive $\ttot$, then we appropriately adjust $\tint$ to match (explain) this possibility. This approach leads to confidence intervals on $\tliv$ that have the largest possible width. Other choices for modeling $\tint$ can produce intervals more stringent either at their lower or their upper edge, but they cannot produce more stringent $\textit{overall}$ (i.e., when considering both their edges) constraints. The implementation of our model for $\tint$, defined as $\ptint(\tildetint)$ with $\tildetint$ being a random realization of $\tint$, depends on the particular method PV/SMM versus ML, and is described separately below.

For constructing CIs on $\tliv$ with PV and SMM, we use a similar approach as in
  Eq.~\ref{eq:pvsmm_conf}. However, instead of using as a pivotal
  quantity $\Err=\tne-\ttot$, we now use
\begin{eqnarray}
\Err' & = & \tne - \tliv  \nonumber \\
          &=& \tne - \ttot + \tint \nonumber \\
          &=& \Err + \tint .
\end{eqnarray}
If we define the PDF of $\Err'$ as $P_{\Err'}(\err')$, where $\err'$ is  a random realization of $\Err'$, and if \mbox{$q'_{(1-CL)/2}$} and \mbox{$q'_{(1+CL)/2}$} are its
\mbox{$(1-CL)/2$} and \mbox{$(1+CL)/2$} quantiles,
 then starting from \mbox{$CL= Pr(q'_{(1-CL)/2}<\Err'<q'_{(1+CL)/2})$}, we derive a CI on $\tliv$ of confidence level CL:
 \begin{eqnarray}
 \label{eq:CI_on_LIV}
 CL &=& Pr(LL'<\tliv<UL') \\
   &=& Pr(\hat{\tau}_{n}-q'_{(1+CL)/2}<\tliv<\hat{\tau}_{n}-q'_{(1-CL)/2}). \nonumber
\end{eqnarray}

Similarly to the CI on $\ttot$ which depends on the quantiles of $P_\Err$ (approximated by $f_r$), the CI on $\tliv$ depends on the quantiles of $P_{\Err'}$. Assuming that the two components of $\Err'$ ($\Err$ and $\tint$) are independent, $P_{\Err'}$ is given by the convolution of their PDFs:
\begin{equation}
  P_{\Err'}(\err')= \int_{-\infty}^{\infty} f_r(\err'-\tildetint)\ptint(\tildetint) d\tildetint. \label{eq:Peps}
  \end{equation}

Up to now, we have described a general way to produce CIs on $\tliv$, independent of the particular choice of $\ptint$. We mentioned above that we would like to choose a model for $\tint$ such that it matches any part of the parameter space for $\ttot$ not excluded by the data. From the expression of the lower and upper limits for $\ttot$ (Eq.~\ref{eq:pvsmm_conf}) we observe that a large (say) positive tail in the $f_r$ distribution implies that our observations are compatible with (they cannot exclude) a symmetrically negative part of the parameter space of $\ttot$, and vice versa. Based on this observation, we choose to model $\ptint(\tildetint)$ as $f_r(-\tildetint)$. With this choice, Eq.~\ref{eq:Peps} becomes:
\begin{equation}\label{eq:AC}
    P_{\Err'}(\err') =\int_{-\infty}^{\infty} f_r(\err'-\tildetint)f_r(-\tildetint)d\tildetint
= P_{\rm AC}(\err'),
   \end{equation}
 where $P_{\rm AC}(\err')$ is defined as the autocorrelation function
  of $f_r$ with argument $\err'$. As an autocorrelation function, $P_{\Err'}(\err')$ is an
  even function with maximum at zero. Because of this symmetry, its $(1+CL)/2$ and $(1-CL)/2$ quantiles,
 $q'_{(1+CL)/2}$ and $q'_{(1-CL)/2}$, respectively, are equal. Thus, the confidence interval in
Eq.~\ref{eq:CI_on_LIV} is symmetric around the observed value
 $\hat{\tau}_{n}$. Finally, since $\mean{\tint}$ was chosen to be zero, in addition to $\tne$ being our best estimate for $\ttot$, it is also our best estimate for $\tliv$. A demonstration of the application of this method for GRB~090510, PairView, and $n=1$ is shown in Sec.~\ref{sec:Results}, in Fig.~\ref{fig:090510_AC_example}.

The confidence interval on $\tliv$ is wider than the one
 calculated on $\ttot$ by a
 degree that depends on the width and shape of the possible variations in
 $\tint$ (and thus of $f_r$). In the simple case of $f_r$ following a Gaussian
 distribution, then the width would increase by a factor of
 $\sqrt{2}$. In our case, the function $f_r$ does not always follow a
 Gaussian, hence the increase is not in general equal to $\sqrt{2}$.

 For the case of the ML method, we follow the same main idea (i.e.,
 assume a $\ptint$ following our observational uncertainty on
 $\ttot$ and produce confidence intervals on $\tliv$)
 but apply it a different way. In this case, we run a second
 set of calibration simulations, in which the likelihood function is
 modified to include a not-necessarily-zero delay due to GRB-intrinsic
 effects. Specifically, Eq.~\ref{eq:flux} becomes:
 \begin{eqnarray}\nonumber
 F'(E,t|&&\tliv;\tildetint)=\\
 &&\phi_0\;E^{-\ind}\;e^{-E/\ecut}f(t-\tliv E^n -\tildetint E^n).
 \end{eqnarray}
 In each iteration of the simulation, we
 sample a different random value $\tildetint$ from the assumed $\ptint$ PDF
 and proceed normally to produce a distribution of lower and
 upper limits on $\tliv$, the means of which will define our confidence
 interval on $\tliv$. The $\ptint$ distribution is chosen in a similar way to the PV/SMM case using the distributions of $\tne$ produced during the first set of calibration simulations.
The properties of the generated confidence
 intervals produced with this approach are the same as those
 constructed by the PV/SMM methods.

We would like to add a point on the meaning of the distribution $\ptint$. In general, the properties of the emission from a given GRB depend on two factors: the initial properties describing the GRB's generating system (e.g., mass, rotation speed, environment, redshift, etc.) and the randomness involved in the physical processes involved in producing the emission. We can imagine the $\tint$ quantity as a constant unknown parameter (a ``true parameter'') that describes the range of possibilities for both factors mentioned above, thus $\ptint$ can be considered as its Bayesian prior. We can alternatively imagine the existence of some true parameter $\tau_{\rm int,0}=\mean{\tint}$ (chosen to be zero) that depends solely on the progenitor properties, and that, during a GRB explosion, a random realization $\tildetint$ is produced depending on the $\tau_{\rm int,0}$ of that particular GRB system. In this case, we can imagine $\ptint$ as a frequentist description of the range of possible $\tildetint$ values occurring among an infinite number of GRBs, all initiated by the same initial conditions (i.e., having the same $\tau_{\rm int,0}$). Based on the above, $\ptint$ can be considered as a Bayesian prior or alternatively as a frequentist statement of the possible values of $\tildetint$ across infinite repetitions of a GRB -- the particular choice, however, does not matter.

As a final note we should mention that our approach assumes that the experimental results allow the possibility of $\ttot$ being zero. With some additional assumptions, however, this approach can be generalized to include the case of a detection of a non-zero $\ttot$. For example, we could make the assumption that a detected non-zero total dispersion is merely result of GRB-intrinsic effects, allow for $\mean{\tint}$ to take a non-zero value (with $\tne$ being the most conservative choice), and produce a final confidence interval on a residual $\tliv$ (that would still be consistent with a zero $\tliv$).\footnote{If a non-zero dispersion is detected, it would also be interesting to test the alternative possibility that this dispersion might indicate a non-zero value of $\tliv$, rather than be fully attributed to $\tint$ as assumed in our method. Since most GRB properties vary weakly throughout the burst prompt emission, we may expect $\tint$ to also do so. In such a case, varying the time interval could change the measured value of $\tint$ , while not affecting $\tliv$.}.
It can be said that this method allows us to quantify the degree to which GRB-intrinsic effects reduce our ability to detect a residual LIV-induced dispersion.

\section{\label{sec:Results}Results}

The configuration of our methods is shown in Tab.~\ref{tab_app_details}, in which we report the range (relative to the GBM trigger time) of the analyzed data samples (common to all the methods), the value of SMM's smoothing parameter $\rho$, the numbers of events used with PV and SMM $N_{\rm 100}$, and some quantities relevant to the ML method, namely the fitted index $\gamma$ of the observed spectrum $\SP(E)$, the number of events in the two parts of the data used for fitting the light-curve template $N_{\rm template}$ and for calculating the likelihood $N_{\rm fit}$, and the energy separating these two parts of the data $E_{\rm cut}$.

It is known that the spectra of LAT-detected GRBs typically comprise two spectral components: a Band (two smoothly connected power laws~\cite{Band1993}) plus a power law function. These components do not necessarily have the exact same light curves and their spectra do not evolve in an identical fashion. As a result, an analysis of a data set consisting of events from both of these components might exhibit GRB-intrinsic spectral evolution that may be erroneously interpreted as LIV. This can be an important systematic uncertainty, and is discussed further in Sec.~\ref{subsec:grbintrinsic}. To reduce the influence of this effect, we performed the PV and SMM analyses on a data set starting from 100~MeV (instead of 30~MeV), a choice made \textit{a priori} to reduce the contamination from the Band spectral component\footnote{The spectrum of GRB~080916C comprises just one spectral component (Band). Thus, even though we did not need to reject the 30--100~MeV events for that GRB, we still applied this cut for consistency between the four analyzed data sets.}\footnote{The particular value of 100~MeV is also the minimum energy typically used in LAT science analyses, since the LAT reconstruction accuracy starts to deteriorate below this energy.}. Because of the greater demand for statistics of the ML method, we did not apply such a minimum-energy cut for this method, and instead we used the events from $E_{\rm cut}$ down to 30~MeV for the light-curve template construction. As a result, any differences in the temporal properties of the two spectral components might have affected the ML method more than the other two methods. However, the magnitudes of any such uncertainties are limited by the typically small contribution of the Band component to the analyzed data and are likely smaller than the statistical errors.

We produce constraints for two confidence levels: a 90\% two-sided (or equivalently 95\% one-sided) CL and a 99\% two-sided (or equivalently 99.5\% one-sided) CL. In the following, the ``one-sided'' or ``two-sided'' designations of the CLs may be omitted for brevity.

\begin{table*}
\caption{\label{tab_app_details}Configuration Details}
\begin{ruledtabular}
\begin{tabular}{ccc|cc|cc|ccccc}
GRB &  \multicolumn{2}{c}{Time Range (s)} & \multicolumn{2}{c}{$\rho$} & \multicolumn{2}{c}{\rm $N_{\rm 100}$}&$\gamma$ & \multicolumn{1}{c}{$N_{\rm template}$} & \multicolumn{2}{c}{$N_{\rm fit}$} & $E_\mathrm{\rm cut}$ (MeV)\\
         & \multicolumn{2}{c}{All Methods} & \multicolumn{2}{c}{SMM} &\multicolumn{2}{c}{PV \& SMM}&\multicolumn{5}{c}{Likelihood} \\
 & $n=1$ & $n=2$ &$n=1$ &$n=2$ &$n=1$ &$n=2$ & $n=\{1,2\}$&$n=\{1,2\}$ & $n=1$ & $n=2$ &$n=\{1,2\}$\\ \hline

  080916C & 3.53--7.89 &3.53--7.80  &  50 & 30&59 &59 &2.2 & 82 & 59 &59 &100\\
  090510 & -0.01--3.11 &-0.01--4.82 & 50 & 70&157 &168 &1.5 &148 &118 &125 &150 \\
  090902B & 5.79--14.22 &5.79--14.21 & 80 &80 &111 &111 &1.9 & 57&87 & 87&150 \\
  090926A & 8.92--10.77 &9.3--10.76&  25 & 30&60 &58 &2.2~\footnote{The spectral model for this GRB also includes an exponential cutoff with pre-set $e$-folding energy $\ecut$=0.4\,GeV in accordance with Ackermann et al.~\cite{GRB090926A:Fermi}.} & 53 &48 & 47&120 \\
\end{tabular}
\end{ruledtabular}
\end{table*}

\begin{figure}[ht!]
\includegraphics[width=0.9\columnwidth]{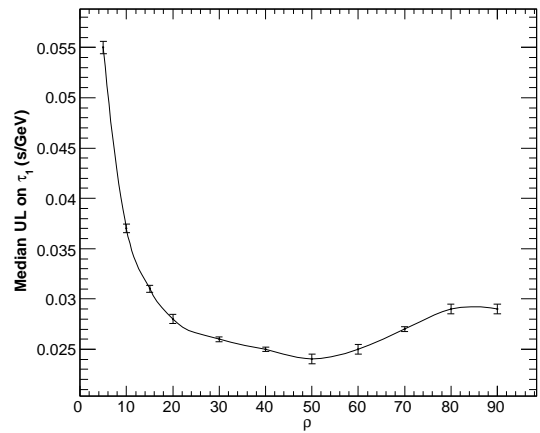}
\caption{\label{fig:090510_ul_vs_kappa}The median of the distribution of 99\% CL upper limits (generated from simulated data sets inspired by the detected light curve) versus $\rho$. The error bars show the 1$\sigma$ statistical uncertainty (arising from the finite number of simulated data sets). This distribution is used for choosing the value of SMM's $\rho$ parameter for the GRB~090510 $n=1$ application.}
\end{figure}

An example plot used for choosing SMM's $\rho$ parameter, here for the case of GRB~090510 and $n=1$, is shown in Fig.~\ref{fig:090510_ul_vs_kappa}. For this case, we chose the value of $\rho$=50, corresponding to the minimum of the curve. The flatness of the curve around the minimum implies a weak dependence of the method's sensitivity on $\rho$ (in the vicinity of the minimum).

\begin{figure*}[ht!]
\includegraphics[width=0.4\textwidth]{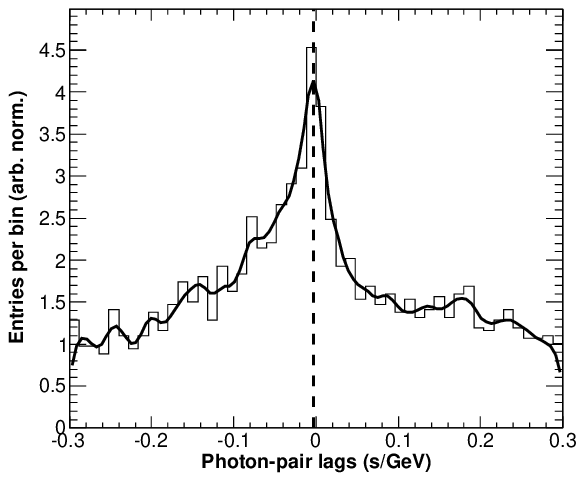}\includegraphics[height=0.308\textwidth,width=0.4\textwidth]{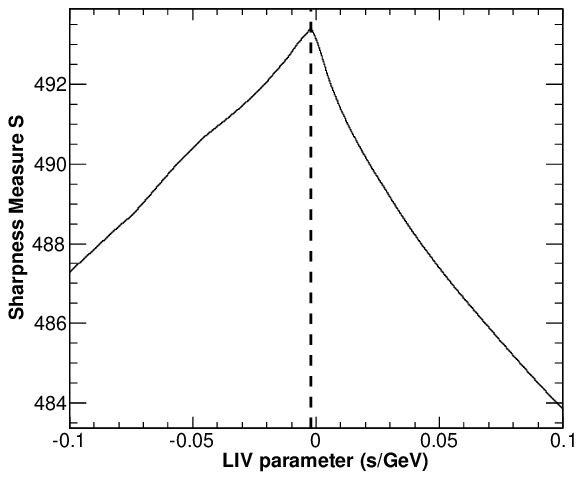}

\includegraphics[width=0.4\textwidth]{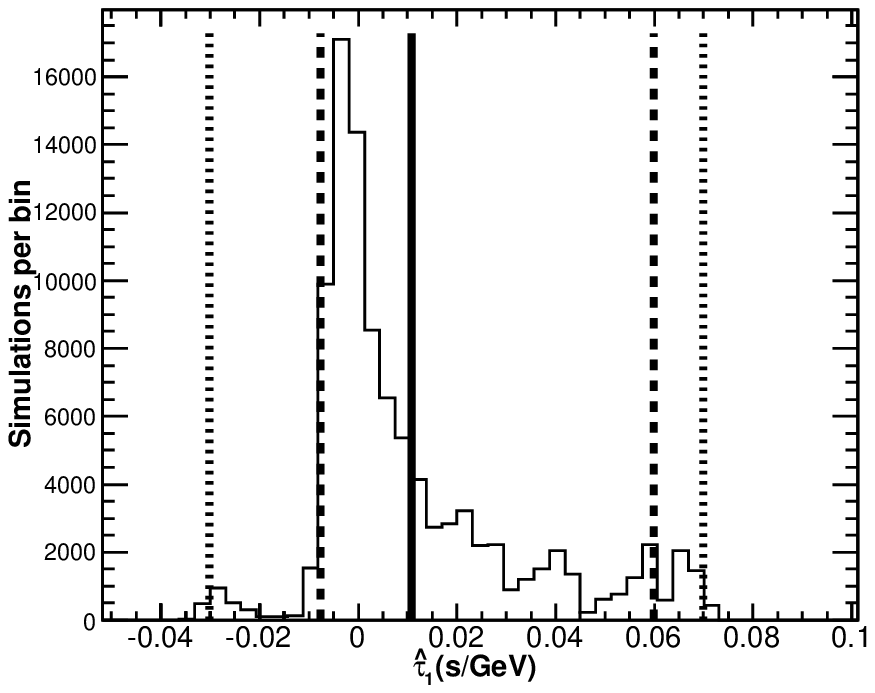}\includegraphics[width=0.4\textwidth]{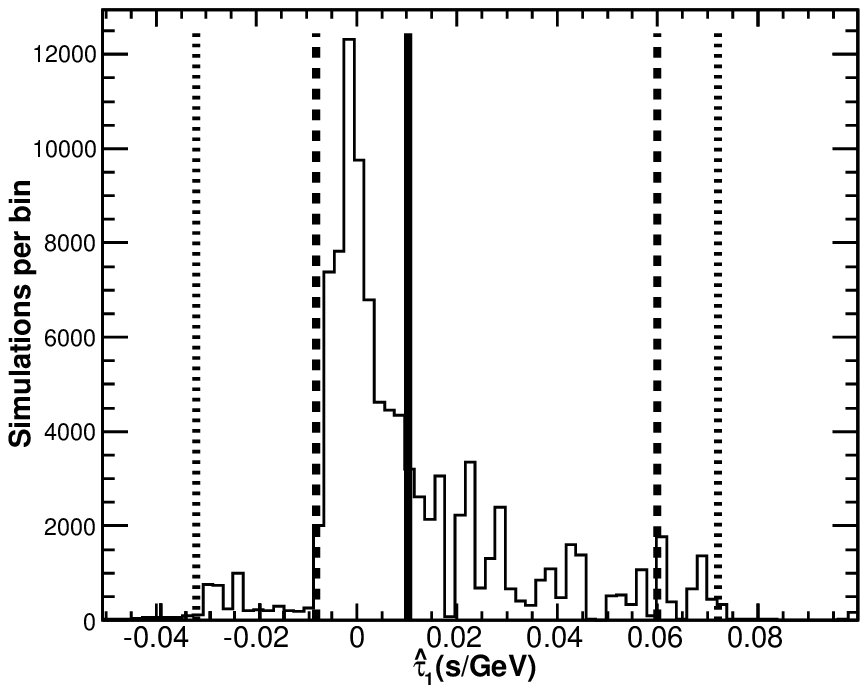}
\caption{\label{fig:090510_result_plots}Plots demonstrating the application of PV (left column) and SMM (right column) on GRB~090510 for $n=1$. Top left: distribution of photon-pair lags (histogram), KDE of the distribution (thick curve), location of the KDE's maximum used as $\tne$ by PV (vertical dashed line). Top right: sharpness measure $\mathcal{S}$ versus trial LIV parameter $\ttot$ (histogram), location of the curve's maximum used as $\tne$ by SMM (vertical dashed line). Bottom row: distributions $f_r$ of the best estimates of the LIV parameter of the randomized data sets (histograms), 5\% and 95\% quantiles (dashed lines), 0.5\% and 99.5\% quantiles (dotted lines), and average value (central solid line).}
\end{figure*}

Figure~\ref{fig:090510_result_plots} demonstrates the application of the PV and SMM methods on GRB~090510 for $n=1$. The top panels show how the best estimate of the LIV parameter is measured, specifically from the location of the maximum of the KDE of the
photon-pair lag distribution for PV (left column) and from the location of the maximum of the sharpness measure $\mathcal{S}$ for SMM
(right column). The bottom panels show the distributions $f_r$ of the best LIV parameters in the randomized data sets, used for constructing the CIs. Their asymmetry and features (inversely) follow the shape of the analyzed light curves. The mean value of $f_r$ can be used as an estimate of the bias of $\tne$. Except for GRB~090510, the magnitude of the bias is considerably smaller than the variance of $f_r$ (i.e., up to $\sim10\%$ of the variance); for GRB~090510, it increases up to 50\% of the variance. The absolute value of the median of $f_r$ is for all cases smaller than $\sim10\%$ of the variance. We correct $\tne$ for biases by subtracting from it the mean value of the $f_r$ distribution. The verification simulations of PV/SMM (described in Appendix~\ref{appendix:PVSMM}) show that the coverage of the produced CIs is approximately proper even for asymmetric or non-zero-mean $f_r$ distributions, such as the ones shown.

\begin{figure}[ht]
\includegraphics[width=0.8\columnwidth]{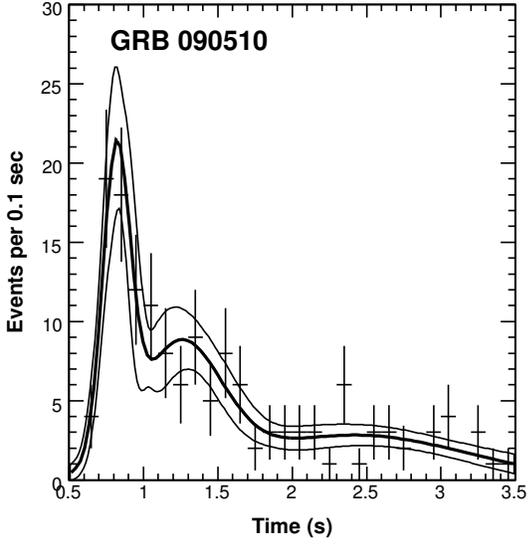}
\caption{\label{fig:likelihood_templates} Template light-curve fit for GRB~090510 used by the ML method. Actual data (histogram), fit template (thick curve), $\pm 1 \sigma$ error ranges of the fit (external thin curves). }
\end{figure}

The light-curve template for GRB~090510 used by the ML method is shown in Fig.~\ref{fig:likelihood_templates}. Any statistical errors involved in the generation of the light-curve templates are properly included in the calibrated CIs of the ML method, as described in Appendix~\ref{appendix:likelihood}.

\begin{figure}[ht]
\includegraphics[width=1.0\columnwidth]{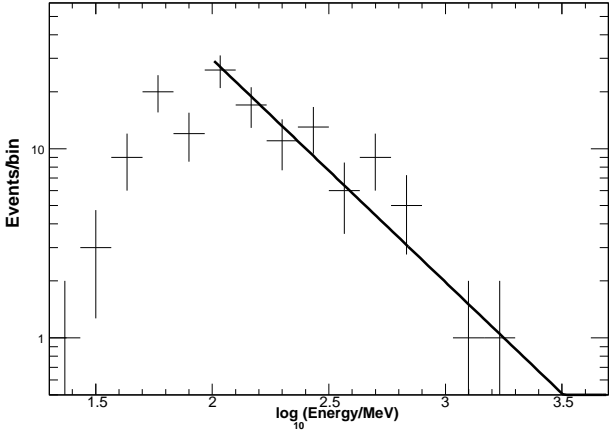}
\caption{\label{080916C_spec_example} Fit of the spectrum of detected events from GRB~080916C used to produce the index $\gamma$ of the $\SP(E)$ function used by the ML method.}
\end{figure}

We show the spectral fit of the \textit{observed} events from GRB~080916C in Fig.~\ref{080916C_spec_example}, which is used to calculate the spectral index $\gamma$ used by the ML method. The drop in the spectrum at low energies is caused by the sharp decrease of the LAT effective area at those energies. In all cases, we choose $E_{\rm cut}$ to be larger than the energy that this instrumental cutoff becomes important. This ensures that the spectral index $\gamma$ of the observed events is a good approximation of the index $\Gamma$ of the incoming GRB flux (within statistical errors). It allows us to considerably simplify the ML analysis by not having to deconvolve the instrument's acceptance from the observed data or having to include the instrument's response in the likelihood function.

\begin{figure}[ht!]
\includegraphics[width=0.5\columnwidth,trim= 20 10 30 20]{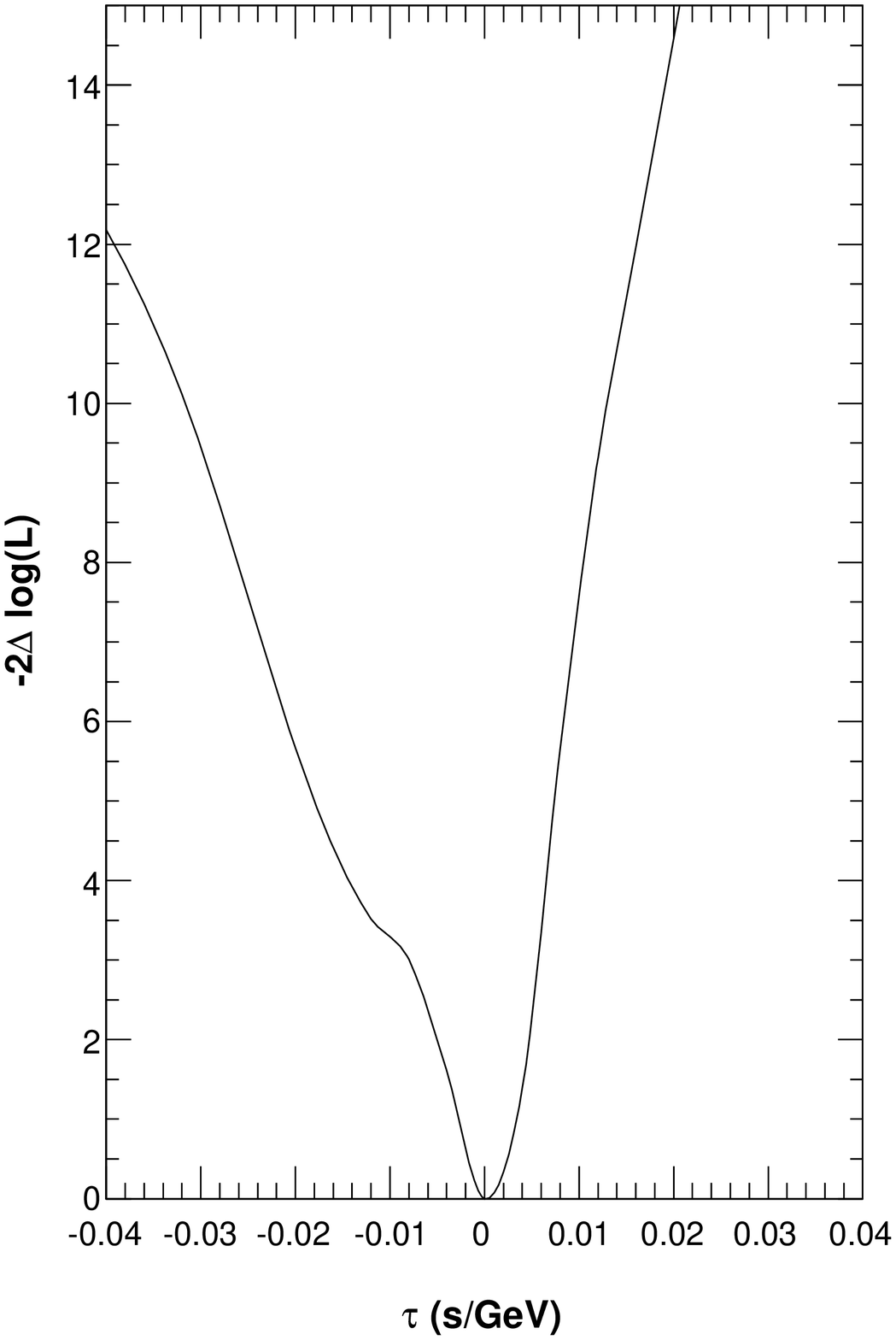}\includegraphics[width=0.5\columnwidth,trim= 20 10 30 20]{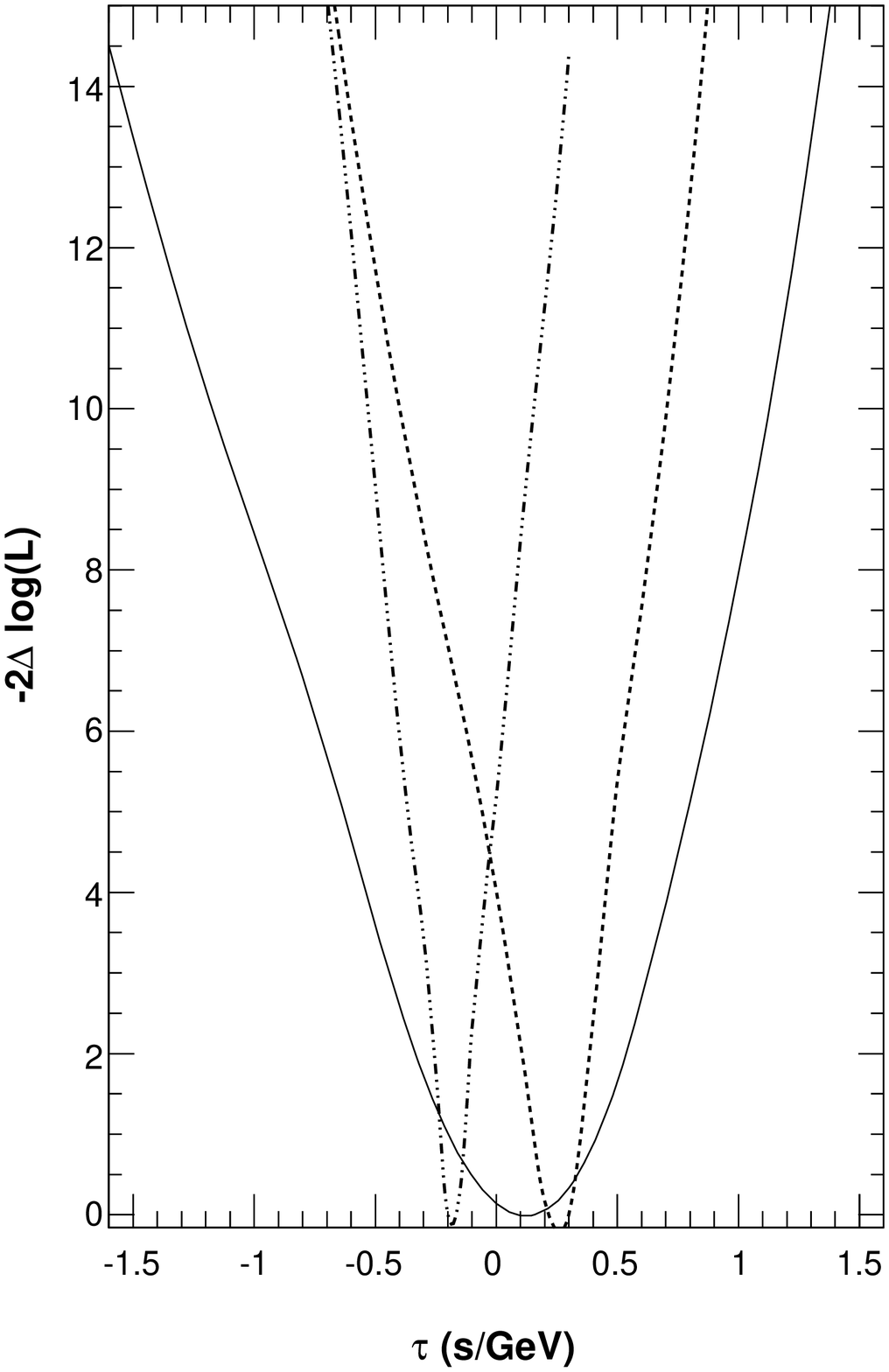}
\includegraphics[width=0.5\columnwidth,trim= 20 10 30 20]{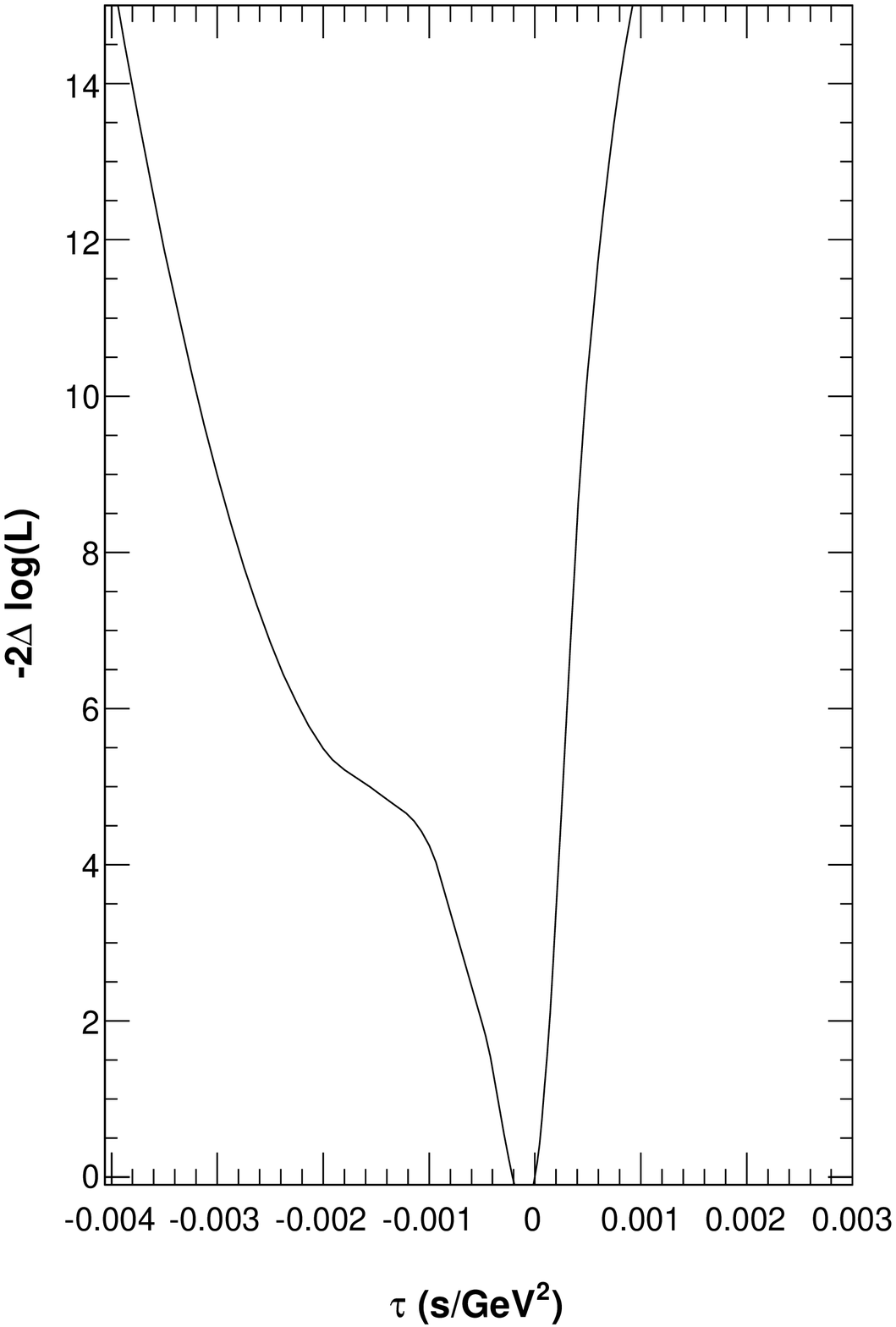}\includegraphics[width=0.5\columnwidth,trim= 20 10 30 20]{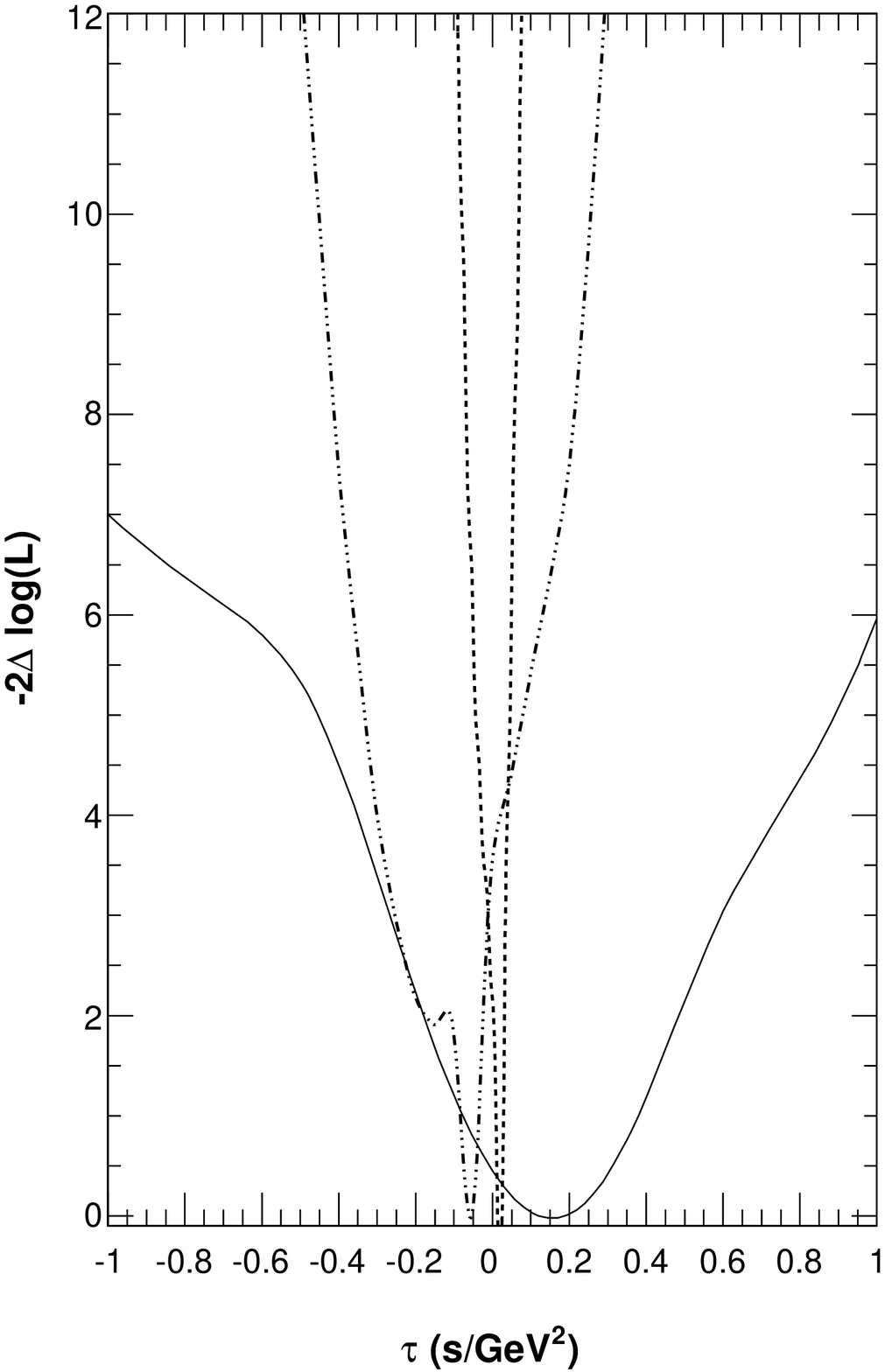}
\caption{\label{fig:like_plots} -2$\Delta$ln($\like$) curves for linear (top row) and quadratic (bottom row) dependence of the speed of light on energy. Left column: GRB 090510. Right column: GRBs 080916C (full line), 090902B (dotted line) and 090926A (dashed double-dotted line).}
\end{figure}

Finally, Fig.~\ref{fig:like_plots} demonstrates the application of the ML method, showing all the \mbox{-2$\Delta \rm{ln}(\like)$} curves. We use the locations of the minima and the shapes of these curves to produce the best estimates and the (obtained directly from the data) CIs on $\ttot$, respectively. These curves are not exactly parabolic (and/or a transformation to a parabolic shape is not always possible). Therefore, any CIs produced based solely on their shape do not have an exactly proper coverage. The calibrated ML CIs (described in Appendix~\ref{appendix:likelihood}) have by construction proper coverage, and are the ones used to constrain the quantities of the LIV models.

\subsection*{Constraints on the Total Degree of Dispersion, $\ttot$}
\begin{figure*}[ht!]
\includegraphics[width=0.5\textwidth]{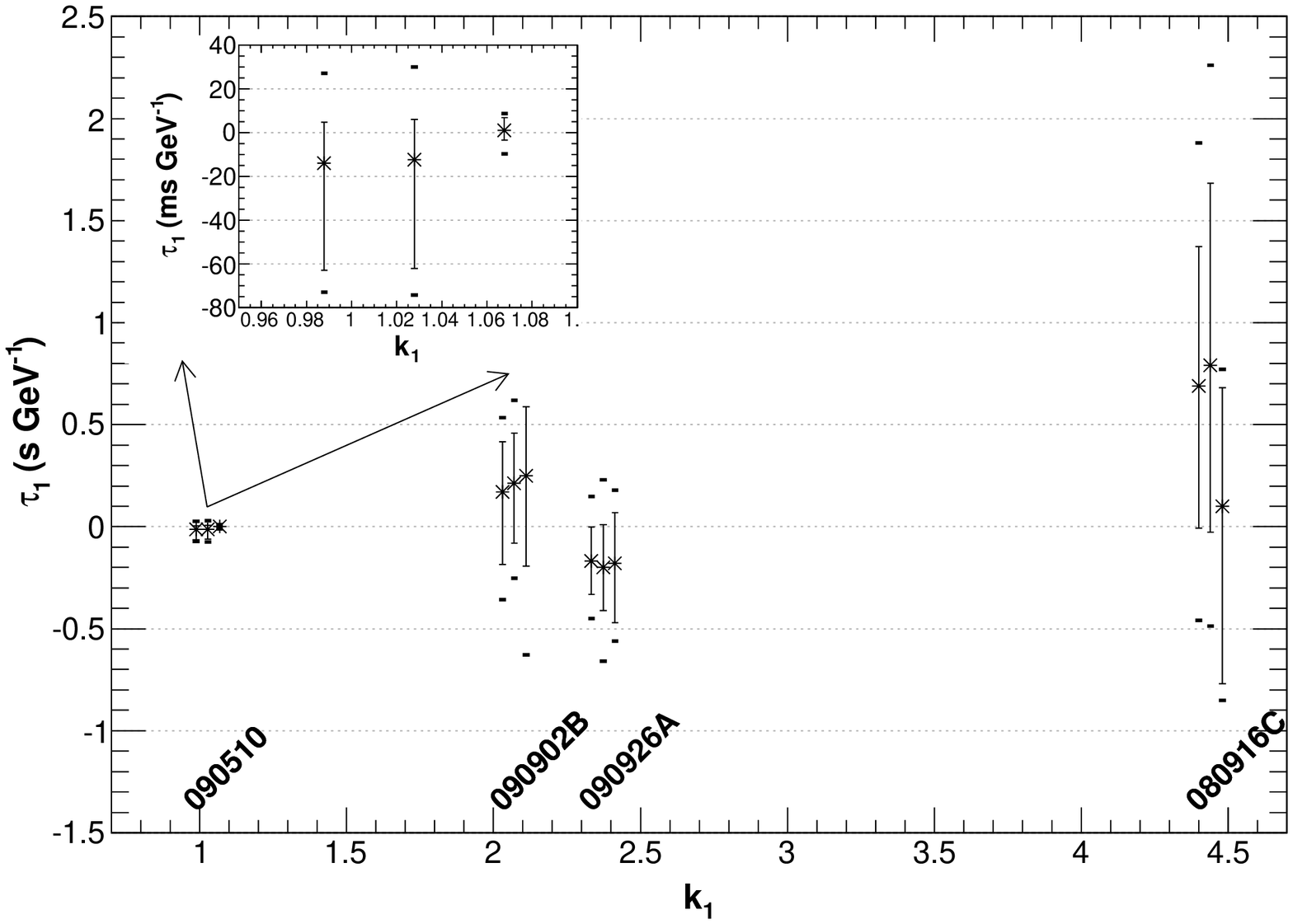}\includegraphics[width=0.5\textwidth]{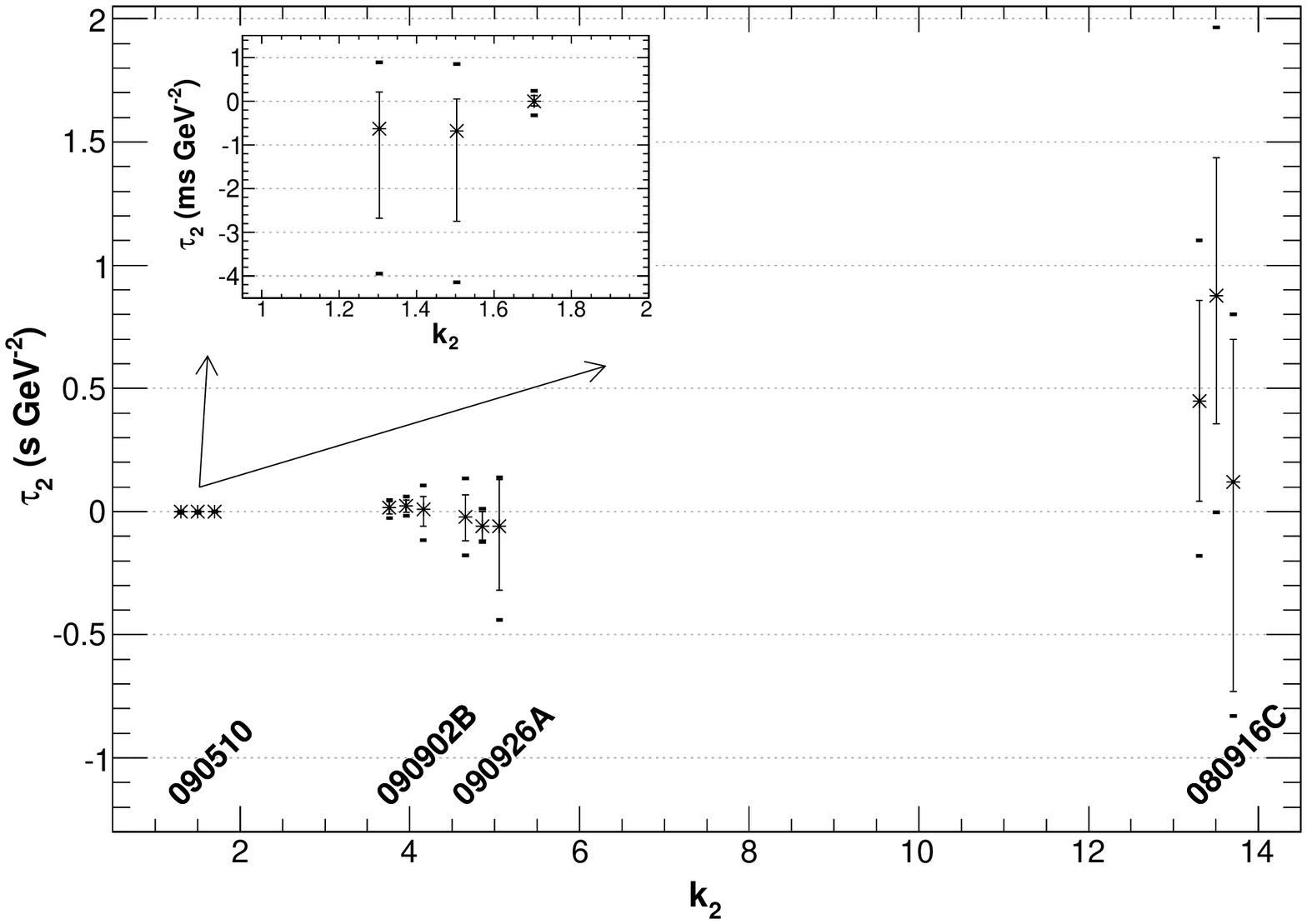}
\caption{\label{fig:ULs_Graph}Our CIs on the total degree of dispersion in the data $\ttot$, obtained without taking into account any source-intrinsic effects, for linear (left panel) and quadratic (right panel) LIV. Each triplet of intervals corresponds to one GRB and shows, left to right, the results of PV, SMM, and ML (calibrated). The PV and ML points are drawn offset for visualization purposes. We present results for two CLs: a 90\% (two-sided) CL denoted by the lines, and a 99\% (two-sided) CL denoted by the external pairs of points.}
\end{figure*}

Table~\ref{tab_results_per_grb} reports our constraints on the \textit{total} degree of dispersion, $\tau_n$, and Fig.~\ref{fig:ULs_Graph} shows our CIs on $\tau_n$ plotted versus the distance $\kappa_n$. According to LIV models (i.e., Eq.~\ref{deltatn}), the magnitude of the observed dispersion due to LIV is proportional to $\kappa_n$. Thus, a positive correlation of $\ttot$ and $\kappa_n$ may imply a non-zero LIV effect. In our case and as can be seen from Fig.~\ref{fig:ULs_Graph}, no such correlation is evident. Additionally, all of our 99\% CIs are compatible with a zero $\ttot$. Both features show that a LIV effect, if any, is dominated in this analysis by statistical and systematic (likely arising from GRB-intrinsic effects) uncertainties. Finally, we note that the results of the three methods (for the same GRB) are in good agreement to each other (i.e., they have considerable overlap), evidence in support of the validity of each method.

\begin{table*}
\caption{\label{tab_results_per_grb}Our measurements on the LIV parameter $\tau_n$ describing the total degree of dispersion in the data. The limits are for a two-sided 99\% CL.}
\begin{ruledtabular}
\begin{tabular}{l ccc|ccc|ccc|ccc}
 GRB Name &   \multicolumn{3}{c}{PairView} & \multicolumn{3}{c}{SMM} & \multicolumn{3}{c}{Likelihood} (from actual data) & \multicolumn{3}{c}{Likelihood (Calibrated)\footnote{These are the ML CIs used for subsequently constraining LIV.}}\\
&\multicolumn{12}{c}{(Lower Limit, Best Value, Upper Limit) (s\,GeV$^{-1}$) $n=1$}\\ \hline
080916C   &-0.46 & 0.69 & 1.9  &-0.49 & 0.79 & 2.3  &-0.75 & 0.1 & 0.72  &-0.85 & -- & 0.77  \\
090510   ($\times 10^{3}$)&-73 & -14 & 27  &-74 & -12 & 30  &-25 & 1 & 6  &-9.8 & -- & 8.6  \\
090902B   &-0.36 & 0.17 & 0.53  &-0.25 & 0.21 & 0.62  &-0.25 & 0.25 & 0.55  &-0.63 & -- & 0.96  \\
090926A   &-0.45 & -0.17 & 0.15  &-0.66 & -0.2 & 0.23  &-0.45 & -0.18 & 0.02  &-0.56 & -- & 0.18  \\
&\multicolumn{12}{c}{(Lower Limit, Best Value, Upper Limit) (s\,GeV$^{-2}$) $n=2$}\\ \hline
080916C   &-0.18 & 0.45 & 1.1  &-0.0031 & 0.88 & 2  &-0.9 & 0.12 & 1.1  &-0.83 & -- & 0.8  \\
090510   ($\times 10^{3}$)&-3.9 & -0.63 & 0.88  &-4.1 & -0.68 & 0.85  &-2.5 & -0.1 & 0.3  &-0.32 & -- & 0.23  \\
090902B   ($\times 10^{3}$)&-26 & 17 & 48  &-18 & 24 & 60  &-60 & 10 & 45  &-120 & -- & 110  \\
090926A   &-0.18 & -0.021 & 0.13  &-0.12 & -0.06 & 0.012  &-0.38 & -0.06 & 0.11  &-0.44 & -- & 0.14  \\
\end{tabular}
\end{ruledtabular}
\end{table*}

\begin{figure*}[ht!]
\includegraphics[width=0.5\textwidth]{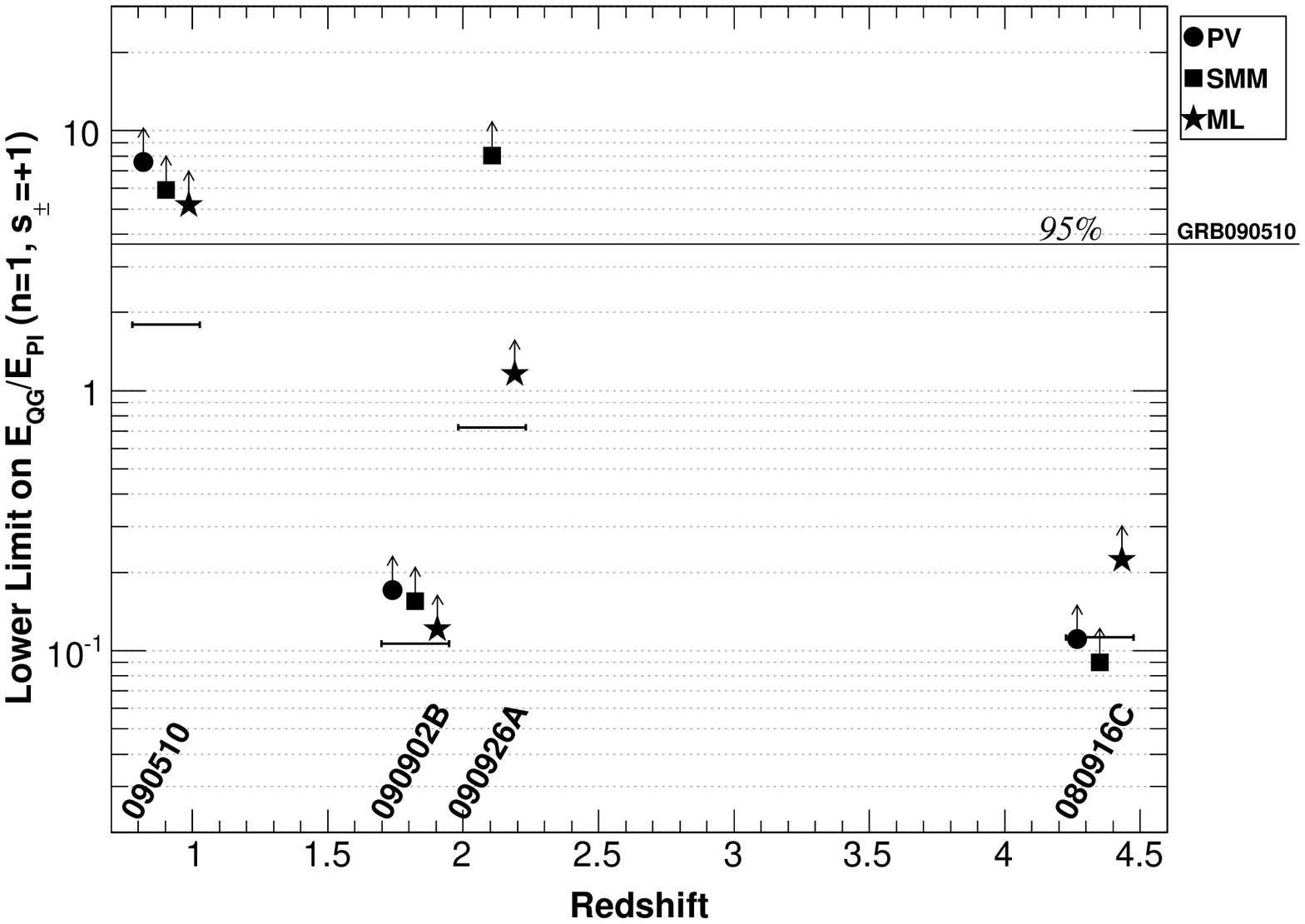}\includegraphics[width=0.5\textwidth]{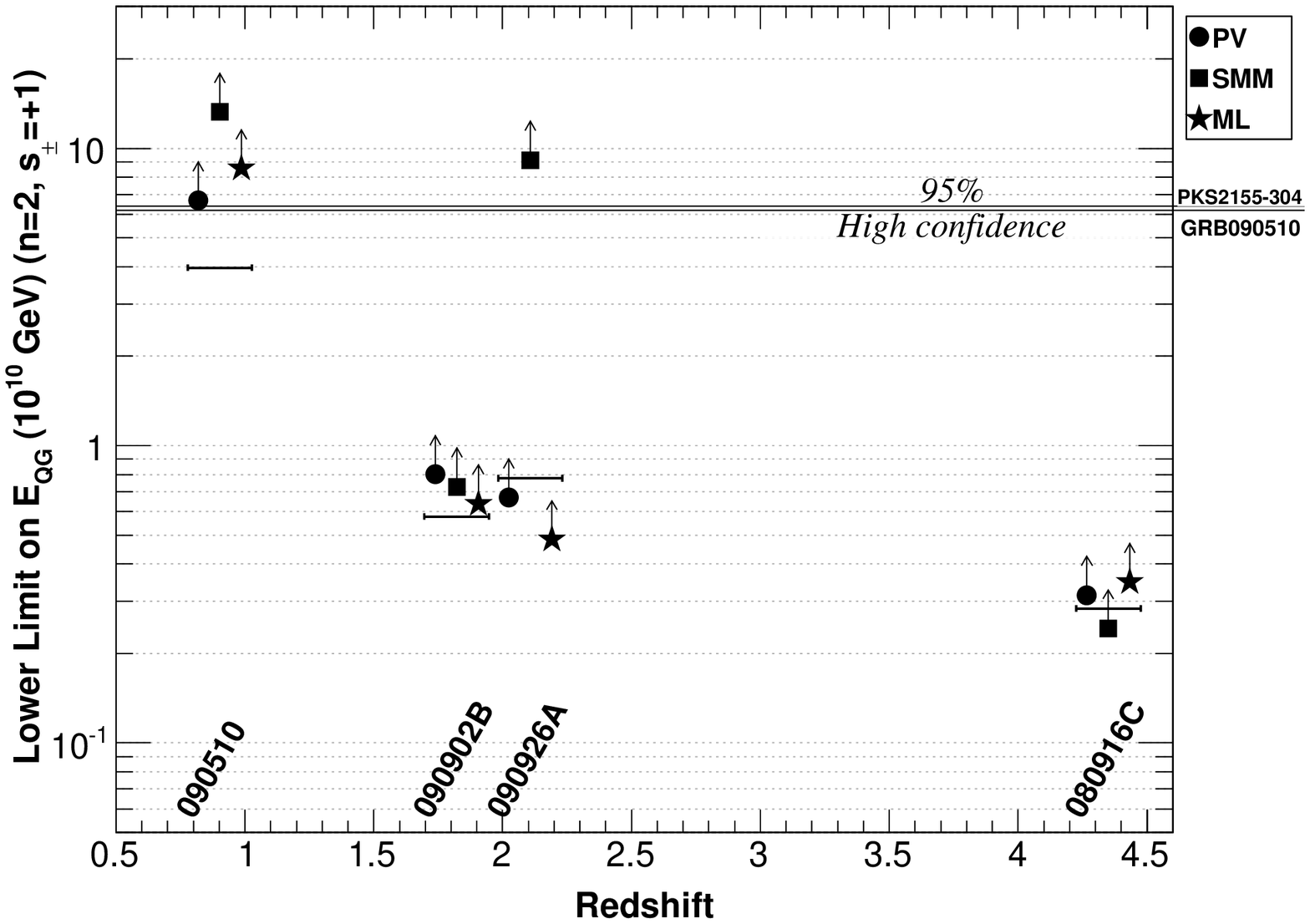}
\includegraphics[width=0.5\textwidth]{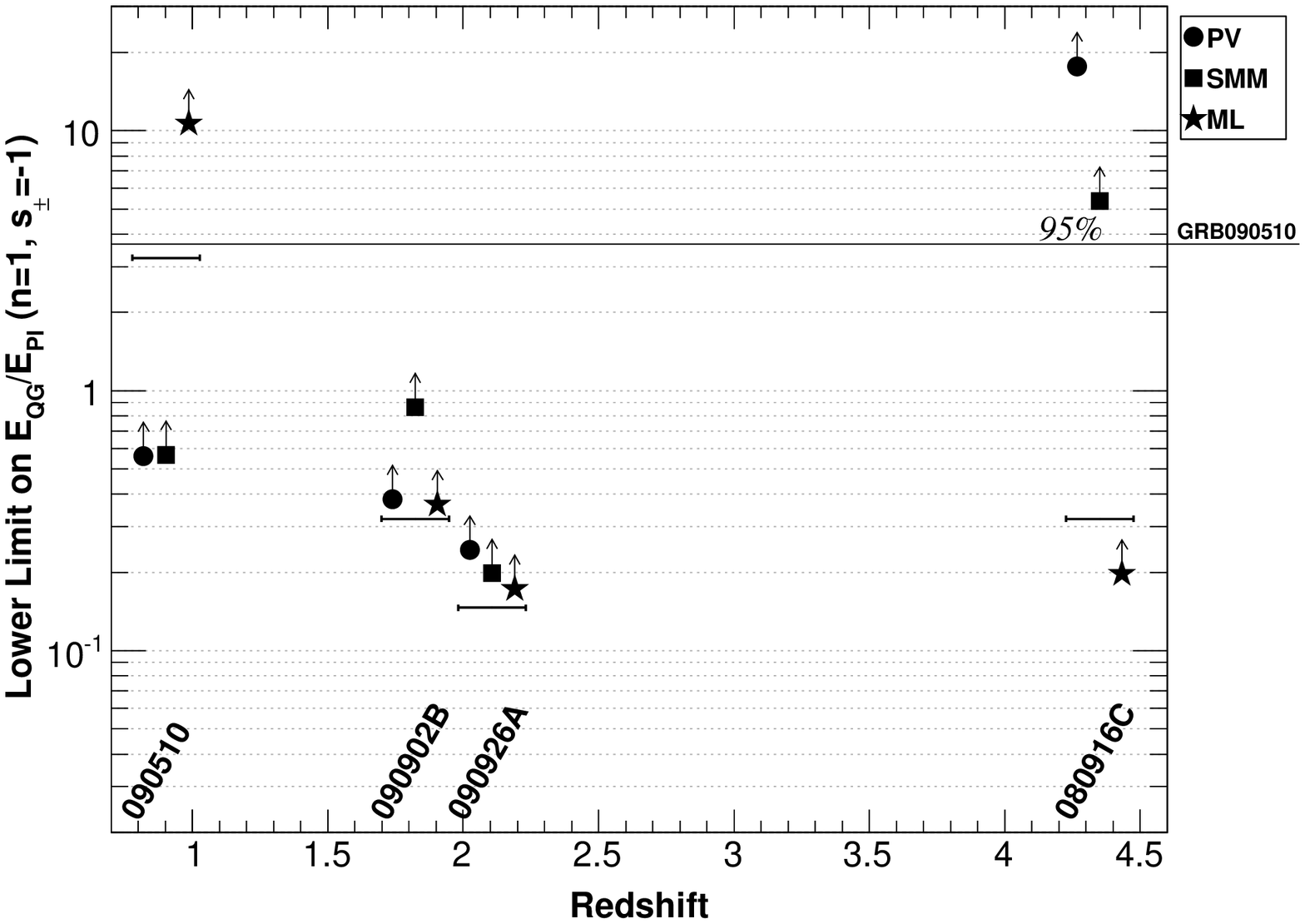}\includegraphics[width=0.5\textwidth]{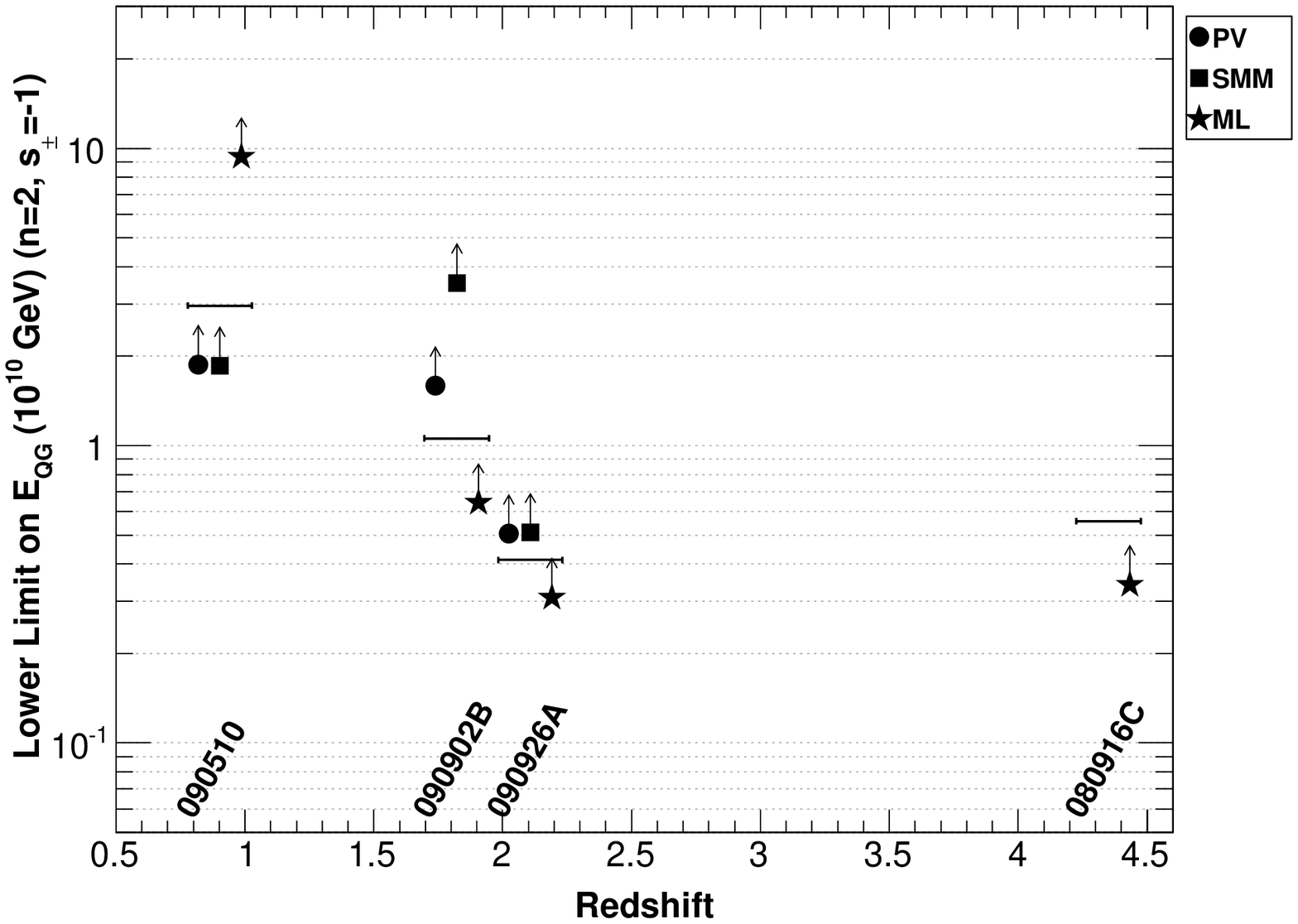}
\caption{\label{fig:ULs_Graph_z}Our 95\% (one-sided) CL lower limits on $\eqgl$ (left) and $\eqgq$ (right) LIV versus the redshift for $\spm=+$1 (top; subluminal case) and $\spm=-$1 (bottom; superluminal case). Similarly to Fig.~\ref{fig:ULs_Graph}, each triplet of markers corresponds to one GRB and shows limits calculated using the constraints on $\ttot$ (i.e., without taking into account any source-intrinsic effects). The horizontal bars correspond to the averaged over the three methods lower limits on $\eqg$ produced using the constraints on $\tliv$  (i.e., after accounting for GRB-intrinsic effects). On the left-hand plots we denote with the horizontal line the limit obtained by \textit{Fermi} on GRB~090510 (DisCan; 95\% limit obtained from paper's Supplementary Information)~\cite{2009Natur.462..331A}. On the top right plot we also denote the ``high confidence'' and ``very high confidence'' limits obtained by \textit{Fermi} on GRB~090510~\cite{2009Natur.462..331A} and the 95\% CL limit from H.E.S.S. study on PKS~2155-304~\cite{hesslike}.}
\end{figure*}

Table~\ref{tab_results_mqg_per_grb} presents lower limits on $\eqg$ calculated using our constraints on $\ttot$. The 95\% lower limits are also plotted versus the redshift in Fig.~\ref{fig:ULs_Graph_z}. These limits do not take into account any GRB-intrinsic spectral evolution. Thus, while they are maximally constraining, they may not be as robust with regards to the presence of such intrinsic systematic uncertainties.

\begin{table*}
\caption{\label{tab_results_mqg_per_grb}Lower Limits on $\eqg$ for linear ($n$=1) and quadratic ($n$=2) LIV for the subluminal ($\mathit{s_\pm}$=+1) and superluminal ($\mathit{s_\pm}$=-1) cases. The CL values are one-sided. These limits were produced using the total degree of dispersion in the data, $\ttot$.}
\begin{ruledtabular}
\begin{tabular}{lcc|cc|cc}
GRB Name & \multicolumn{2}{c}{PairView} & \multicolumn{2}{c}{SMM} & \multicolumn{2}{c}{Likelihood\footnote{Calculated using the calibrated limits.}} \\
& \multicolumn{6}{c}{$n$=1, $\mathit{s_\pm}$=+1 (E$_{Pl}$ units)}\\
& 95\%& 99.5\% & 95\%& 99.5\% & 95\%& 99.5\% \\ \hline
080916C   &0.11 & 0.081 &0.09 & 0.067 &0.22 & 0.2  \\
090510   &7.6 & 1.3 &5.9 & 1.2 &5.2 & 4.2  \\
090902B   &0.17 & 0.13 &0.15 & 0.11 &0.12 & 0.074  \\
090926A   &-- & 0.55 &8 & 0.35 &1.2 & 0.45  \\
& \multicolumn{6}{c}{$n$=1, $\mathit{s_\pm}$=-1 (E$_{Pl}$ units)}\\
& 95\%& 99.5\% & 95\%& 99.5\% & 95\%& 99.5\% \\ \hline
080916C   &18 & 0.33 &5.4 & 0.31 &0.2 & 0.18  \\
090510   &0.56 & 0.48 &0.57 & 0.48 &11 & 3.6  \\
090902B   &0.38 & 0.2 &0.86 & 0.28 &0.37 & 0.11  \\
090926A   &0.24 & 0.18 &0.2 & 0.12 &0.17 & 0.15  \\
& \multicolumn{6}{c}{$n$=2, $\mathit{s_\pm}$=+1 ($10^{10}$~GeV units)}\\
& 95\%& 99.5\% & 95\%& 99.5\% & 95\%& 99.5\% \\ \hline
080916C   &0.31 & 0.28 &0.24 & 0.21 &0.35 & 0.33  \\
090510   &6.7 & 3.3 &13 & 3.3 &8.6 & 6.4  \\
090902B   &0.8 & 0.72 &0.73 & 0.64 &0.64 & 0.49  \\
090926A   &0.67 & 0.48 &9.1 & 1.6 &0.48 & 0.47  \\
& \multicolumn{6}{c}{$n$=2, $\mathit{s_\pm}$=-1 ($10^{10}$~GeV units)}\\
& 95\%& 99.5\% & 95\%& 99.5\% & 95\%& 99.5\% \\ \hline
080916C   &-- & 0.69 &-- & 5.2 &0.34 & 0.32  \\
090510   &1.9 & 1.5 &1.9 & 1.5 &9.4 & 5.4  \\
090902B   &1.6 & 0.97 &3.5 & 1.2 &0.64 & 0.46  \\
090926A   &0.51 & 0.42 &0.51 & 0.5 &0.31 & 0.26  \\
\end{tabular}
\end{ruledtabular}
\end{table*}

Indeed, as we observed from, e.g., Fig.~\ref{fig:ULs_Graph_z}, some of our 90\% CL CIs are offset to a degree that their edges (i.e., limits) are very close to zero (e.g., GRB~090926A). For those CIs, the corresponding limits on $\eqg$ are constraining to a suspicious degree, given the considerably larger width of their CIs. It would be more acceptable if any very constraining limits were associated with correspondingly narrow CIs, contrary to what happens with some of the GRBs in our study. This feature required further scrutiny, hence, we examined our data and results in detail, and concluded that the CIs are offset likely because of GRB-intrinsic spectral evolution effects.

\begin{figure}[ht!]
\includegraphics[width=0.5\columnwidth]{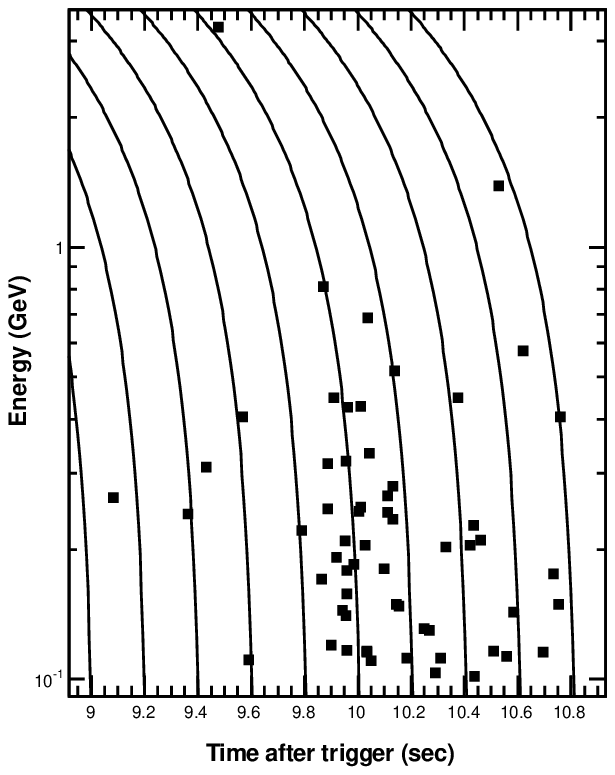}\includegraphics[width=0.5\columnwidth]{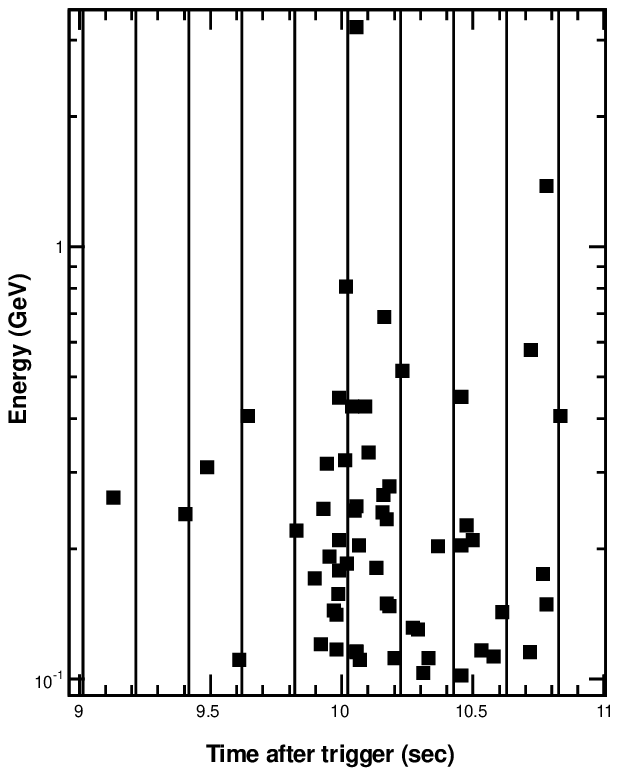}
\includegraphics[width=0.5\columnwidth]{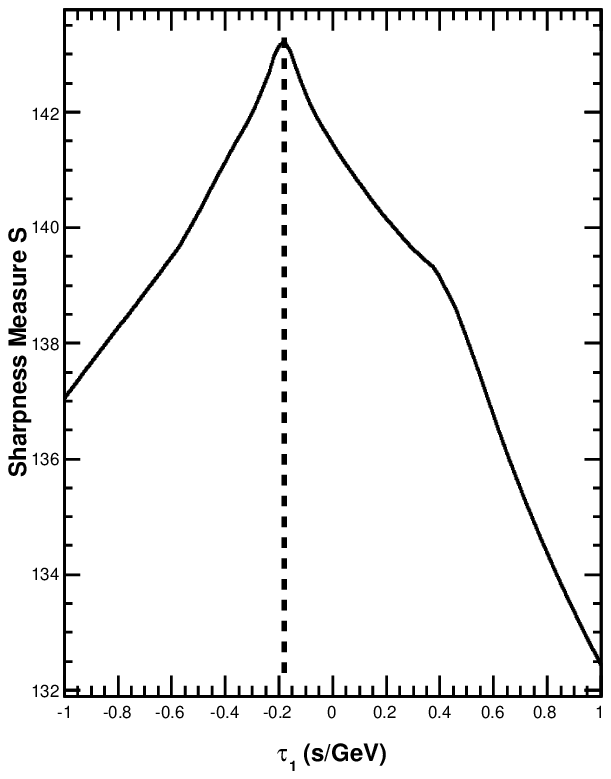}\includegraphics[width=0.5\columnwidth]{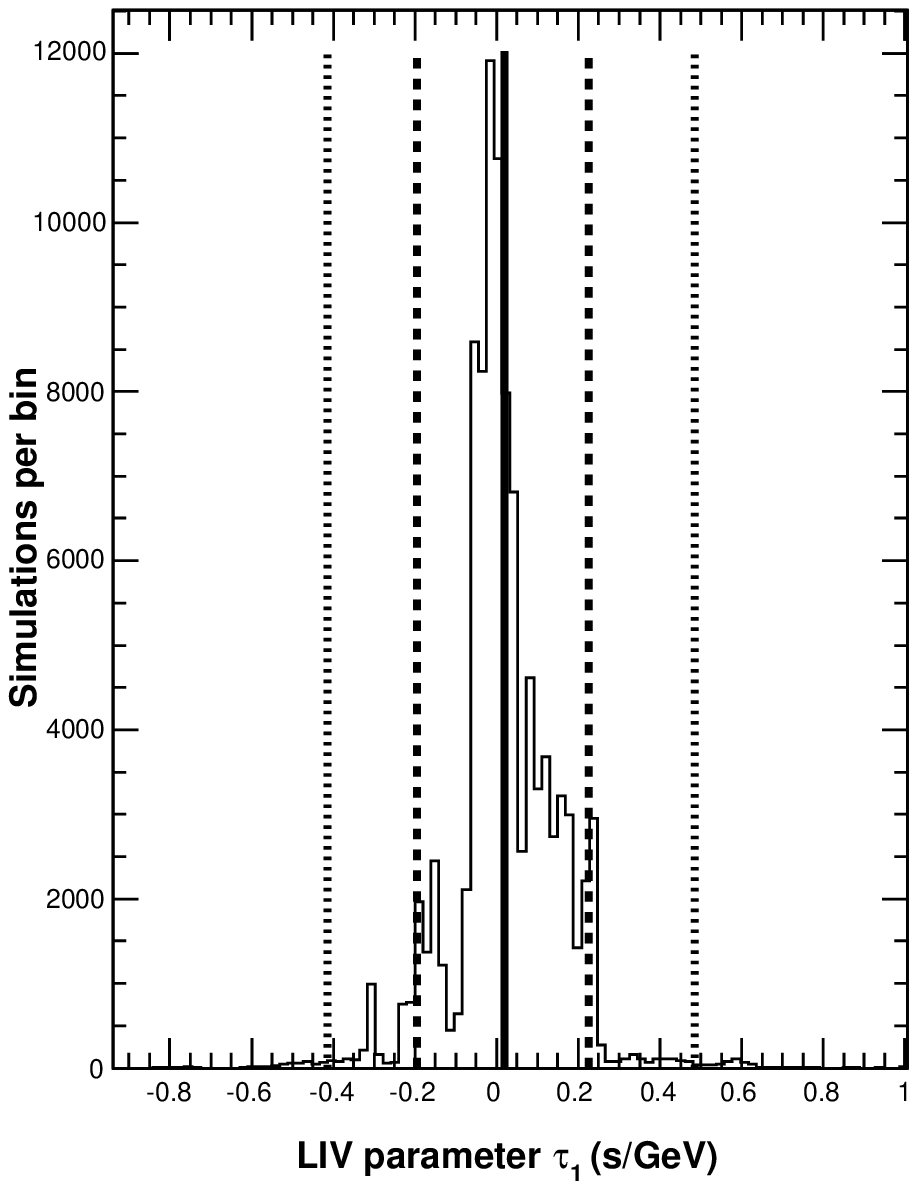}
\caption{\label{fig:090926_case} SMM's results for GRB~090926A and $n=1$. Top left: photons detected in the default time interval, top right: SMM's maximally-sharp version of these events. The curves/lines in the top row act as guides to the eye for the effects of a dispersion equal to the best measured value of $\hat{\tau}_{\rm 1}$=-0.17\,s/GeV (left figure) and zero (right figure). The bottom figures are of the same type as the figures in the right column of Fig.~\ref{fig:090510_result_plots}.}
\end{figure}

For the case of GRB~090926A, the 90\% CL CIs on $\tau_1$ from our three methods, and the CIs on $\tau_2$ from SMM for both CLs are either not consistent with zero or considerably offset towards negative values (something which produces spuriously stringent upper limits on $\ttot$). For example, the $n=1$ CIs (not shown in the tables) are \mbox{(-0.33, -0.17, -0.0010)~s/GeV}\footnote{(lower limit, best estimate, upper limit)} for PV, \mbox{(-0.41, -0.20, 0.010)~s/GeV} for SMM, and \mbox{(-0.25, -0.18, -0.13)~s/GeV} for ML (from data). As a result, the 95\%  CL lower limits on $\eqgl$ for the subluminal case ($\mathit{s_\pm}$=+1) are either suspiciously strong (SMM) or they could not be calculated at all (PV). The top left panel of Fig.~\ref{fig:090926_case} shows the E$>$100~MeV events from GRB~090926A processed by PV and SMM. As can be seen, the highest-energy photon in the data has an energy of $\sim$3~GeV and is detected $\sim$0.5~s before the main pulse. Our three methods
predict that this event was most likely initially emitted in coincidence with the main pulse, and that it had been subsequently advanced by LIV to be detected before it. This case, shown in the top right panel of Fig.~\ref{fig:090926_case}, implies a \mbox{$\hat{\tau}_{1}\simeq$-0.5\,s/3~GeV=-0.17~s/GeV}, in accordance with the measured values.
In the simulations performed for PV and SMM, such relatively small values were rare. Specifically, they occurred in a fraction of the iterations approximately equal to the ratio of the number of photons detected at least as early as the 3~GeV photon (4) over the total number of photons (58 for $n=1$), i.e. only 5-6\%. This resulted in our 95\% (one-sided) CL upper limits on $\tau_1$ being negative or too small.

The physical reason for these too negative CIs and $\hat{\tau}_1$ values may be GRB-intrinsic spectral evolution effects, likely associated with the presence of spectral cutoff $\ecut \simeq$0.4~GeV during the main bright pulse~\cite{GRB090926A:Fermi}. If this cutoff did not exist, more GeV photons might have been detected during this bright pulse, while if the cutoff also existed right before this pulse, the 3~GeV photon might have not been detected. Both cases would correspond to a $\hat{\tau}_{1}$ closer to zero, and weaker, though, less spurious constraints.
We conclude that our results from GRB~090926A are likely affected by a GRB-intrinsic spectral evolution, artificially strengthening (weakening) our limits on $\eqg$ produced using $\ttot$ for the subluminal (superluminal) case.

Contrary to the case of GRB~090926A, for which the results hint towards negative $\tau_{1}$ values, the results from GRB~080916C hint towards positive values. This either does not allow us to calculate lower limits on $\eqg$ for the superluminal case (PV and SMM for $n=2$; 95\% CL) or produces spuriously constraining results (PV and SMM for $n=1$ at 95\% CL, and SMM $n=2$ at 99.5\% CL).

A likely physical explanation for this positive lag is the progressive hardening of the prompt-emission spectrum of GRB~080916C at LAT energies. According to broadband time-resolved spectroscopic studies~\cite{2009Sci...323.1688A}, that spectrum can be adequately described by a Band function, the high-energy component of which, $\beta$, is initially very soft at a value of $-2.63\pm0.12$ during \mbox{[0.004--3.58]~s}, hardens considerably to a value of $-2.21\pm0.03$ during \mbox{[3.58--7.68]~s}, after which it stays constant (within statistics) to a value of $-2.16\pm0.03$ up to at least 15.87~s. Based on this pattern, some soft-to-hard spectral evolution is expected at least for the beginning of our analyzed intervals (\mbox{[3.53--7.89]~s} for $n=1$ and \mbox{[3.53--7.80]~s} for $n=2$). Similarly to the GRB~090926A case, we conclude that our GRB~080916C constraints on $\eqg$ (produced using $\ttot$) might also be affected by GRB spectral evolution, artificially strengthening our superluminal-case limits and weakening our subluminal-case limits for PV and SMM.

Finally, we notice that both of the calibrated ML lower limits on $\tau_n$ for GRB~090510 are considerably more constraining by about an order of magnitude than those from PV/SMM. We feel that this difference can be explained by the reduced sensitivity of the PV/SMM methods for constraining lower limits of the LIV parameter in the presence of long tails of the emission after the main peak, a feature of our chosen data set from GRB~090510. This effect was demonstrated in the one-to-one comparisons of the three methods described in Appendix~\ref{appendix:comparison} and illustrated in the left panel of Fig.~\ref{fig:comparison_1d}. Therefore, we attribute it to differences between the methods' sensitivities.

\subsection*{Constraints Using the LIV-Induced Degree of Dispersion, $\tliv$}

\begin{figure}[t!]
\includegraphics[width=1.0\columnwidth]{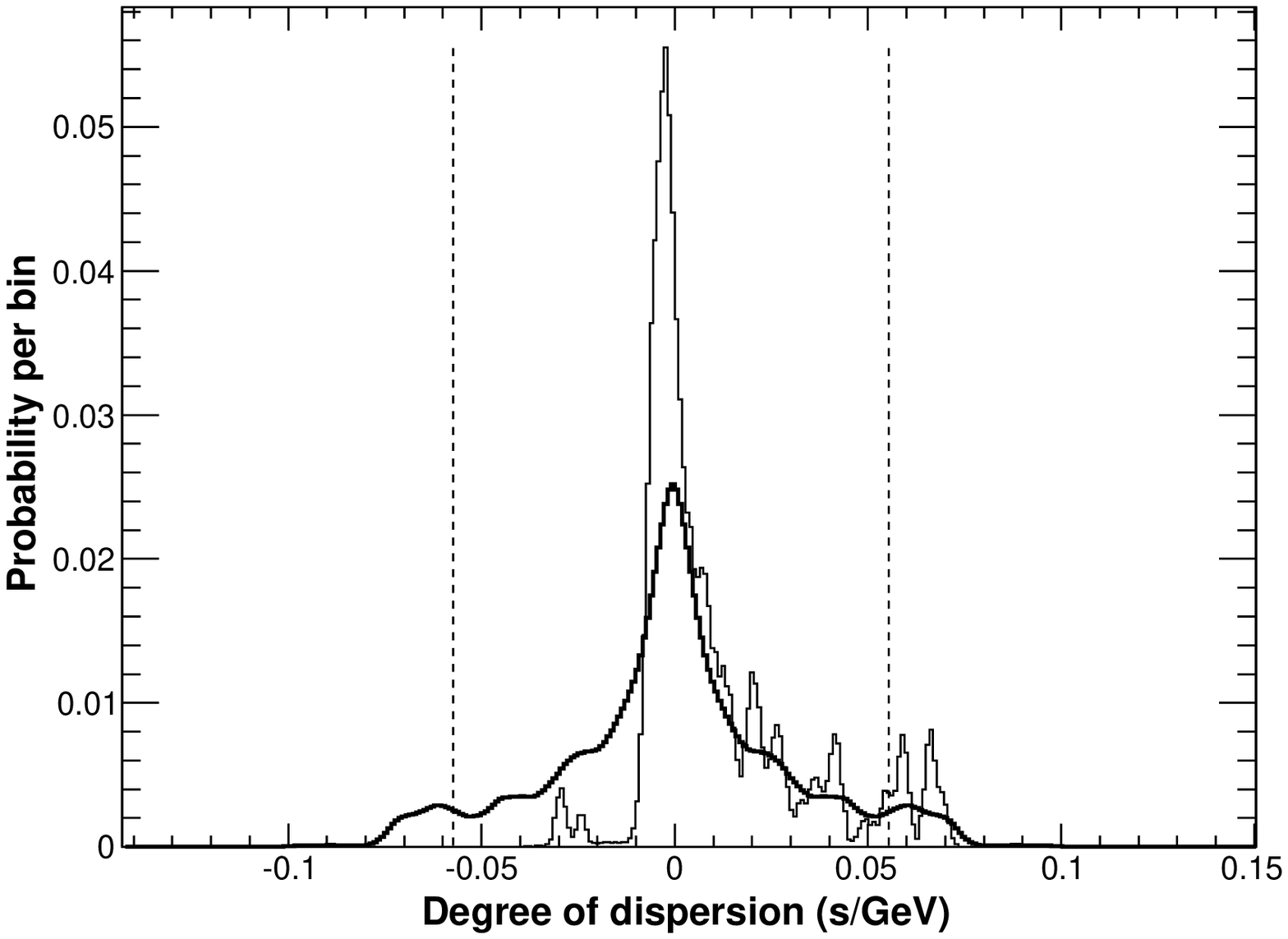}
\caption{\label{fig:090510_AC_example}Demonstration of the generation of CIs on $\tliv$ for GRB~090510, PairView, and $n=1$. The thin curve shows the normalized distribution $f_r$ used to approximate $P_\Err(\err)$ (the same distribution is also shown with different binning in the bottom left panel of Fig.~\ref{fig:090510_result_plots}), the thick curve shows the autocorrelation function $P_{\Err'}(\err')$ calculated using Eq.~\ref{eq:AC}, and the dashed lines show its 5\% and 95\% quantiles used for constructing the 90\% CL CI on $\tliv$.}
\end{figure}

The spuriously strong limits mentioned above imply that our sensitivity actually reaches the level of GRB-intrinsic effects. This motivated us to produce an additional set of constraints, this time on $\tliv$, taking into account intrinsic effects and according to the methodology in Sec.~\ref{subsec:intrinsic_corrections}. As an illustration of this method, Fig.~\ref{fig:090510_AC_example} shows the intermediate plots involved the calculation of the CI on $\tliv$ for GRB~090510, PairView, and $n=1$.

For simplicity we do not report our CIs on $\tliv$. Instead, we just report the final limits on the LIV-model quantities, after averaging over the three methods. Tables~\ref{tab:eqg_limits_corrected} and \ref{tab:sme_limits_corrected} show our new 95\% CL limits on $\eqg$ and on the SME coefficients, respectively. Our lower limits on $\eqg$ are also illustrated with the horizontal bars in Fig.~\ref{fig:ULs_Graph_z}, along with those produced without correcting for intrinsic effects (from $\ttot$; shown with the markers). As can be seen, the limits produced using $\tliv$ are considerably weaker than those produced using $\ttot$. The biggest difference is for the cases of GRBs~090926A and 080916C, which had some spuriously strong limits that we attributed above to source-intrinsic effects.

\begin{table*}
\caption{\label{tab:eqg_limits_corrected}Our 95\% CL lower limits on $\eqg$, averaged over the three methods and calculated using the CIs on $\tliv$ (i.e., taking into account GRB-intrinsic effects).}
\begin{ruledtabular}
\begin{tabular}{l|cccc}
GRB Name &\multicolumn{2}{c}{$n=1$ ($E_{Pl}$)}&\multicolumn{2}{c}{$n=2$ (10$^{10}$~GeV)}\\
         &$\mathit{s_\pm}=+1$ & $\mathit{s_\pm}=-1$ & $\mathit{s_\pm}=+1$ &$\mathit{s_\pm}=-1$ \\ \hline
080916C & 0.11 & 0.32 & 0.28 & 0.56 \\
090510 & 1.8 & 3.2 & 4.0 & 3.0 \\
090902B & 0.11 & 0.32 & 0.58 & 1.1 \\
090926A & 0.72 & 0.15 & 0.78 & 0.41 \\
\end{tabular}
\end{ruledtabular}
\end{table*}

\begin{table*}
\caption{\label{tab:sme_limits_corrected}Our 95\% CL limits on the SME coefficients, averaged over the three methods and calculated using the CIs on $\tliv$ (i.e., taking into account GRB-intrinsic effects).}
\begin{ruledtabular}
\begin{tabular}{l|c|c|cccc}
Model & Source & Quantity &  Lower Limit ($10^{-20}$~GeV$^{-2}$) & Upper Limit ($10^{-20}$~GeV$^{-2}$)\\ \hline
Vacuum & 080916C & $\sum_{jm}{_0Y_{jm}}(145^\circ,120^\circ)c^{(6)}_{(I)jm}$ &-8.7 &20 \\
 &090510 & $\sum_{jm}{_0Y_{jm}}(117^\circ,334^\circ)c^{(6)}_{(I)jm}$  &-0.31 &0.16 \\
&090902B & $\sum_{jm}{_0Y_{jm}}(63^\circ,265^\circ)c^{(6)}_{(I)jm}$  &-3.4 &5.2 \\
&090926A & $\sum_{jm}{_0Y_{jm}}(156^\circ, 353^\circ)c^{(6)}_{(I)jm}$ &-11 &5.2 \\
Vacuum isotropic& 080916C & $c^{(6)}_{(I)00}$ &-31 &70 \\
 &090510 & $c^{(6)}_{(I)00}$  &-1.1 &0.57 \\
&090902B &$c^{(6)}_{(I)00}$  &-12 &18 \\
&090926A &$c^{(6)}_{(I)00}$ &-37 &19 \\
\end{tabular}
\end{ruledtabular}
\end{table*}

\section{Systematic Uncertainties}
\label{sec:systematics}

In this section, we discuss several systematic effects potentially influencing our results, namely those originating from the source and those having instrumental origins. Any dispersion induced by non-GRB standard physical processes is expected to be negligible compared to the dispersion produced by LIV~\cite{0264-9381-21-12-L01}.

\subsection*{Systematic Uncertainties from GRB-Intrinsic Effects}
\label{subsec:grbintrinsic}
GRB-intrinsic effects that can cause systematic uncertainties in our results fall into two main categories:
\begin{itemize}
 \item the presence of multiple spectral components in the data not evolving with temporal coincidence, and
 \item spectral evolution during the course of the burst or during each individual pulse.
\end{itemize}
 A full physical modeling of the emission processes occurring in the GRBs considered here is beyond the scope of this paper. Instead, we utilize published time-resolved spectral analyses to estimate the influence of any observed spectral evolution on our results. In the initial \Fermi papers on the GRBs analyzed in this study ~\cite{2009Sci...323.1688A,GRB090510:ApJ,GRB090902B:Fermi,GRB090926A:Fermi}, the prompt-emission spectra were fitted in relatively coarse time
bins from keV to GeV energies with the combination of the empirical Band function with a
high-energy power law. It was found that typically the Band component peaks at $\lesssim$MeV energies whereas the power-law component becomes dominant at LAT energies (i.e., above $\sim$100~MeV).

In the case of GRB\,080916C, the spectrum was well fitted by a Band function only, while the significance of the existence of an additional power-law component was found to be small. Some soft-to-hard spectral evolution could be present in the beginning of our analyzed intervals, as was discussed in the previous section. The broadband keV--GeV spectrum of the other three bursts is best represented by a combination of both spectral components:
\begin{itemize}
\item in GRB~090510, the high-energy power law starts from the onset of the main emission in the LAT (at $\sim$0.5~s post-trigger) and dominates the Band component at energies above $\sim$100\,MeV after $\sim$0.7~s post-trigger.
\item In GRB~090902B, the high-energy power law is detected from the trigger time, and completely dominates the Band component in the LAT energy range. The spectral hardness of the emission in the LAT energy range is relatively constant during the time interval analyzed in this study.
\item In GRB~090926A, the high-energy power-law starts at the time of the bright pulse observed at $\sim$10~s post-trigger and persists until $\sim$22~s. Our analyzed time interval corresponds to the main bright pulse, during which the power law component dominates the emission in the LAT energy range, while exhibiting a high-energy spectral break with a cutoff energy $\ecut\sim$0.4~GeV.
\end{itemize}

Since the two spectral components may be possibly originating from different physical regions of the burst and/or may be generated by physical processes evolving in different time scales, one might not necessarily expect them to be detected with exact temporal coincidence. This might lead to spurious signals originating from intrinsic effects rather than LIV. There is only one case (GRB~090510) for which the LAT data during the analyzed time intervals cannot be sufficiently approximated to contain a single spectral component, discussed in detail below.

Using the spectral fits published in Ref.~\cite{GRB090510:ApJ}, we estimate that about half of the LAT-detected events from GRB~090510 below $\sim$100-200\,MeV can be attributed to the Band component during the main episode observed around $\sim$0.8~s post-trigger (comprising the bulk of the events in our analyzed time interval). This non-negligible fraction can potentially affect the ML method, which essentially compares the time profiles between the low (below $\sim$100--150~MeV) and high energy emissions in the LAT. On the other hand, its effect on the PV and SMM methods is weaker because these methods analyze a subset of the data produced with a higher-energy cut (E$>$100~MeV), for which only $\lesssim$15\% of the events are estimated to be associated with the Band component. Fortunately, evidence from cross checks performed in this work and from previously published results show that this effect has likely a negligible influence on our results. Specifically, a cross-correlation analysis between the time profiles of the keV--MeV emission (dominated by the Band component) and of the $>$100\,MeV emission (dominated by the power-law component) of GRB\,090510 from 0.6~to~0.9~s~\cite{GRB090510:ApJ} did not show any evidence of a time lag between the two spectral components. Furthermore, as shown in Appendix~\ref{appendix:Checks}, the PV and SMM CIs produced using the data above 30~MeV are in good agreement with the results produced with the default cut of $E>100$~MeV. We conclude that the inclusion of events related to the Band component for GRB~090510 did not cause any significant distortions in any of our analyses.

Another potential source of systematic uncertainties is the spectral evolution detected with high significance in most LAT GRBs. One of its manifestations is the E$>$100~MeV emission detected by the LAT having a systematically delayed onset with respect to the keV/MeV emission detected by the GBM~\cite{Piron2012}. Even though this spectral evolution can manifest as LIV, it happens so rapidly that typically only a very small fraction of the emission is detected during this transition. Furthermore, after the emission in the LAT is established, it usually continues with a relatively stable degree of spectral hardness, at least according to the coarsely binned time-resolved spectral analyses mentioned above.

For example, for the case of GRB~090510, cross-correlation analyses between the GBM-detected keV/MeV and LAT-detected E$>$100~MeV emissions revealed that the onset of the E$>$100~MeV emission trailed the onset of the keV/MeV emission by $\sim$0.2--0.3~s~\cite{GRB090510:ApJ}. This offset implies the existence of a delay between the LAT data below and above 100~MeV, something that can potentially affect our results. However, the number of events detected during the onset of the LAT emission ($\sim$0.5--0.75~s) is negligible. Specifically, only $\sim$8\% of the events above 30~MeV and within the default $n=1$ interval were detected during the onset of the LAT emission. Furthermore, and as mentioned above, once the GRB~090510 emission is establishes in the LAT energy range, its spectral hardness remains relatively stable. We conclude that spectral evolution during the course of the emission of GRB~090510 affects only a very small fraction of the analyzed events. Thus, it is not expected to have a considerable influence on our results.

GRB\,090926A is a peculiar case, as a strong spectral variability has been observed even after the onset of the high-energy emission in the
LAT~\cite{GRB090926A:Fermi}. From the trigger time and until $\sim$10~s, the high-energy power-law component is not detectable and the emission is well described by a single Band component. At $\sim$10~s, a bright pulse appears (comprising the bulk of the events in our analyzed time interval), during which the power-law component becomes very bright dominating the emission and exhibiting a spectral cutoff at high energies. After the bright pulse, the two components become comparable in flux, while the cutoff of the power-law component appears to be increasing in energy. Clearly, the results of a LIV analysis on an interval wide enough to include all these spectral-evolution effects would be strongly affected by them. By focusing only on a narrow snapshot of the GRB~090926A's emission (i.e., the main bright pulse), during which the GRB spectrum is assumed not to vary too much\footnote{It should, however, be noted that even though an increase of the cutoff energy within the pulse could not be significantly detected due to the limited LAT statistics at GeV energies, an interpretation of this cutoff as arising from internal-opacity effects does predict an associated evolution during the course of the spike~\cite{2008ApJ...677...92G,2012MNRAS.421..525H}.}, we considerably reduced our exposure to such effects.

At shorter time scales, the spectral hardness of GRB pulses is known to be correlated to their intensity and fluence at keV--MeV energies~\citep{ryde02}.
Due to the difficulty of measuring the GRB spectrum on a pulse-per-pulse basis with the limited photon statistics available to the LAT, there has been no evidence that this correlation extends to higher energies. However, the light curves of the GRBs analyzed in this study exhibit sharp peaks and fast variability, thus the presence of any
spectro-temporal correlations at high energies might, in principle, affect our results.
This incomplete knowledge of GRB properties at high energies constitutes an intrinsic limitation of our model (e.g., it is unclear if the factorization in \eqc{eq:flux} holds at LAT energies at short time scales) and represents a major source of systematic uncertainty in any GRB-based study of LIV.

\subsection*{Systematic Uncertainties from Instrumental Effects}
The probability of the LAT detecting an event of some energy depends on many factors, including the off-axis angle of the photon, with the probability
being approximately inversely dependent on the off-axis angle. As a result, a constant-spectrum source observed at progressively larger (smaller) off-axis angles
will correspond to a data set of a progressively harder (softer) average reconstructed energy. Such a data set may erroneously appear as containing a non-zero degree of
spectral evolution. Fortunately, this effect is negligible for our observations since for the time scales under consideration the off-axis angles of
the GRBs were almost constant.

The energy-reconstruction accuracy of the LAT depends primarily on the true energies of the events. For the analyzed data set, about 90\% of photons with
energy greater than 1\,GeV are predicted to have a reconstructed energy within $\pm \sim$20\% of their true energy~\cite{FermiPass7}, which can be used to produce a rough estimate of the error on the produced limits on $\eqg$ of up to 20\% (90\% CL). To verify this rough estimate we generated a collection of data sets derived from GRB~090510 by smearing the detected energies according to the energy dispersion function of the instrument. For simplicity, during the production of the data sets we assumed that the detected energy was the true one. The 90\% and 99\% CL upper and lower limits varied by a factor of $\sim$10\% ($n=1$) and $\sim$15\% ($n=2$)
(1$\sigma$), in agreement with the rough estimate.

The effective area of the LAT, corresponding to the P7\_TRANSIENT\_V6 selection used in this work, is typically an increasing function of the energy up to $\sim100$~GeV. It starts from a zero value at few MeV and increases with increasing energy at a rate that is initially rapid but then gradually flattens above $\sim100$~MeV. In the ML analysis, we have ignored the dependence of the effective area on the energy and approximated the spectrum of incoming events with the spectrum of detected events (i.e., $\gamma\simeq\Gamma$). Because of this dependence, the spectrum of detected events appears slightly harder (less steep) than the spectrum of incoming events. This could affect the results of the ML analysis, depending on how sensitive it is on using an exactly correct spectral index. However, we have verified that the difference between the two spectral indices is always smaller than the statistical error of our measured spectral index, i.e., $|\gamma-\Gamma|<\sigma_{\Gamma}$. Thus, any systematic uncertainties by this approximation are dominated by the statistical uncertainty of determining the true source spectrum.

The effects from background contamination are expected to be negligible, since the background rate for our data selection is very low, of the order of 0.1~(10$^{-3}$)~Hz above 0.1~(1)~GeV.

The errors on the redshifts have a negligible effect on the lower limits on $\eqg$. A 1$\sigma$ change in the redshift of GRB~080916C causes
a relative change of about $10^{-2}$ on the final limits. For the other GRBs in our sample, the relative change is also negligible, at the level of
$10^{-3}$ or smaller. The errors on the cosmological parameters give an error of $\sim$3\%.

\section{Discussion/Conclusion}
\label{sec:Conclusion}

We derive strong upper limits on the total degree of dispersion, $\ttot$, in the data of four LAT-detected GRBs. We use three statistical methods, one of which (PV) was developed as part of this study. The previously published most stringent limits on $\ttot$ (at 95\% CL; subluminal case) have been obtained for $n=1$ by the \Fermi GBM and LAT collaborations using GRB~090510 ($\eqgl>$3.5\,$\epl$; DisCan; Supplementary Information of Ref.~\cite{2009Natur.462..331A})\footnote{That work also reported lower limits on $\eqgl$ as high as 10~$\epl$. These limits, however, were not associated with a well quantified confidence level, but rather with a degree of confidence (``very high'' to ``medium''). Thus, they cannot be directly compared to the exact-CL limits produced in this work.} and for $n=2$ by H.E.S.S. using the bright flares of PKS~2155-304 ($\eqgq>$6.4$\times$10$^{10}$\,GeV; ML~\cite{hesslike}). Our results from GRB~090510, namely $\eqgl >$7.6\,$\epl$ (PV) and $\eqgq>$1.3$\times 10^{11}$\,GeV (SMM), improve these constraints by a factor of $\sim$2.~~\footnote{At the 99\% level, we improve on the $\Fermi$ limits  $\eqgl>$1.2~$\epl$ (DisCan) by a factor of $\sim4$.}

In the above comparisons we do not consider other more constraining limits that were either produced in a very model-dependent manner or are of a moderate statistical significance. Specifically, Chang et al.~\cite{2012APh....36...47C} tried to take into account intrinsic GRB time delays by estimating them using the magnetic-jet GRB-emission model. However, our knowledge of GRB physics is not complete enough to be able to predict such intrinsic lags with sufficient certainty. Thus, even though such an approach proceeds in the right direction, it is highly sensitive to the particular choice and configuration of the employed model.
Nemiroff et al.~\cite{2012PhRvL.108w1103N} took an innovative approach with which they zoomed in on the micro-structure of the burst's emission above 1~GeV to produce very stringent constraints that were based, however, on observables of low statistical significance.\footnote{They identified two pairs and one triplet of E$>$1~GeV photons in a 0.17~s interval of GRB~090510, with each photon being detected within $\sim$1~ms of each other. The triplet, which contained the 31~GeV photon, was used to place a stringent constraint. They calculated a probability of $\sim$3~$\sigma$ for such a bunching of photons to arise by chance from a uniform emission in time. However, this significance is overestimated since it doesn't account for the number of trials taken. Moreover, it does not reflect the confidence of their limit, since it strongly relies on associating the emission time of the 31~GeV photon with a tentative ms ``pulse''.}

To investigate why our GRB~090510 results are more constraining than the previous \Fermi analysis of the same GRB, we applied the PV method to the same exact data used in the original \Fermi publication. We used identical energy, time, and event selection cuts (as reported for the DisCan method), and obtained again more constraining results than the original \Fermi publication by a factor of $\sim$2--4 (depending on the CL). Additionally, we repeated our PV and SMM analyses using the configuration determined in this paper (i.e., time interval, energy range, $\rho$) but using the P6\_V3\_TRANSIENT event selection of the previous \Fermi work. The resulting constraints were again of equal or higher strength (see Appendix~\ref{appendix:Checks}). These results show that the methods employed in this work are more sensitive than the previous \Fermi analyses.

Our measurements are compatible with a zero degree of total dispersion in all the analyzed GRBs (at 99\% CL). However, among these results, there are some spuriously strong limits on the total degree of dispersion, which we interpret as a product of GRB-intrinsic spectral-evolution effects.

Using a maximally conservative set of assumptions to account for GRB-intrinsic effects, we constrain any residual dispersion in the data attributed solely to LIV, $\tliv$. The resulting CIs on $\tliv$ are less stringent than those on $\ttot$, albeit more robust with respect to the presence of GRB-intrinsic effects, and thus, more appropriate for constraining LIV. Our assumptions describe the worst-case scenario for GRB-intrinsic effects, and, as such, correspond to a maximum overall decrease in sensitivity. Our best constraints in the linear/subluminal case at 95\% CL are \mbox{$\eqgl\gtrsim$2~$\epl$} for GRB~090510 and \mbox{$\eqgl\gtrsim$0.1~$\epl$} for the other three GRBs. We obtain results of similar strength in the linear/superluminal case.

As a final note we would like to mention that we considered combining the results from the four GRBs to produce a single result that would be more constraining and/or less affected by any GRB-intrinsic spectral-evolution effects (hoping that they might average out). However, we noticed that our GRB~090510 measurement is overall considerably more constraining than the other three cases. Thus, a combined result would not be very different from that of GRB~090510. Additionally, we do not expect that the intrinsic spectral-evolution effects for short GRBs (i.e., GRB~090510) are similar to those in long GRBs (other three cases). Thus, a combined analysis aimed at producing more robust results would have to be performed on short and long GRBs separately. Also, because our sample contains only a small number of long GRBs, we do not expect the average of their intrinsic effects to be an accurate representation of the typical long-GRB intrinsic evolution. Therefore, a combined result obtained using the three long GRBs, would still be considerably less robust compared to each of the maximally conservative CIs on $\tliv$ we produced here. e conclude that there are no sufficient sensitivity or robustness benefits that a combined analysis of this limited data set can bring.

There are many theoretical indications that Lorentz invariance may well break down at energies
approaching the Planck scale. They come from the need to cut off the UV divergences in quantum field theory and black hole entropy calculations~\cite{1996PhRvL..77.3288R}, from various quantum gravity scenarios such as in loop
quantum gravity~\cite{1991PhRvD..44.1740A}, some string theory and M-theory scenarios, and non-commutative geometry models. There is one way to prescribe Lorentz invariance; there are many ways to violate Lorentz invariance.
Kinematic tests of Lorentz invariance violation in QED depend on the possibility that the
Lorentz violating terms can be different for electrons and photons~\cite{2004PhRvL..93b1101J}. It becomes even more complicated when hadronic interactions are considered. Many of
these other tests, while quite sensitive, depend on the differences between the individual maximum attainable velocities
of various particle species~\citep{1999PhRvD..59k6008C}. In the context of effective field theories \cite{2003PhRvL..90u1601M}, birefringence tests have already produced very strong constraints on LIV \cite{2011PhRvD..83l1301L,2011APh....35...95S}. Photohadronic interactions have also provided some powerful constraints~\cite{2009NJPh...11h5003S}.

One particular model inspired by string theory concepts presents the prospect that only photons
would exhibit an energy-dependent velocity~\cite{2008PhLB..665..412E}. This model envisions
a universe filled with a gas of point-like D-branes that only interact with photons. It predicts that vacuum has an energy-dependent index of refraction that causes only a retardation.
Since all photons are retarded, there is no associated vacuum birefringence effect, even though
the degree of retardation has a first order dependence on the photon energy. The absence of an associated birefringence and the low-order ($n=1$) dependence of the predicted delay on the photon energy, render our results particularly unique for testing such a model
\footnote{It has been argued that the D-brane model in Ref.~\cite{2008PhLB..665..412E} would suppress pair production interactions of ultrahigh energy (UHE) photons with cosmic microwave background photons, resulting in a flux of UHE photons in conflict with observations~\cite{2010PhRvL.105b1101M}. This would appear to be an independent argument against it. However, in Ref.~\cite{2010PhLB..694...61E}, it was argued that because electrons are not affected by the D-brane medium and because the pair production interaction involves an internal electron at the tree level, the resulting LIV effect in pair production is suppressed. Thus, this model is not ruled out by constraints on the UHE photon flux.}. Indeed, our constraints obtained using the total degree of dispersion, $\ttot$, reiterate and support the previously-published results from $\textit{Fermi}$~\cite{2009Natur.462..331A}, strongly disfavoring by almost an order of magnitude this model, and, in general, any class of models requiring $\eqgl \lesssim \epl$. Our maximally-conservative set of constraints obtained using $\tliv$ also support the above statement.

More GRB observations at high energies will allow us to improve GRB models and produce robust predictions on GRB-intrinsic delays (i.e., on $\ptint$), which will lead to more constraining CIs on $\tliv$. Additionally, a larger collection of CIs on $\ttot$ can be used for disentangling LIV-induced delays, which have a predicted dependence on the redshift, from the source-frame value\footnote{The degree of intrinsic dispersion at the source is smaller than the observed degree of (intrinsic) dispersion at the Earth by a factor of $(1+z)^{n+1}$ due to the relativistic expansion of the universe causing time dilation and redshift.} of GRB-intrinsic delays, which can be assumed to not have a redshift dependence or at least to have a different dependence than $\tliv$ (see for example the approach in Refs.~\cite{LamonIntegral, 2008ApJ...676..532B, 2006APh....25..402E}).
Future simultaneous observations of GRBs at MeV/GeV energies with \textit{Fermi}-LAT and at GeV energies with HAWC~\cite{2012APh....35..641A} will have considerably increased statistics at GeV energies and a lever arm that extends to an even higher energy than this work, properties that can provide uniquely constraining results.

\begin{acknowledgments}
The \textit{Fermi} LAT Collaboration acknowledges generous ongoing support
from a number of agencies and institutes that have supported both the
development and the operation of the LAT as well as scientific data analysis.
These include the National Aeronautics and Space Administration and the
Department of Energy in the United States, the Commissariat \`a l'Energie Atomique
and the Centre National de la Recherche Scientifique / Institut National de Physique
Nucl\'eaire et de Physique des Particules in France, the Agenzia Spaziale Italiana
and the Istituto Nazionale di Fisica Nucleare in Italy, the Ministry of Education,
Culture, Sports, Science and Technology (MEXT), High Energy Accelerator Research
Organization (KEK) and Japan Aerospace Exploration Agency (JAXA) in Japan, and
the K.~A.~Wallenberg Foundation, the Swedish Research Council and the
Swedish National Space Board in Sweden. We also would like to thank French "GDR PCHE" (Groupement de Recherche - Ph\'enom\`enes Cosmiques de Hautes Energies) for their support during the preparation of this paper.This research has made use of NASA's Astrophysics Data System. J.G thanks R. J. Nemiroff for useful conversations about his recent work on this topic. The authors would also like to thank Jan Conrad for reading the manuscript and giving suggestions.
\end{acknowledgments}

\bigskip
\appendix
\section{\label{appendix:PVSMM}PV and SMM Verification Tests}

We thoroughly tested the PV and SMM methods using a large number of simulated data sets to check for biases on the best estimates of the LIV parameter, to verify the proper coverage of the produced CIs, and to examine the robustness of the techniques (e.g., to find which properties of the data could alter the validity and accuracy of the results).

We performed the verification tests on a variety of collections of data sets, with each collection corresponding to a different light-curve and spectrum template, and to a different LIV parameter. The data sets of some collection represented the possible outcomes of the observation of the same exact source
by a large number of identical detectors. By comparing the fraction of produced CIs that included the true value of the LIV parameter to their CL, we verified the coverage of these CIs. By repeating this exercise on a diverse collection of data sets (produced with, e.g., different statistics, number of pulses, light-curve asymmetry, pulse shape, spectral properties, degree of dispersion), we verified the robustness of the techniques.

Our verification tests were performed on collections comprising ten thousand simulated data sets, with each of these sets being constructed in two steps: first its photon energy--time pairs were randomly sampled from the light-curve and spectral template of the particular collection, and then a common degree of dispersion was applied.

\begin{figure}[ht!]
\includegraphics[width=1.0\columnwidth]{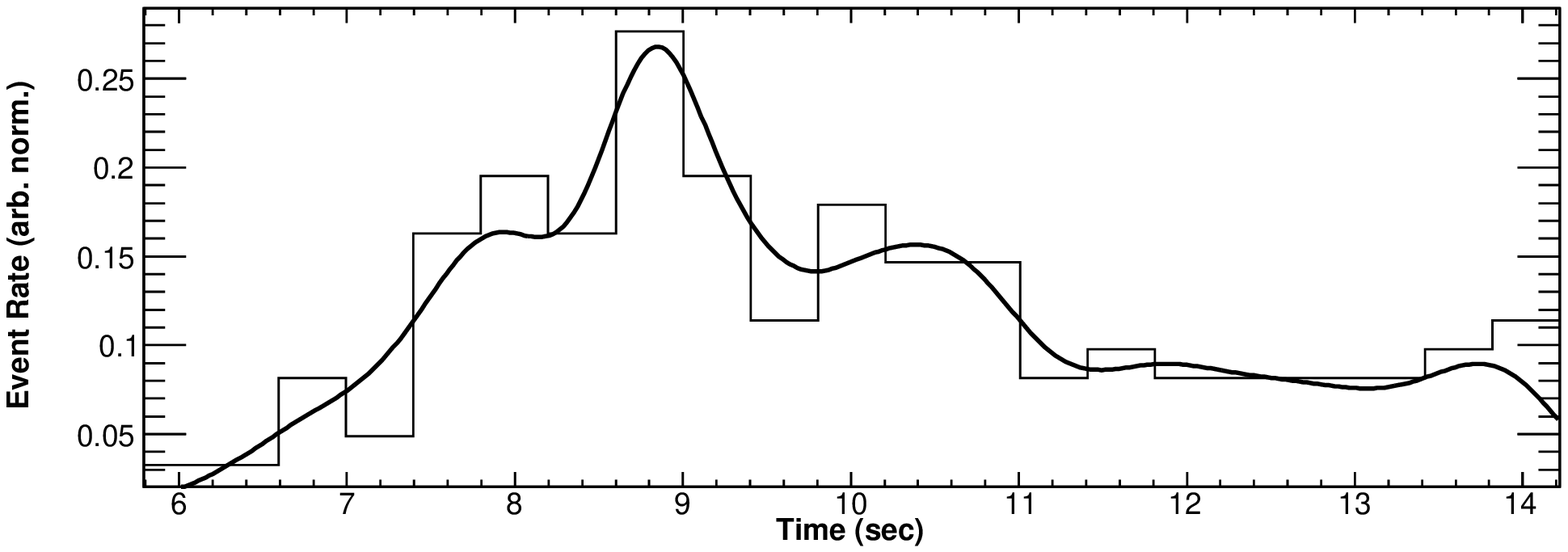}
\includegraphics[width=1.0\columnwidth]{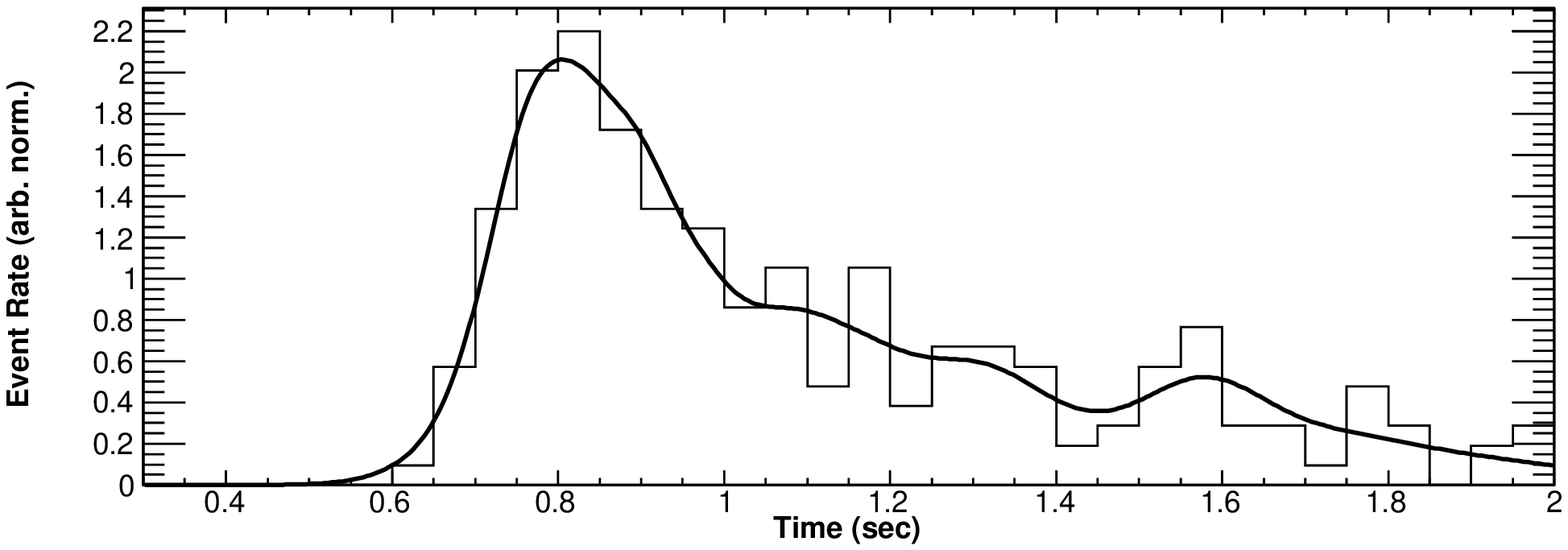}
\caption{\label{fig:lc_templates}Two of the light-curve templates used in our verification tests, one inspired by GRB~090902B (top) and one by GRB~090510 (bottom). The histograms are the GRB light curves (as actually detected), and the thick curves are our templates (produced by a KDE of the histograms).}
\end{figure}

We used two kinds of functional templates for the light curve. We started with simple synthetic templates composed of superpositions of Gaussian pulses of different widths, amplitudes, and means, and continued with templates inspired from the actually-observed GRB light curves. As an example, we show in Fig.~\ref{fig:lc_templates} two of the light-curve templates used in our simulation. Both were inspired by actual detections, namely GRBs~090902B (top panel) and 090510 (bottom panel), representative cases of a long and short GRB, respectively. We obtained the light-curve templates using kernel density estimation (shown with the curve) of the actual light curves (histograms).

In all tested configurations, the energy spectrum of the non-dispersed data sets followed a power law, and extended from 100\, MeV to 40\, GeV. For the GRB-based data sets we used an identical number of events as in the actual observations, and for the synthetic ones we simulated a range that was similar to that typically observed.

We chose the maximum range of tested LIV parameters so that the simulated degree of dispersion did not distort the tested data too much. This way, we avoided the unrealistic possibility of having the highest-energy photons be disjoint and external from the bulk of the emission. To accomplish this, the magnitude of the tested LIV parameter was not considerably larger than about the light-curve half-width divided by the highest simulated energy raised to the $n$ power.

We did not include any energy and temporal reconstruction instrumental effects (i.e., it was assumed that all photons were detected with the same energy-independent and constant-in-time efficiency).
A full simulation including the LAT response to the GRB signal would also model any effects from a time-dependent effective area and of any inaccuracies in the event-energy reconstruction. The dependence of the results on both factors is expected to be very small, as discussed in Sec.~\ref{sec:systematics}.

As a demonstration of the verification process we present some of the diagnostic plots produced using the GRB~090510 light-curve template shown in the bottom panel of Fig.~\ref{fig:lc_templates} and a zero LIV parameter.

\begin{figure}[ht!]
\includegraphics[width=1\columnwidth,trim=20 30 50 0,clip=true]{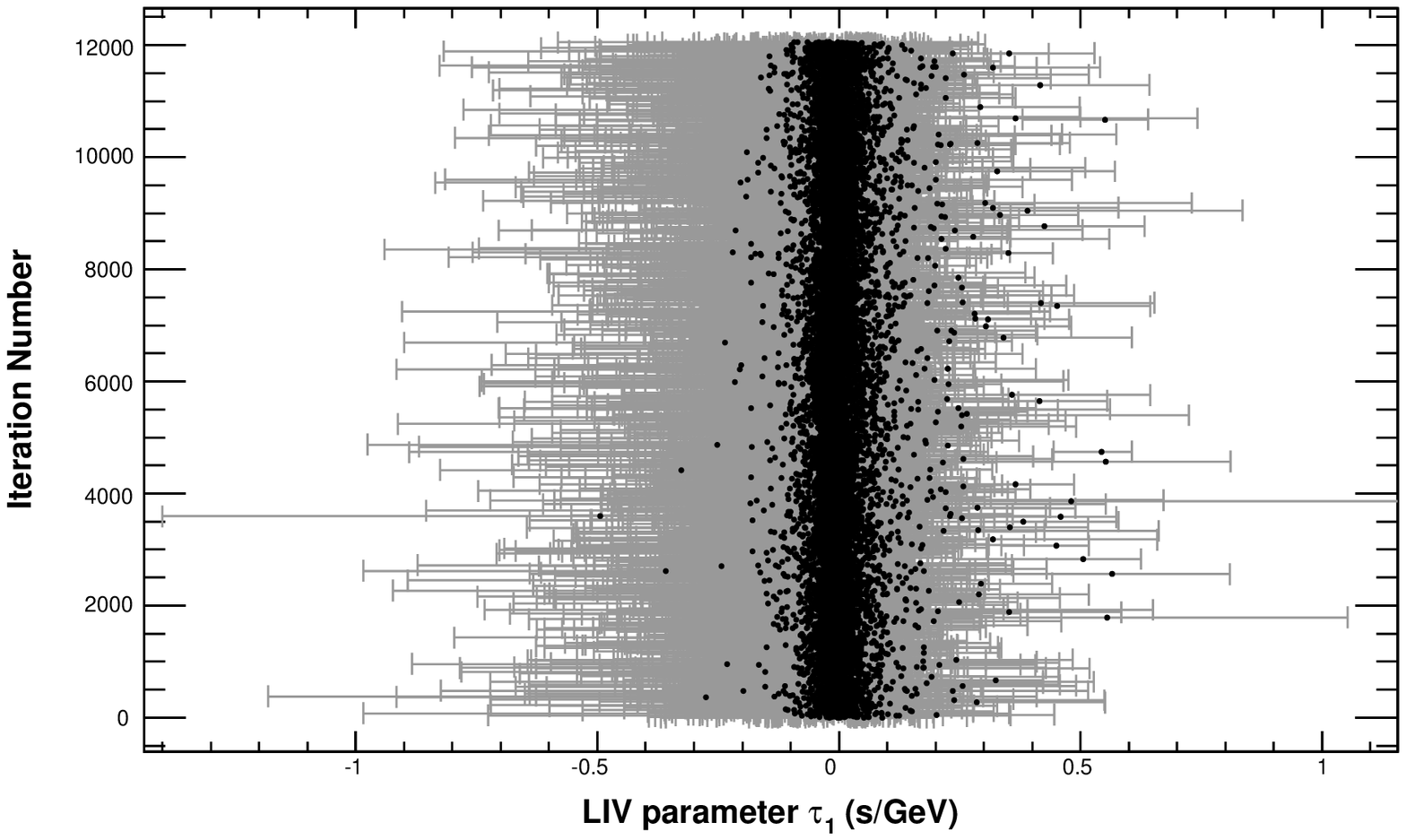}
\includegraphics[width=1\columnwidth,trim=20 0 50 30,clip=true]{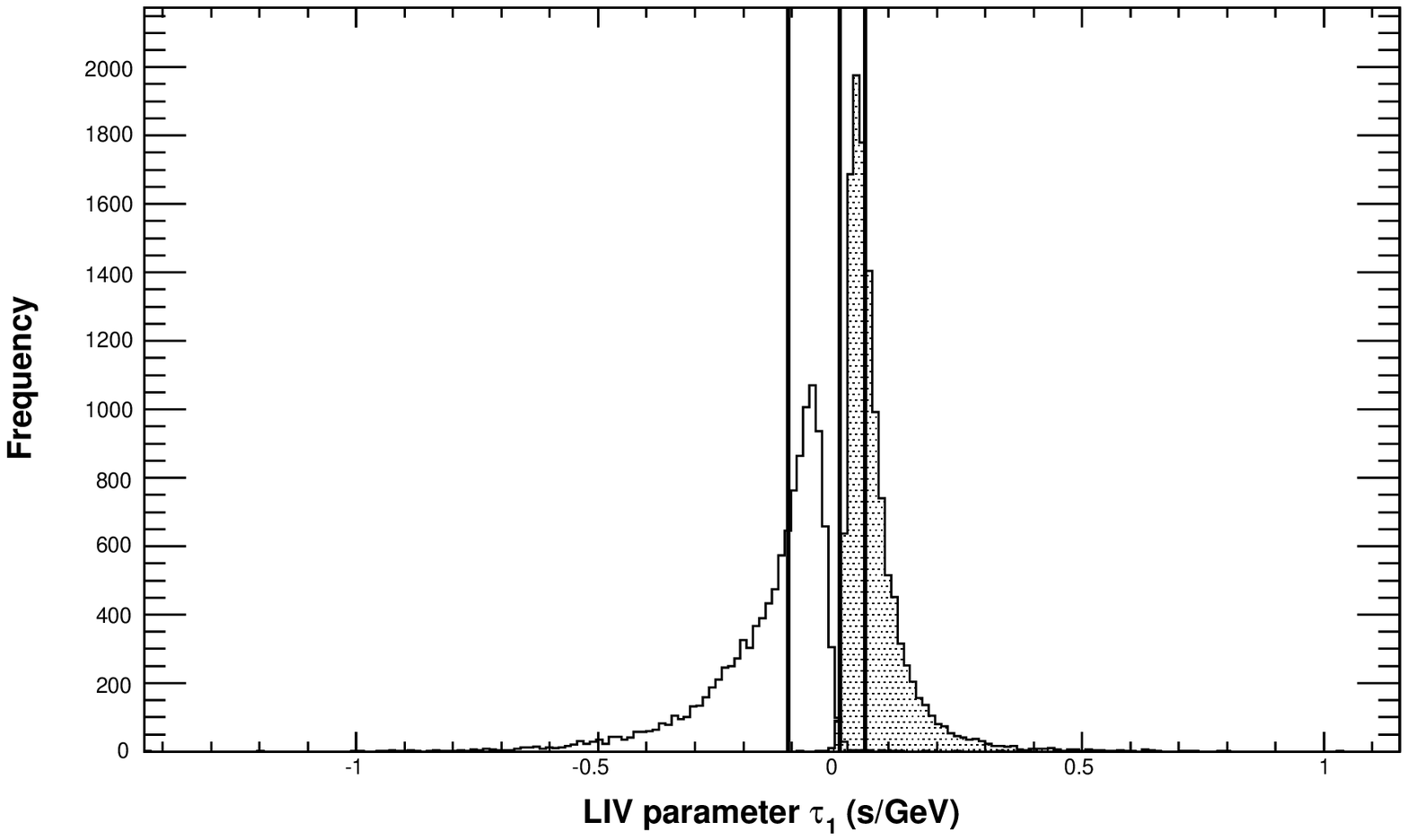}
\caption{\label{fig:PV_CIs}Top: 99\% CL CIs produced by the application of PV on the GRB~090510-inspired simulated data set for $\tau_1$=0~s/GeV. The black dots denote the best estimates of the LIV parameter. Bottom: distributions of the lower (left) and upper limits (right) of these confidence intervals. The two external vertical lines denote the medians of the two distributions, and the middle vertical line denotes the mean
of the best estimates.}
\end{figure}

One of the first steps after a collection of data sets was constructed was to examine its distribution of associated confidence intervals. The top
panel of Fig.~\ref{fig:PV_CIs} shows a stack of the confidence intervals produced by PV, and the bottom panel shows the distributions of lower and
upper limits corresponding to these confidence intervals. The two external vertical lines in the latter figure denote the distribution medians, and the middle vertical line shows the mean of the best estimates. By comparing the mean of the best estimates to the
actual LIV parameter we checked for the presence of biases in the best estimates.

\begin{figure}[ht!]
\includegraphics[width=1\columnwidth]{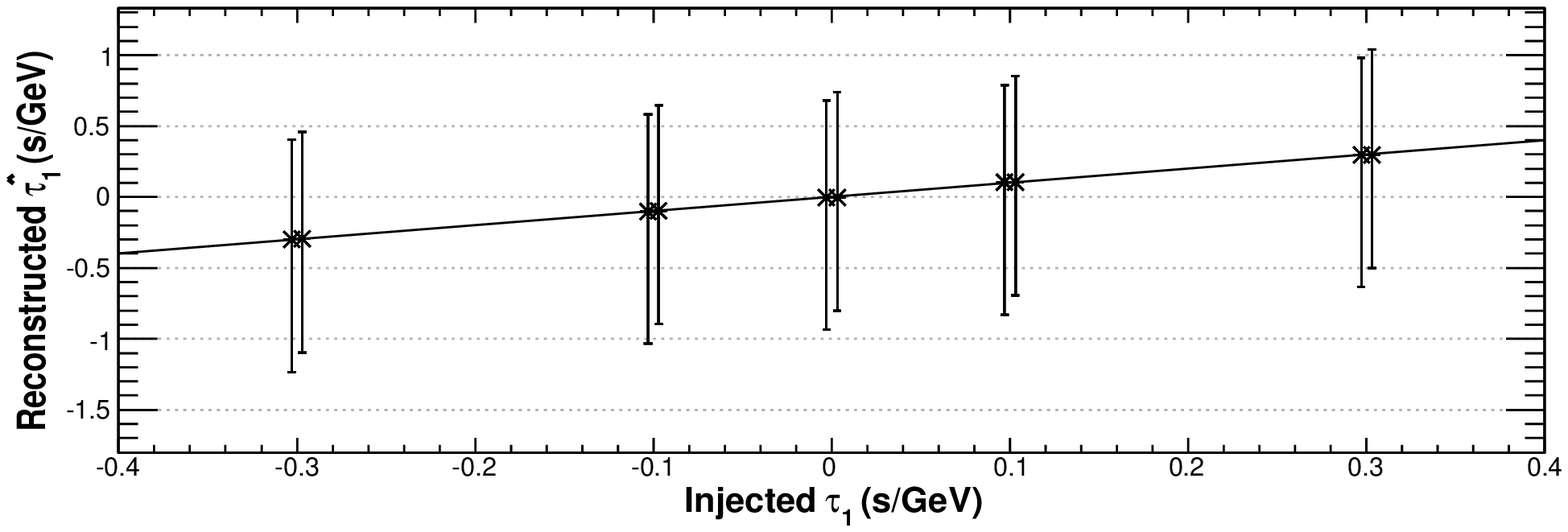}
\includegraphics[width=1\columnwidth]{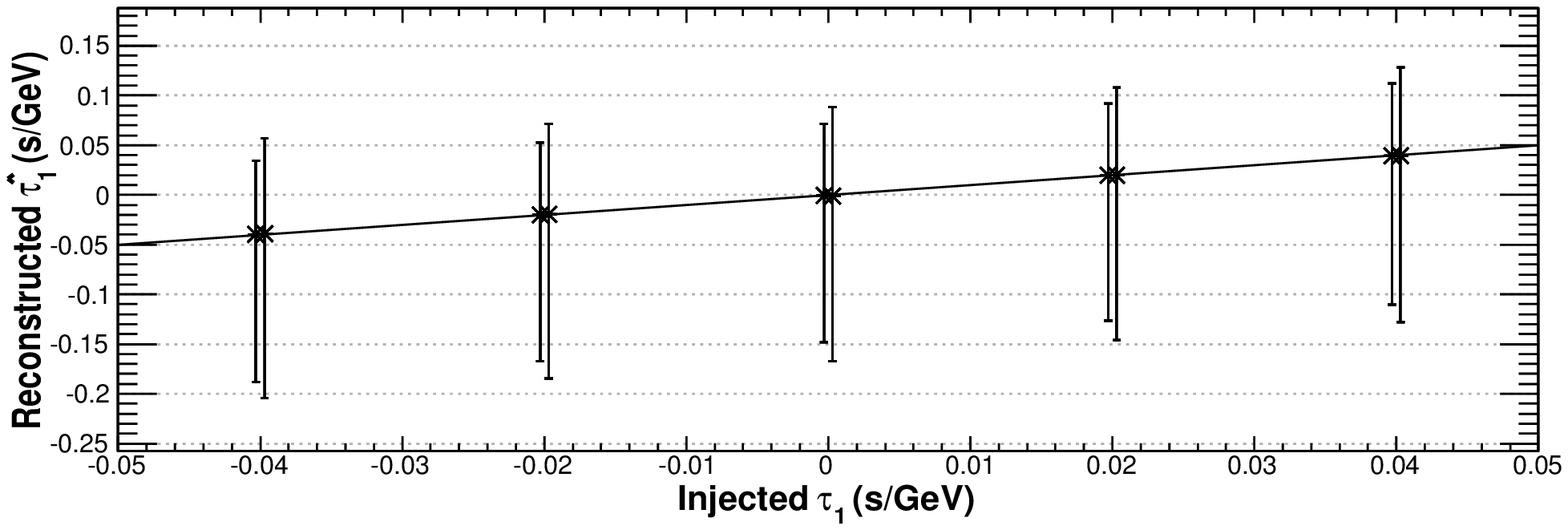}
\caption{\label{fig:Calibs}Calibration plots demonstrating some of the simulation results from the GRB~090902B- and 0900510-inspired data sets, top and bottom respectively. Each pair of intervals corresponds to a different value of the true (injected) LIV parameter $\tau_{1}$, and shows the results from PV (left interval) and SMM (right interval). The markers show the means of the best estimates, and the edges of the intervals correspond to the means of the upper and lower 99\% CL limits on $\tau_{1}$.}
\end{figure}

Figure~\ref{fig:Calibs} shows two calibration plots produced by our simulations. These plots show the average best estimate and upper/lower limits on the LIV parameter for different injected values of $\ttot$. As can be seen, the methods properly measure the injected value with negligible bias. Furthermore, their sensitivity does not have a considerable dependence on the injected degree of dispersion.

As mentioned in Sec.~\ref{subsec:PV_and_SMM}, the distribution $f_r$ is used as an approximation of the PDF of the measurement error of $\Err$, $P_{\Err}$. Since $\Err$ is a random variable (taking different values $\err$ across the simulated data sets), the quantity $C(\err)=\int_{0}^{\err}P_{\Err}(\tilde{\err}) d\tilde{\err}$ is also a random variable. $C(\err)$ behaves similarly to a p-value, hence, $C\sim U(0,1)$. We use the theoretical expectation of the uniformity of the PDF of $C$, to verify whether the distribution $f_r$ (produced using our randomization simulations described in Sec.~\ref{sec:pvsmm_conf}) is a good approximation of $P_{\Err}$, an approximation that is a cornerstone of our CI-construction procedure. If this is indeed valid, then the empirical distribution, $P_{C_{emp}}$, of the quantity $C_{emp}(\err_i)=\int_{0}^{\err_i}f_{r,i}({\tilde{\err}}) d\tilde{\err}$, where $\err_i$ and $f_{r,i}$ are the realizations of $\Err$ and $f_r$ in the i-th simulated data set, should also be distributed as a $U(0,1)$.

\begin{figure}[ht!]
\includegraphics[width=1.0\columnwidth]{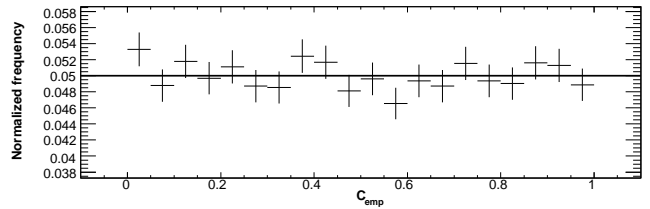}
\caption{\label{fig:aCDF}Normalized empirical distribution $P_{C_{emp}}$ (histogram) obtained from the application of PV on the GRB~090510-inspired data set for $tau_1=0$~s/GeV superimposed on its theoretical expectation (uniform distribution, shown with the horizontal line). The fact that the empirical distribution follows a uniform distribution supports the validity of the assumptions behind our CI construction.}
\end{figure}

Figure~\ref{fig:aCDF} shows a normalized version of $P_{C_{emp}}$ produced using the GRB~090510 inspired simulated data sets, PV, and $\tau_1=0$~s/GeV. As can be seen, the empirical distribution is indeed uniform, supporting the validity of our approximation $f_r\sim P_\Err$.

$P_{C_{emp}}$ is also used for verifying the coverage of the produced CIs, for any CL.\footnote{We also performed the simple test of counting the fraction of CIs of a collection of data sets that included the true (injected) value of the LIV parameter to verify that the fraction was, as expected, equal to their CL, for two different values of CL: 90\% and 99\%}. Since the quantiles of the $f_r$ distribution are used for constructing our CIs, any erroneous distortions of $f_r$ (and equivalently any deviations of $P_{C_{emp}}$ from uniformity) will be associated with an improper coverage of the CIs. For example, if the CIs were erroneously under-covering, then $P_{C_{emp}}$ would acquire a V shape. On the other hand, if the CIs were erroneously wide (over covering), then $P_{C_{emp}}$ would acquire a $\Lambda$ shape. By verifying the uniformity of $P_{C_{emp}}$ across its full range of values, we effectively tested the proper coverage of the CIs across the whole range of CLs (to the degree that the available statistics permitted).

Using the verification tests mentioned above, we also found that
\begin{itemize}
 \item  the sensitivities of both methods depend on the asymmetry on the light curve. Specifically, the longer the tail in the rising or falling side of the light curve is, the smaller the sensitivity of setting an upper or lower limit on $\ttot$, respectively, becomes. The coverage, however, remains proper even for highly asymmetric light curves (e.g., like the one shown in the bottom panel of Fig.~\ref{fig:lc_templates}).
\item Miscoverage and bias can increase and sensitivity can decrease if the light curve includes separated bright pulses, due most likely to some form of interference between the individual pulses. This systematic uncertainty becomes more prominent with the SMM method and when using large values of the $\rho$ parameter. Our default data selection always includes a single bright pulse, so this problem does not affect our results.
\item Bias and miscoverage is larger for strongly spectrally distorted light curves, i.e., those produced with a LIV parameter large enough that the highest-energy component is temporally disjoint from the bulk of the emission. The actual data did not appear to be spectrally distorted to the degree required for this systematic uncertainty to appear.
\item Tests performed on synthetic light curves comprising several pulses of a different spectral index (so as to simulate a GRB-intrinsic spectral evolution) revealed that this evolution is typically picked up by our methods as a non-zero LIV parameter. something that reflects perhaps the most important irreducible uncertainty in our results. It is, however, fortunate that the additional GRB-intrinsic spectral evolution did not always dominate the simulation results, and that while there may be some non-zero bias, the miscoverage of the CIs was typically not severe.
\end{itemize}

\section{\label{appendix:likelihood}Maximum Likelihood Method Tests and Calibrations}

\subsection*{Verification Tests}
We verified the ML method using Monte Carlo simulations, in a similar fashion to the PV/SMM methods, as described in the previous appendix. We performed tests on simple synthetic data sets as well as on data sets closely resembling the four GRBs in our sample. One of the main tests was the construction of calibration curves, in which we verified whether an injected LIV parameter was properly measured by the method with a reasonable degree of statistical accuracy.

\begin{figure}[ht!]
\includegraphics[width=1.0\columnwidth]{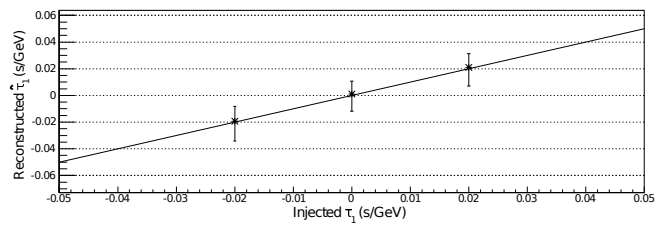}
\caption{\label{fig:090510_ML_calib}A calibration plot obtained by the ML method on an GRB~090510-inspired data set. The markers show the means of the best estimates of $\tau_1$ and the intervals correspond to the mean upper and lower 99\% CL limits on $\tau_{1}$.}
\end{figure}

As an illustration of these tests, we show in Fig.~\ref{fig:090510_ML_calib} a calibration plot demonstrating the simulation results from a GRB~0900510-inspired data set. The markers and the intervals show the average best estimates and 99\% CL CIs on $\tau_1$, respectively. These averages were calculated across the different simulated realizations of the GRB emission. Our tests did not reveal any significant biases or other systematics.

\subsection*{Calibrated Confidence Intervals}

We construct calibrated CIs on $\ttot$ by first generating several thousand simulated data sets having the same exact statistics as the actual data but with event energies and times randomly sampled from the fitted spectral and light-curve templates (e.g., such as from the templates shown in Figs.~\ref{080916C_spec_example} and \ref{fig:likelihood_templates}, respectively). We then apply the ML method to each one of them, using the same configuration as its application on the actual data, and calculate a CI and a best estimate on $\ttot$ for each one of them. After all of the data sets have been processed, we calculate the average of the produced low and upper limits. Since we do not apply any spectral dispersion to the simulated data sets, we shift the mean low and upper limit values by the value of $\tne$ as measured from the actually detected data set, to finally produce a single calibrated CI.\footnote{In this last step and for simplicity, we make the assumption that the sensitivity of the method has a small dependence on $\ttot$, at least for the small possible values $\ttot$ is expected to have (given past observations). Thus, we effectively assume that our simulating a zero-LIV-parameter data set and then offsetting the mean upper and lower limits is equivalent to simulating a $\tne$ LIV-parameter data set and constructing a calibrated CI directly from the mean lower and upper limits.}

The CIs are constructed using a pair of thresholds on ${-2\Delta \rm{ln}(\like})$ common to all the simulated data sets, and chosen to ensure the proper coverage of the produced CIs. Specifically, these thresholds are chosen so that exactly a fraction $(1-CL)/2$ of the simulated lower limits and a fraction $(1+CL)/2$ of the simulated upper limits are greater or smaller, respectively, than the value of $\ttot$ in the simulated data sets (equal by construction to zero).

The calibration procedure includes the re-fitting of a light-curve template for each simulated sample. Thus, the produced CIs properly include the systematic uncertainties arising from the light-curve template generation procedure. On the other hand, for computational simplicity, we do not refit a spectral template, and instead use the one obtained from the actual data. Thus, the calibrated CIs do not include uncertainties from the spectral fit. These are, however, negligible, since, as we have seen from toy Monte Carlo simulations and from the calculations described in Sec.~\ref{subsec_Likelihood}, the final results depend weakly on the spectral index, contrary to their stronger dependence on the light-curve template.

\begin{figure}[ht]
\includegraphics[width=0.5\columnwidth]{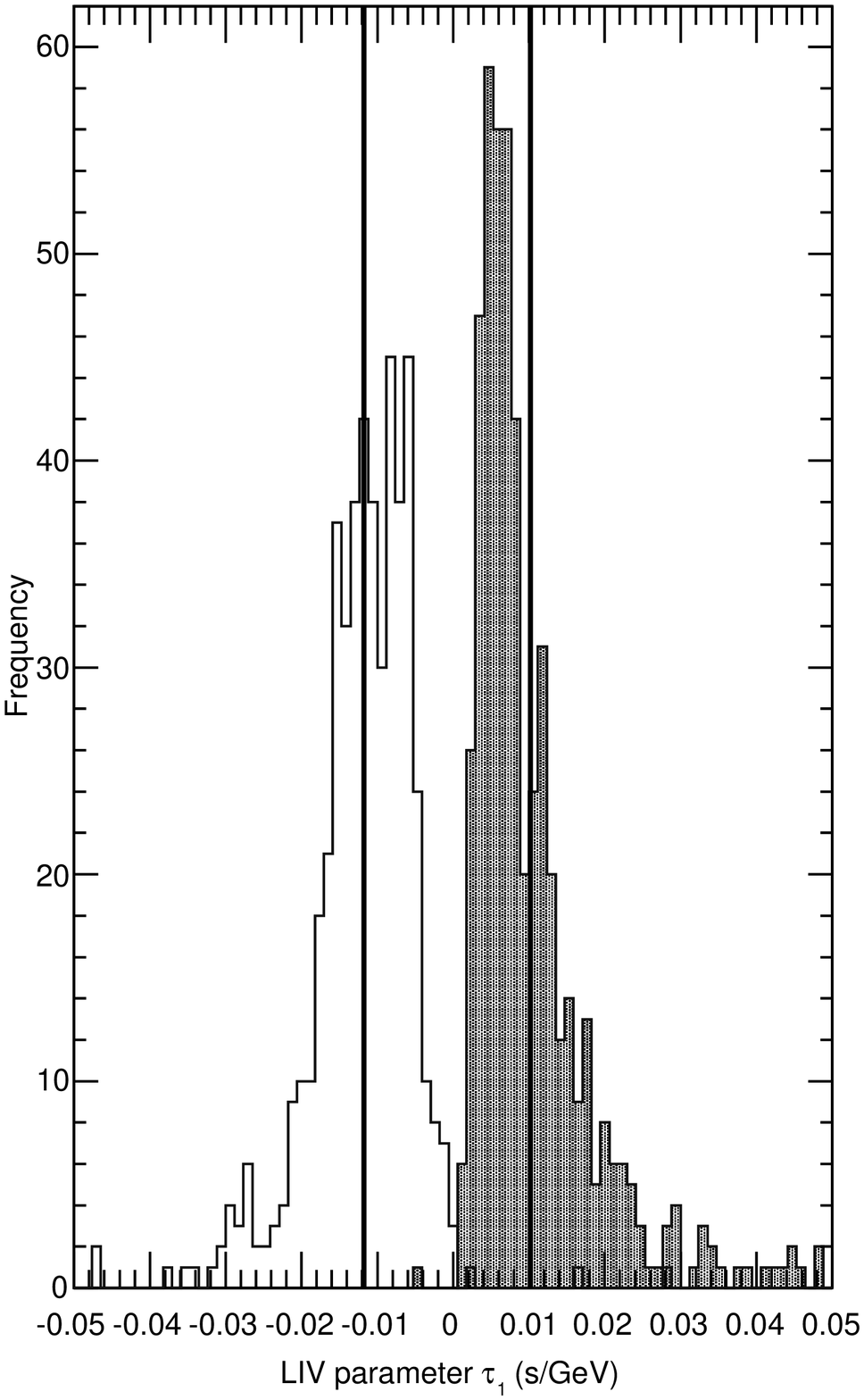}\includegraphics[width=0.5\columnwidth]{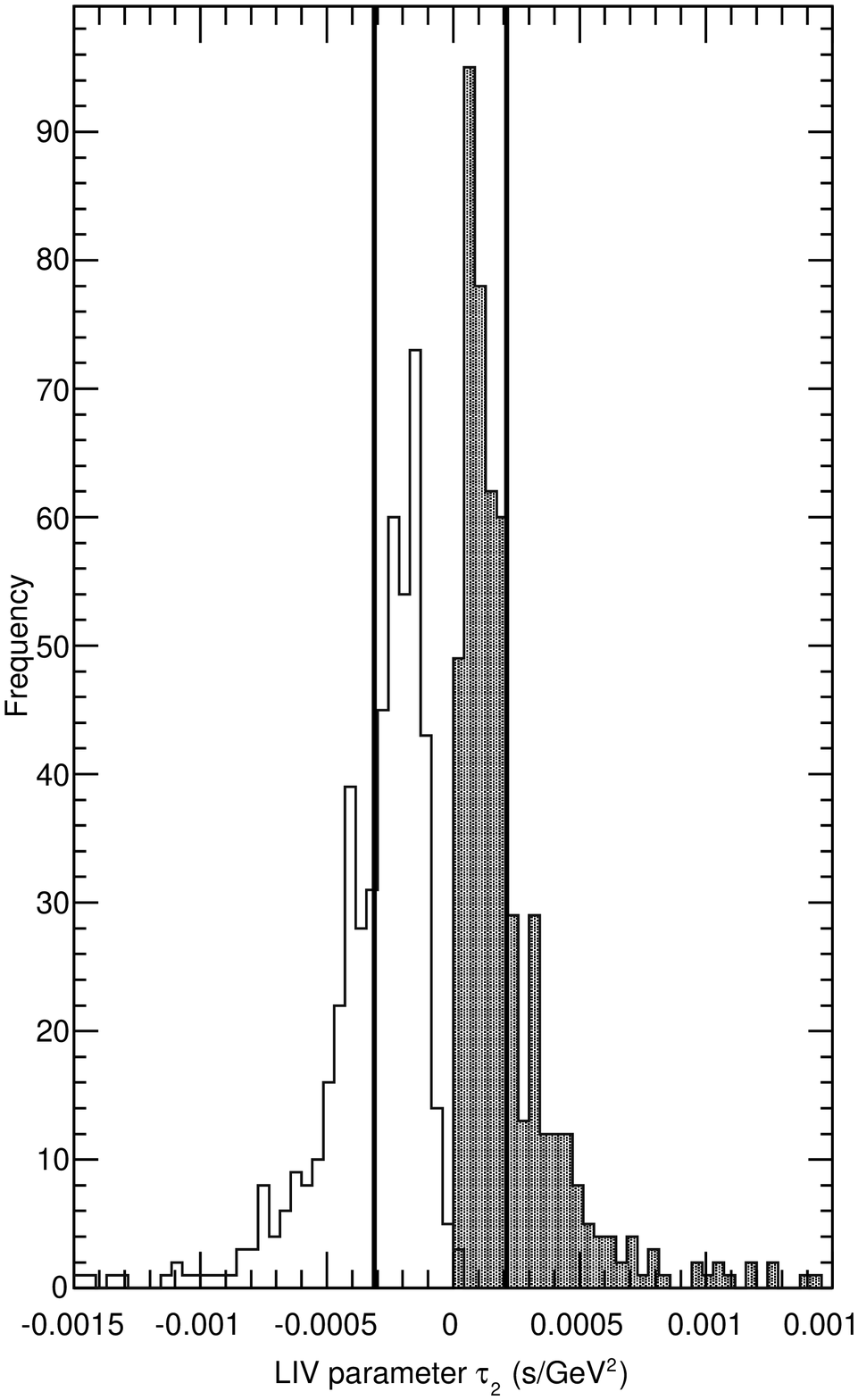}
\caption{\label{fig:calib_like_randomized_limits}Distributions of the lower and upper 99\% CL limits for $n=1$ (left pad) and $n=2$ (right pad) for GRB~090510. The vertical lines denote the means of the distributions, used for constructing the calibrated CIs.}
\end{figure}
\begin{figure}[ht]
\includegraphics[width=0.5\columnwidth]{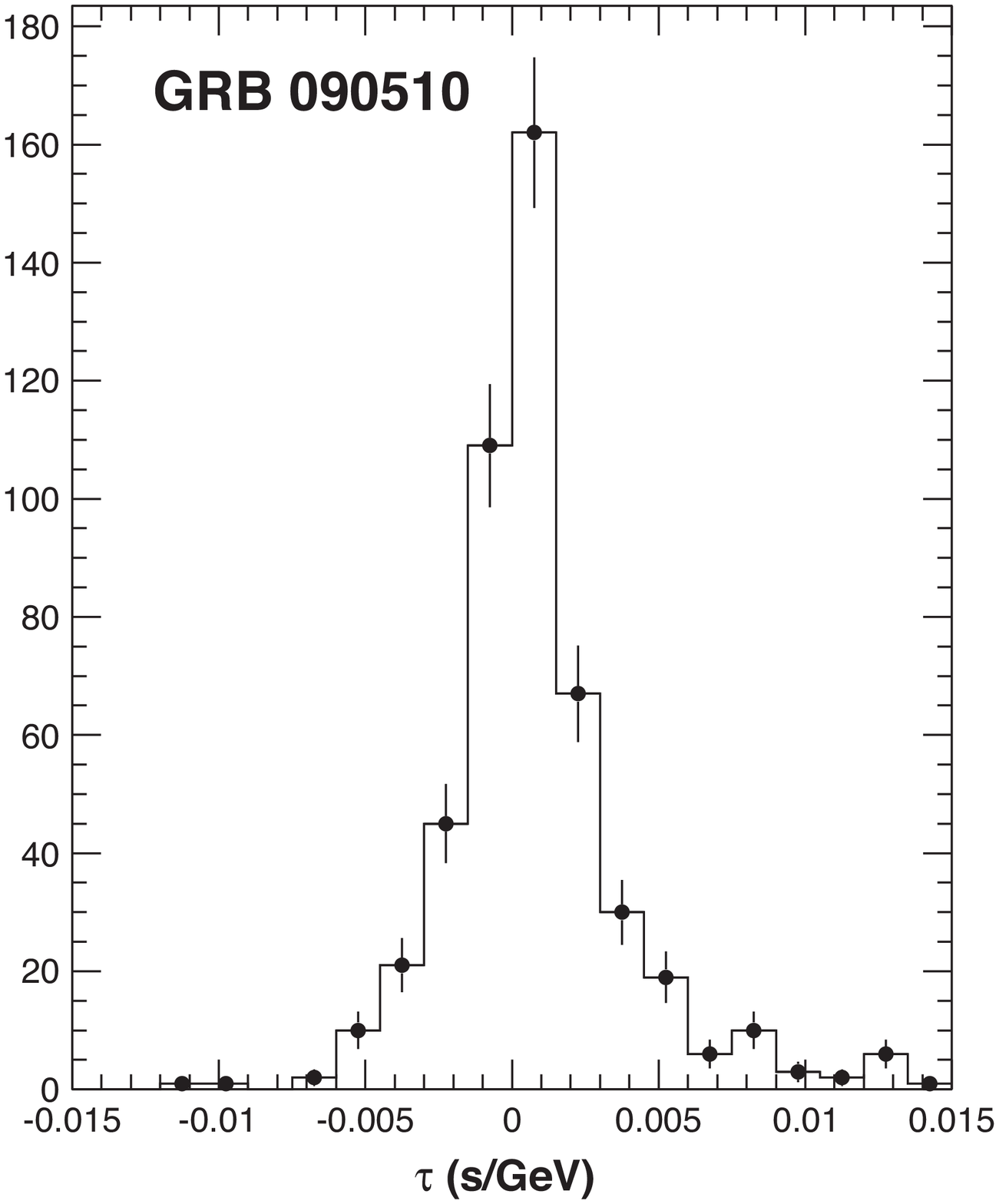}\includegraphics[width=0.5\columnwidth]{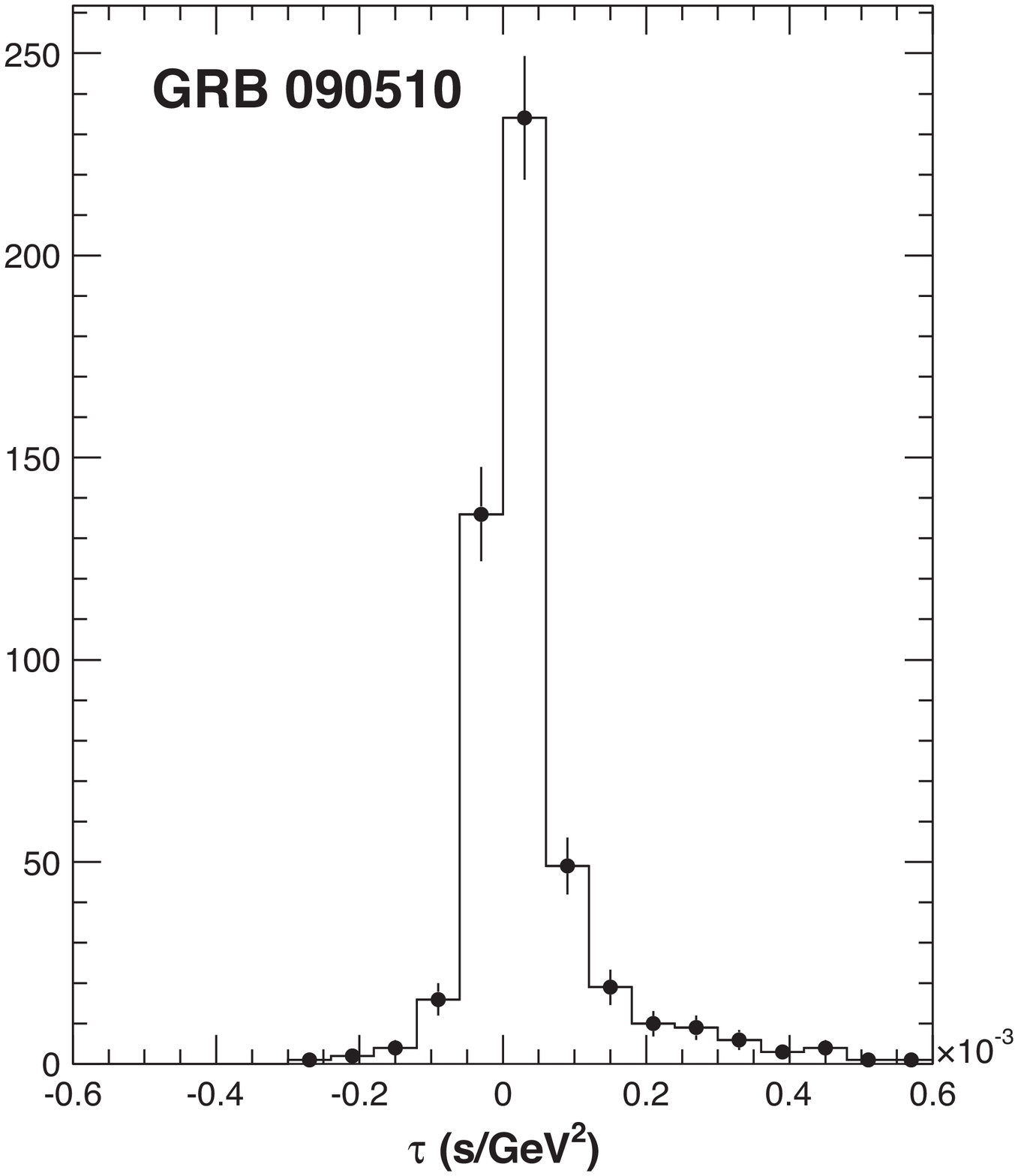}
\caption{\label{fig:calib_like_randomized_best_lags}Distributions of the best LIV parameters obtained from each simulated data set for $n=1$ (left) and $n=2$ (right) for GRB~090510. Since there was no spectral dispersion applied to the simulated data, these curves should peak near zero.}
\end{figure}

To illustrate the method, we show some intermediate results from its GRB~090510 application. Figure~\ref{fig:calib_like_randomized_limits} shows the distributions of low and upper limits obtained from the simulated data sets. The mean values of these distributions are offset by $\tne$ to produce our single calibrated upper and lower limits (i.e., those shown in the last column of Tab.~\ref{tab_results_per_grb}). From the mean of the simulated best estimates of the LIV parameter (see, e.g., Fig.~\ref{fig:calib_like_randomized_best_lags}) we estimated the bias of $\tne$. In all cases, the bias was negligible with respect to the root mean square of the simulated best estimates.
%

%

\section{\label{appendix:comparison}Comparison of the Methods}
We compared the three methods by applying them to the same collection of simulated data sets to verify their validity and to help us explain any discrepancies observed in their application on the actual data. The simulated data of this test were produced using the GRB~090510-inspired light-curve template shown in Fig.~\ref{fig:lc_templates} of Appendix~\ref{appendix:PVSMM} and no extra applied dispersion ($\ttot$ was zero by construction).

For the ML method we used CIs calculated directly from the data (instead of from calibration simulations). However, we adjusted the two threshold values of -2$\Delta \rm{ln}(\like)$ used to produce its lower and upper limits, to ensure a proper coverage (evaluated across the simulated data set).

\begin{figure*}[ht!]
\includegraphics[width=0.33\textwidth]{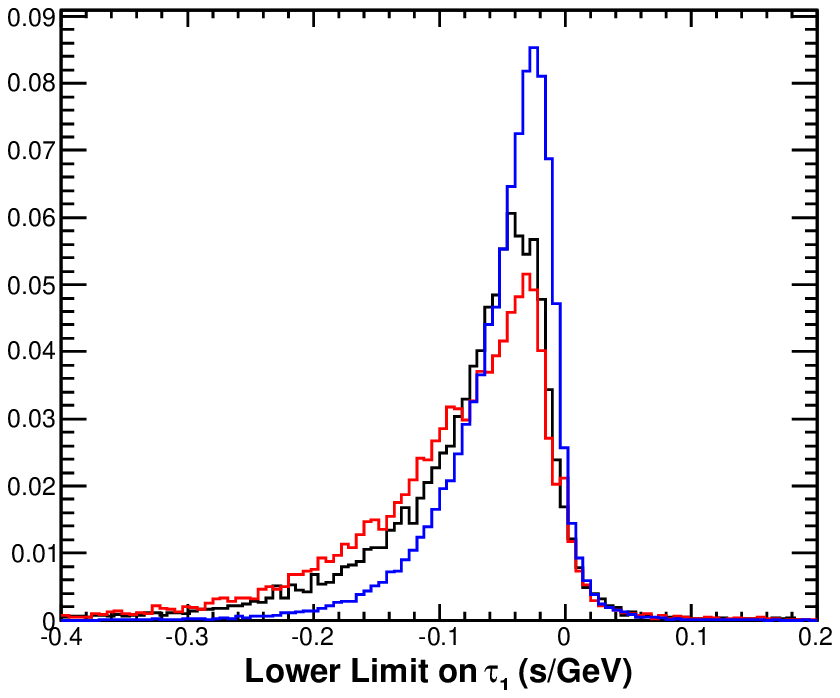}\includegraphics[width=0.33\textwidth]{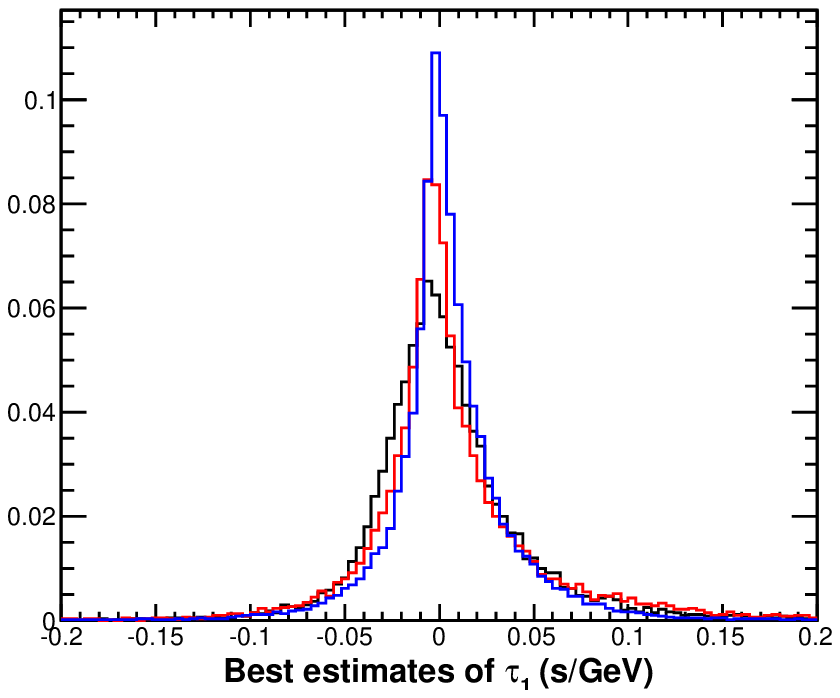}\includegraphics[width=0.33\textwidth]{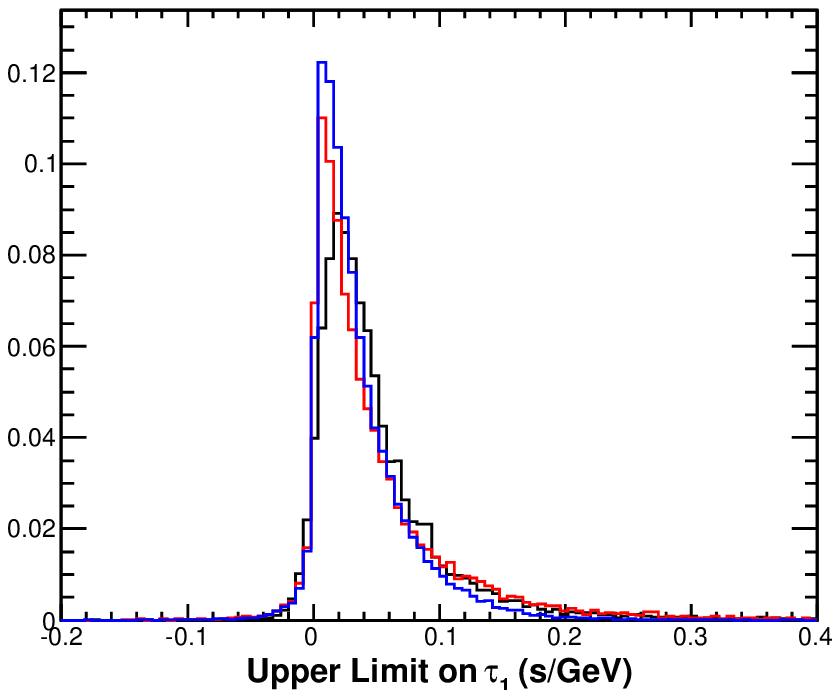}
\caption{\label{fig:comparison_1d} Histograms of the 95\% (one-sided) CL lower limits (left panel), best estimates (middle panel), and upper limits (right panel) of the LIV parameter $\tau_1$ produced by PV (black), SMM (red), and ML (blue) on simulated data sets.}
\end{figure*}

The first and third panel of Fig.~\ref{fig:comparison_1d} show the obtained distributions of lower and upper limits, respectively. As can be seen, the sensitivities of the three methods are very similar. In the first panel, we also see that the ML method is slightly more sensitive when producing lower limits. We used this finding to explain in Sec.~\ref{sec:Results} why the ML method produced more constraining limits than the other two methods on $\tau_1$ and GRB~090510.

The histograms of the best estimates of the LIV parameter (middle panel of Fig.~\ref{fig:comparison_1d}) peak, as expected, near the true value of the LIV parameter, set equal to zero. The PV and SMM best-lag distributions peak at slightly more negative values than the approximately zero position of the ML method's distribution. This can be attributed to the increased asymmetry of the PV and SMM distributions (skewness $\sim$0.75) compared to the asymmetry of the likelihood distribution (skewness $\sim$0.39), which moves the mode to lower values than the mean or the median. However, for considerations regarding the bias of the best estimates, the important fact is that both the median and the mean of these distributions are negligible compared to their root mean square. Thus, the effect of any biases on the coverage of the produced CIs is expected to be negligible (as has been verified by the dedicated simulation tests).

\begin{figure*}[ht!]
\includegraphics[width=0.33\textwidth]{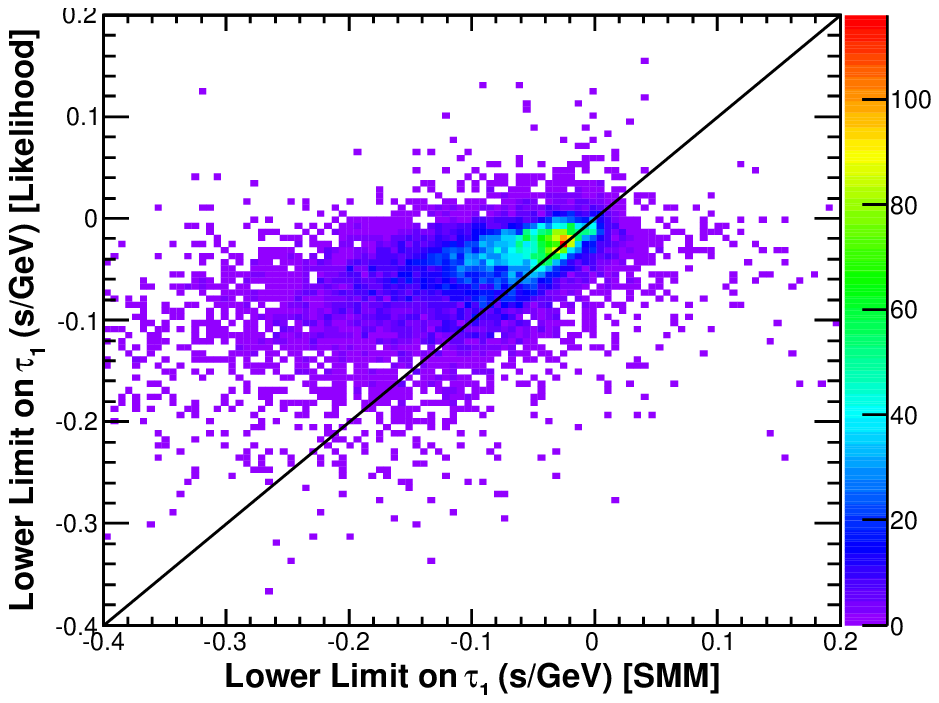}\includegraphics[width=0.33\textwidth]{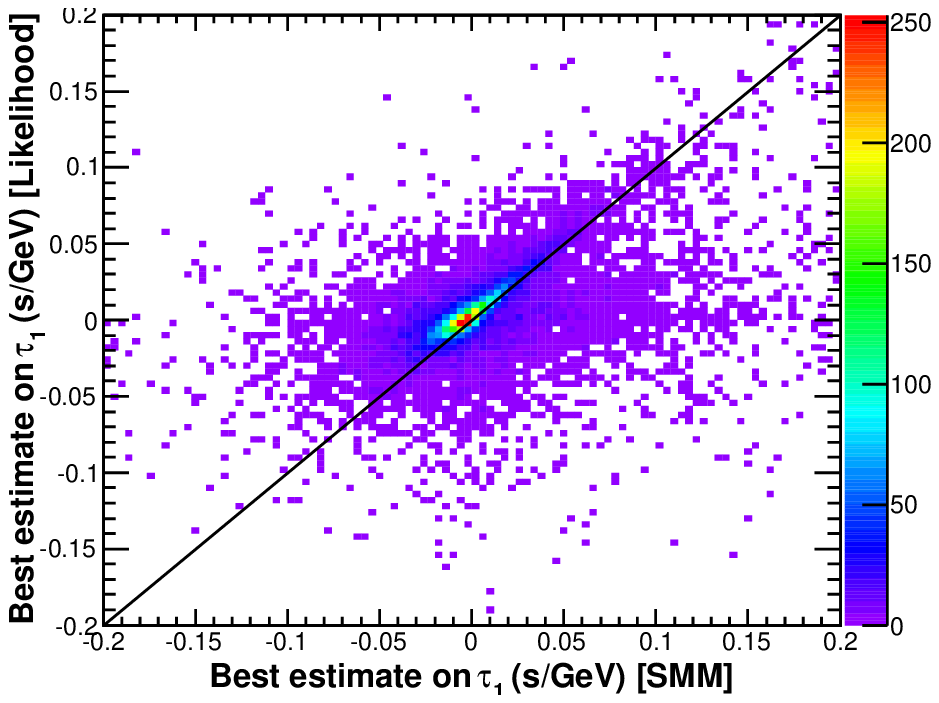}\includegraphics[width=0.33\textwidth]{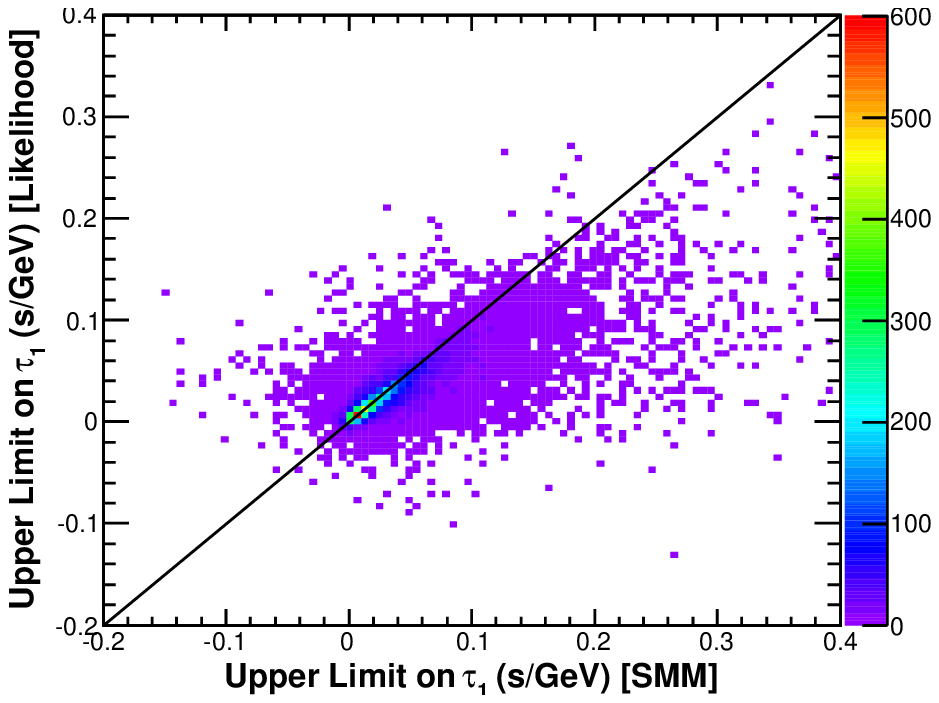}

\includegraphics[width=0.33\textwidth]{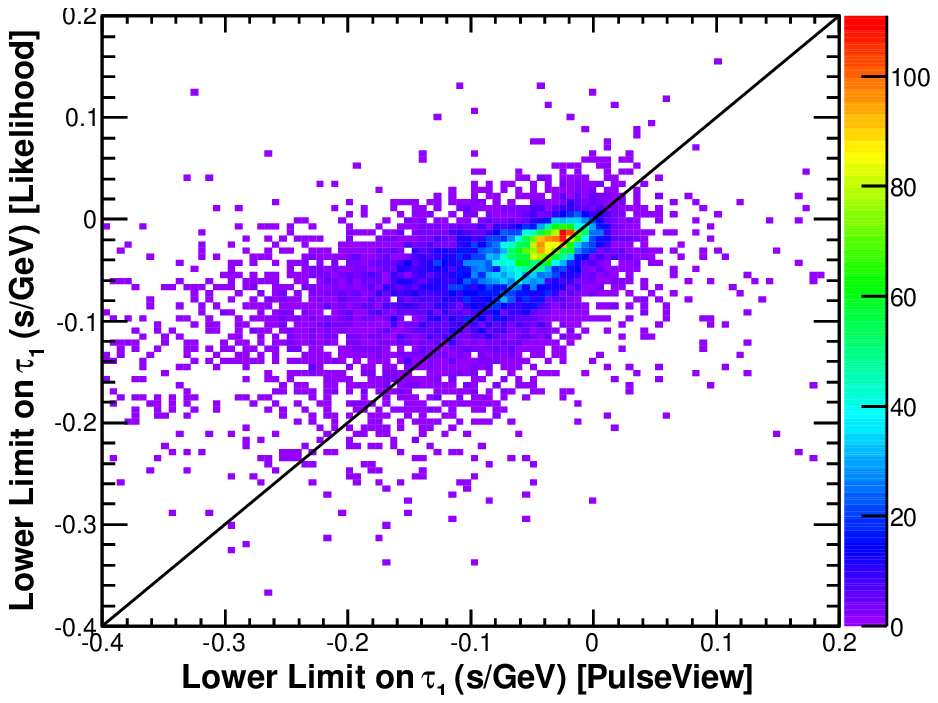}\includegraphics[width=0.33\textwidth]{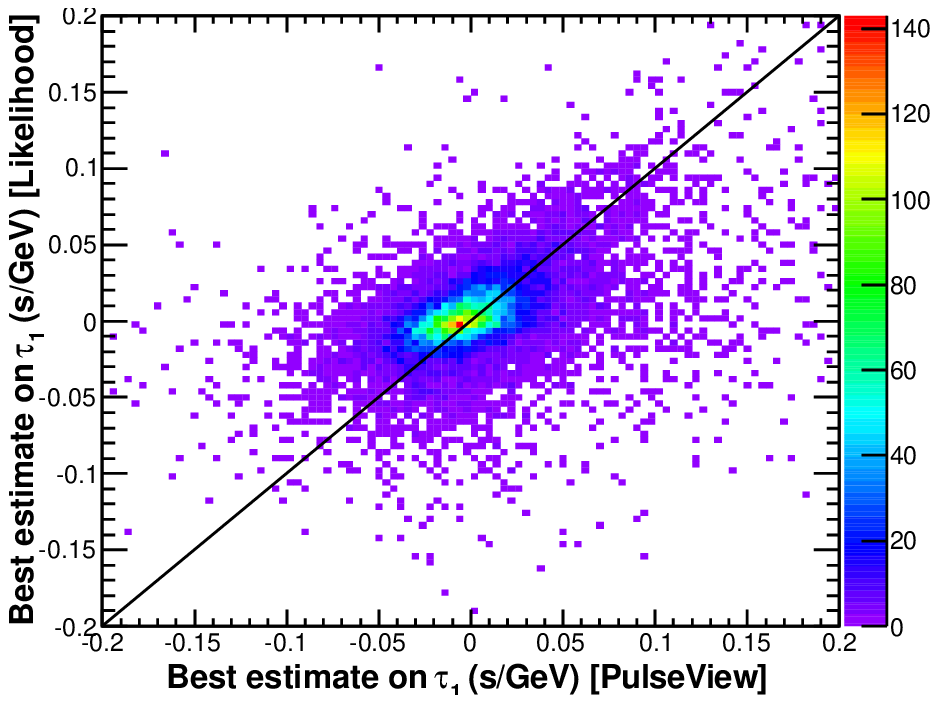}\includegraphics[width=0.33\textwidth]{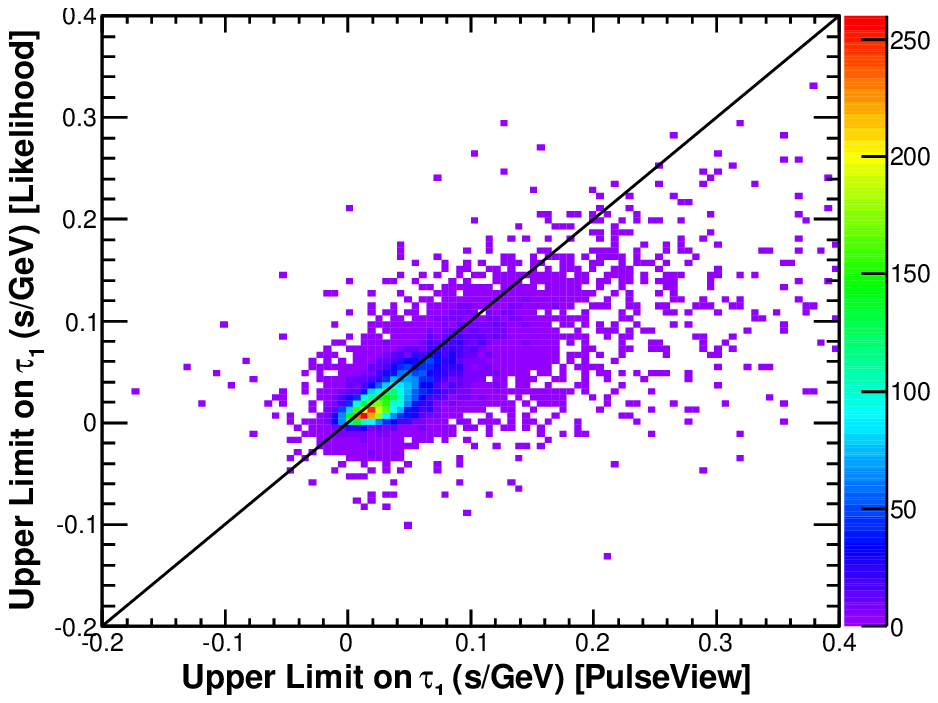}

\includegraphics[width=0.33\textwidth]{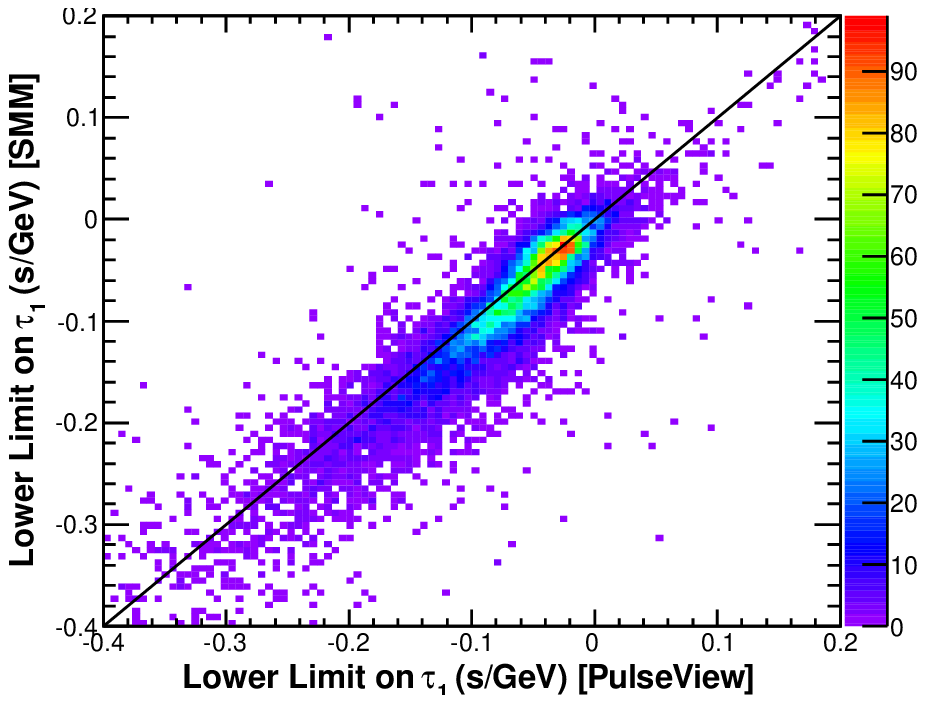}\includegraphics[width=0.33\textwidth]{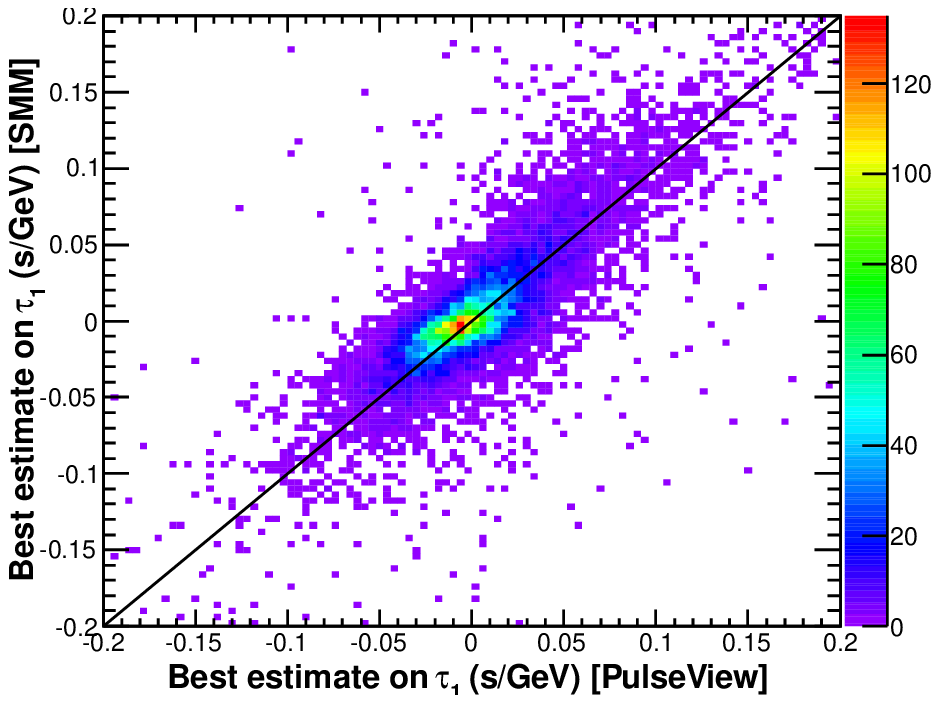}\includegraphics[width=0.33\textwidth]{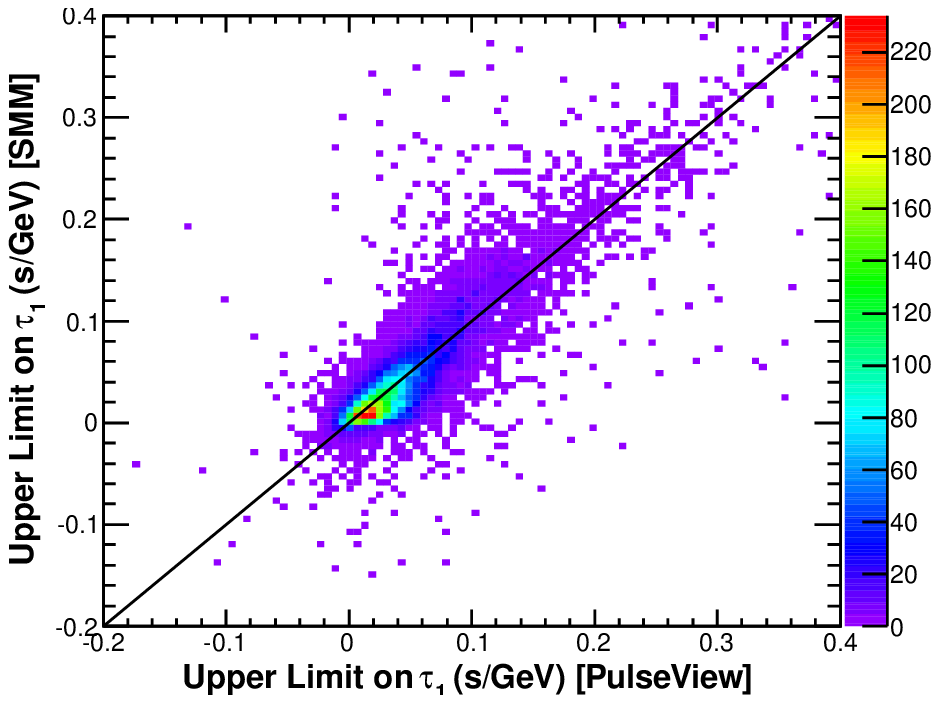}

\caption{\label{fig:comparison_2d} One to one comparisons of the 95\% (one-sided) CL lower and upper limits (left and right columns respectively) and of the best estimates (middle column) on the LIV parameter $\tau_1$ produced by our three methods on simulated data sets: ML vs SMM (top row), ML vs PV (middle row),  SMM vs PV (bottom row).}
 \end{figure*}

The 2D histograms in Fig.~\ref{fig:comparison_2d} provide a deeper view of how our methods compare. For the majority of the simulated data
sets, there is a close correspondence between their results. The PV and SMM results are the most similar, implying that these two methods probe the data in a similar fashion. The existence of differences between the methods' results highlights their complementarity.

Finally, we note that more than 99\% of the examined triplets of 90\% CL CIs (one per simulated data set) are overlapping. This fraction is even larger (more than 99.9\%), if CIs of a higher CL (99\%) are examined (not shown here). This large fraction of overlapping CIs shows that the troubling possibility of the three methods not allowing a common part of the parameter space is extremely unlikely.

\section{\label{appendix:Checks}Analysis Cross-Checks}

We examined how the 99\% CIs on $\ttot$ vary with respect to changes in the configuration of our methods and the data selection, to cross-check the validity and robustness of the results, and to gain insight on the behavior of our methods. Specifically:
\begin{itemize}
 \item we repeated the analysis excluding the highest-energy photon in the data, since it is expected to provide the most information about LIV dispersion.
\item We applied our methods on an extended time interval extending from the GRB trigger up to the time that the temporal variability has considerably subsided. The time intervals, selected with visual inspection, are 0--20~s for GRB~080916C, -0.01--10~s for GRB~090510, 0--60~s for GRB~090902B, and 0--40~s for GRB~090926A. The extended time intervals allow for maximal statistics, but at the same time potentially include a large degree of GRB-intrinsic spectral evolution that can, however, masquerade as LIV dispersion.
\item we repeated the PV and SMM analyses using data produced using an earlier version of LAT's event selection, P6\_V3\_TRANSIENT~\cite{2009ApJ...697.1071A}, also used by \textit{Fermi} to constrain LIV~\cite{2009Sci...323.1688A,2009Natur.462..331A}.
\item Finally, we repeated the PV and SMM analyses starting from 30~MeV instead of their default 100~MeV. While this change corresponds to increased statistics, it comes, however, with a larger contamination from the Band spectral component, which can increase the GRB-intrinsic systematic uncertainties.
\end{itemize}

For the case of ML, we do not expect the calibrated CIs to vary in a considerably different way than the CIs obtained directly from the data, during the tests mentioned above. Thus, for simplicity, we only present ML results obtained directly from the data.

\begin{figure}[ht!]
\includegraphics[width=1.0\columnwidth]{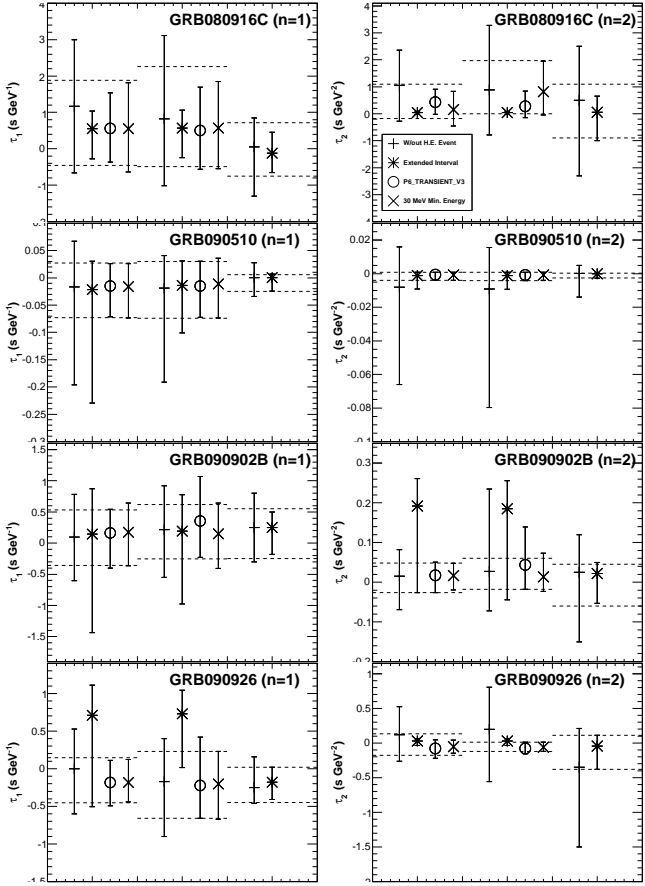}
\caption{\label{fig:robustness} Comparison between the 99\% CL CIs on $\ttot$ obtained with the default configuration (horizontal dashed lines) to those obtained with modified configurations or data sets (vertical error bars). There are three groups of results in each figure, each corresponding to a different method: PV, SMM, ML (from left to right). The ML method CIs shown were produced directly from the data, instead using calibration simulations.}
\end{figure}

The test results are shown in Fig.~\ref{fig:robustness}. In all cases, the CIs produced by different methods (and for the same test) are in agreement with each other (i.e., they have some overlap). Their widths and centers do change somewhat across tests, something expected considering the different statistics and degrees of GRB-intrinsic spectral evolution in the different data sets.

The removal of the highest-energy event, as expected, widened the produced CIs. The magnitude of the increase in their widths is a probe for the degree with which our methods draw information from the single highest-energy event and also for the systematic uncertainty associated with the possibility of that event being background. Because of the very low background contamination in our data sets, the highest-energy photons are typically securely associated to the GRB (see, e.g., the Supplementary Information of Ref.~\cite{2009Natur.462..331A} regarding the association of the 31~GeV photon to GRB~090510). Thus, we do not consider the option of removing the highest-energy photon to increase the robustness of the results warranted.

The changes brought by the use of the extended time interval did not correspond to a specific pattern. They were likely caused by the inclusion of emission of energy considerably higher than that included in the default time interval or the inclusion of significantly more GRB-intrinsic spectral evolution (likely in the case of GRB~090926A). Perhaps the most significant change happened with GRB~080916C and $n=2$ on the PV and SMM results. For this case, the extended interval included at 13~GeV photon detected $\sim$16.54~s post-trigger, which had an almost a decade in energy higher than that of the rest of the photons. As such, it dominated the PV/SMM estimation procedures with the edges of the confidence intervals on $\tau_2$ roughly corresponding to the time difference between its detection time and the edge of the analyzed interval divided by the square of its energy.
The case of GRB~090926A is likely affected by both the inclusion of a very energetic event ($\sim$7 times higher energy than the rest of the events) and the strong spectral evolution observed throughout this burst's emission. We observe that the choice of time interval can significantly affect the final results, and conclude that an \textit{a priori} and carefully chosen selection for the time interval, as in this work, is important for the validity of the results.

Repeating the analysis with the P6\_V3\_TRANSIENT data set did not bring any considerable changes to the produced CIs, supporting the case that the improved limits produced in this work, when compared to past \textit{Fermi} analyses of GRB~090510~\cite{2009Natur.462..331A}, are a result of more sensitive analysis techniques rather than of a more constraining data set.

Finally, repeating the PV/SMM analyses starting from a lower minimum energy (30~MeV) did change the results significantly, implying that the systematic effects induced by the presence of two spectral components in the data are limited.

We also repeated the ML analysis for GRBs~080916C and 090510 after performing some configurational changes affecting the light curve parametrization, such as using asymmetric (instead of symmetric) Gaussian pulses, a larger number of Gaussian pulses, different bin widths for the histogram used in producing the template (e.g., such as the one shown in Fig.~\ref{fig:likelihood_templates}), or different $E_{\rm cut}$ values to split the data. The CIs varied up to a factor of $\sim$2 with respect to CIs obtained with the default configuration.

\clearpage
\bibliographystyle{apsrev4-1}
\bibliography{mnemonic,refs}

\end{document}